\documentclass[10pt,superscriptaddress]{revtex4}

\usepackage{setspace}
\usepackage{latexsym}
\usepackage{graphicx}
\usepackage{rotating}
\usepackage{amsfonts}
\usepackage{amsmath}
\usepackage{amssymb}
\usepackage{braket}
\usepackage{bm}
\usepackage{color}

\newcommand{\der}[2]{\frac{d #1}{d #2}}

\begin{document}

\title[Four-Body Problems]{The Hyperspherical Four-Fermion Problem}

\begin{abstract}
 The problem of a few interacting fermions in quantum physics has
 sparked intense interest, particularly in recent years owing to
 connections with the behavior of superconductors, fermionic
 superfluids, and finite nuclei.  This review addresses recent
 developments in the theoretical description of four fermions having
 finite-range interactions, stressing insights that have emerged from
 a hyperspherical coordinate perspective. The subject is complicated,
 so we have included many detailed formulas that will hopefully make
 these methods accessible to others interested in using them. The
 universality regime, where the dominant length scale in the problem
 is the two-body scattering length, is particularly stressed,
 including its implications for the famous BCS-BEC crossover problem.
 Derivations and relevant formulas are also included for the
 calculation of challenging few-body processes such as recombination.
\end{abstract}

\author{Seth T. Rittenhouse}
\affiliation{Department of Physics and JILA, University of Colorado,
Boulder CO, 80309-0440}
\affiliation{ITAMP, Harvard-Smithsonian Center for Astrophysics, Cambridge, MA 02138}
\author{J. von Stecher}
\affiliation{Department of Physics and JILA, University of Colorado,
Boulder CO, 80309-0440}
\author{J. P. D'Incao}
\affiliation{Department of Physics and JILA, University of Colorado,
Boulder CO, 80309-0440}
\author{N. P. Mehta}
\affiliation{Department of Physics, Grinnell College, Grinnell, IA 50112}
\affiliation{Department of Physics and JILA, University of Colorado,
Boulder CO, 80309-0440}
\author{Chris H. Greene}
\affiliation{Department of Physics and JILA, University of Colorado,
Boulder CO, 80309-0440}


\maketitle

\tableofcontents

\section{Introduction}

The problem of four interacting particles in nonrelativistic quantum
mechanics arises in a number of different physical and chemical
contexts.\cite{HuSchatz2006,ClaryScience1998,Wigner1933,BetheBacher1936}
While tremendous theoretical progress has been achieved in the
three-body problem,\cite{Lin1986review, Fano1983,
  SolovevTolstikhin2001, TolstikhinWatanabeMatsuzawa1996,
  ArcherParkerPack1990, EsryGreeneBurke1999, nielsen1999ler,
  Braaten2006physrep, ParkerWalkerKendrickPack2002,
  QuemenerLaunayHonvault2007, HonvaultLaunay2001} particularly in the
past two decades, the four-body problem remains still in its infancy
by comparison.  Like the three-body problem, the four-body problem
consists of two qualitatively different subcategories, one in which
some of the particles have Coulombic
interactions,\cite{YangXiBaoLin1995,YangBaoLin1996,MorishitaTolstikhinWatanabeMatsuzawa1997,MorishitaLin1998,MorishitaLin1999,DIncao2003,MorishitaLin2003,Madsen2003}
and the other subcategory in which all forces between particles have a
finite range or else have at most a rapidly-decaying multipole
interaction at
long-range.\cite{ClaryScience1998,HuSchatz2006,BrooksClary1990,Clary1991,Clary1992,QuemenerGoulvenBalakrishnanNaduvalath2009,ZhangLight1997,RichardBarnea2009}
The subject of the present review concerns the latter category, which
is particularly relevant to modern day studies of ultracold quantum
gases composed of neutral atoms and/or molecules.  The scope of this
subject is much broader than that of ultracold gases alone, however,
as 4-body reactive processes such as AB+CD$\rightarrow$AC+BD, or
$\rightarrow$A+BCD, or $\rightarrow$A+B+C+D occur in nuclear and
high-energy physics as well as in chemical physics.   The time-reverse
of these processes is also important for understanding the loss rate
in a degenerate quantum gas, notably the process of four-body
recombination which had hardly received any attention until very
recently.

While of course many important advances have been achieved in few-body
physics without the use of hyperspherical coordinates, treatments
using these coordinates have real advantages for a number of
problems. Early on, for instance, Thomas \cite{Thomas1935} proved an
important theorem about the nonzero range of nucleon-nucleon forces,
using an analysis in which the hyperradial coordinate played a crucial
role although he did not refer to it by that name. (See, for instance,
Eq.111c of \cite{BetheBacher1936}.) Further developments in the use of
hyperspherical coordinates in collision problems were pioneered by
Delves\cite{delves1959tag,delves1960tag} and they played a key role in
the derivation of the Efimov effect\cite{ efimov1970ela,
  efimov1971wbs} As we will see below, the advantages accrue not only
in terms of computational efficiency, but also in terms of the
insights and quasi-analytical formulas that can be deduced for
scattering, bound, and resonance properties of the system.  For this
reason, the present review concentrates on the hyperspherical studies
of the four-body problem, concentrating on recent progress and results
that have emerged, and on problems that currently seem ripe for
pursuit in the near future.

In early studies \cite{Fock1958,Klar1985}, hyperspherical coordinates
were viewed as capable of providing a deeper understanding of the
nature of exact bound state solutions, for instance for the helium
atom \cite{Forrey2004}.  And Delves\cite{delves1959tag,delves1960tag}
used these coordinates to discuss rearrangement nuclear collisions
from a formal perspective. But a turning point in the utility of
hyperspherical coordinate methods was introduced by Macek in
1968\cite{MacekJPB1968}, in the form of two related tools: the {\it
  adiabatic hyperspherical approximation} and the (in principle exact)
{\it adiabatic hyperspherical representation}.  Both of these methods
single out a single collective coordinate for special treatment, the
hyperradius $R$ of the $N$-body system, which is handled differently
from all remaining space and spin coordinates, $\Omega$.  The
hyperradius is a positive ``overall size coordinate'' of the system,
whose square is proportional to the total moment of inertia of the
system, i.e. $R^2= \frac{1}{M}\Sigma_i m_i r_i^2$, where $m_i$ is the
mass of the $i$-th particle at a distance $r_i$ from the center of
mass, and $M$ is any characteristic mass which can be chosen with some
arbitrariness.\cite{Fano1981}.

In Macek's adiabatic {\it approximation}, the Hamiltonian is
diagonalized at fixed values of $R$, and the resulting energies
plotted as functions of the hyperradius can be viewed as adiabatic
potential curves $U_\mu(R)$ as in the ordinary Born-Oppenheimer
approximation for diatomic molecules.  The first prominent success of
the adiabatic approximation was the grouping together of He
autoionizing levels having similar character into one such potential
curve.\cite{MacekJPB1968}  Subsequent studies showed that He and H$^-$
photoabsorption is dominated by a small subset of such potential
curves,\cite{CDLReview, SadeghpourGreene90,Domke1991} suggesting that
Macek's adiabatic scheme is much more than just a mathematical
technique for solving the Schršdinger equation, but that it also
provides an insightful physical and intuitive formulation that can be
used qualitatively and semiquantitatively in the same manner as the
Born-Oppenheimer treatment which has been so successful in molecular
physics.

At the same time, however, subsequent applications of the strict
adiabatic hyperspherical approximation showed its
limitations.\cite{KlarKlar1978,JonsellHeiselbergPethick2002prl}  Some
classes of energy levels or low-energy scattering properties could be
described to semiquantitative accuracy, but in other cases it failed
to give a reasonable description of the spectrum, sometimes even
qualitatively.  As this has become more and more appreciated, it has
become increasingly common to treat few-body systems using the
adiabatic hyperspherical {\it representation}, in principle an exact
theory that does not make the adiabatic approximation; in this method
several adiabatic hyperspherical states are coupled together and their
nonadiabatic interactions are treated explicitly.  Implementation of
the adiabatic hyperspherical representation is sometimes carried out
in exact numerical
calculations\cite{TolstikhinWatanabeMatsuzawa1996,BondiConnorManzRomelt1983,Chuluunbaatar2008},
but in many cases semiclassical theories such as the
Landau-Zener-Stueckelberg formulation are sufficiently accurate and
useful.\cite{NikitinBook1974}

In the four-body problem, some initial studies using hyperspherical
coordinates were carried out for the description of 3-electron atoms
such as Li, He$^-$, and H$^{--}$.\cite{clark1980hat,greene_pra1984}
But the method was improved to the point of being a comprehensive
approach by
Refs.\cite{YangXiBaoLin1995,YangBaoLin1996,MorishitaTolstikhinWatanabeMatsuzawa1997,MorishitaLin1998,MorishitaLin1999,DIncao2003,MorishitaLin2003}
Despite our focus in the present review article on four interacting
particles with short-range interactions, we summarize briefly the
headway that has previously been achieved for Coulombic systems. For
three-electron atoms, the topology is of course quite different and
more interesting than for two-electron atoms.  For instance, whereas
one observes one or more two-electron hyperspherical potential curves
that converge at $R\rightarrow \infty$ to every possible one-electron
bound state, the three-electron atom potential curves converge also to
unstable resonance levels of the residual two-electron ion that have a
nonzero autoionizing decay width.  There are multiple families of
potential curves that represent new physical processes such as
post-collision interaction in addition to the triply-excited states
and their decay pathways.  Tremendous technical challenges were
overcome in an impressive series of articles by Lin, Bao, Morishita,
and their collaborators, to enable the calculation of accurate
hyperspherical potential curves for three-electron
atoms.\cite{YangXiBaoLin1995,
  YangBaoLin1996,MorishitaTolstikhinWatanabeMatsuzawa1997,MorishitaLin1998,MorishitaLin1999,DIncao2003,MorishitaLin2003}  For a recent broader review of triply-excited states that also
discusses alternative approaches beyond the hyperspherical analysis,
see \cite{Madsen2003}.

Another theoretically challenging type of four-body problem in
chemical physics has been the dissociative recombination of H$_3^+$
induced by low energy electron collision.  Here the 3 bodies are the
nuclei (augmented by two ``frozen'' 1s electrons that play no
dynamical role at low energies), while the fourth body is the incident
colliding electron.  The solution of this problem, including the
identification of Jahn-Teller coupling as the controlling mechanism,
has been greatly aided by the use of hyperspherical internuclear
coordinates.  They allowed a mapping of the dynamics to a single
hyperradius, in addition to multichannel Rydberg electron dynamics
that could be efficiently handled using multichannel quantum defect
techniques and a rovibrational frame
transformation.\cite{KokooulineGreeneEsry2000,kokoouline2003utt,SantosKokooulineGreene2007}

More relevant to the present review of four-body interactions of
short-range character are some long-standing problems of reactive
processes in nuclear physics and in chemical physics.  Fundamental
groundwork was laid by 
Kuppermann\cite{kuppermann_jpc1997,LepetitWangKuppermann2006} and by
Aquilanti and Cavalli\cite{aquilanti1997qmh}, which concentrated on
developing coordinate systems and useful solutions of the
noninteracting problem, which are the hyperspherical
harmonics. However, whereas hyperspherical harmonics constitute a
complete, orthonormal basis set in general, which have numerous useful
formal properties, in our experience they provide poor convergence
when used alone as a basis set to expand a reactive collision
wavefunction.

The tremendous growth of ultracold atomic physics has stimulated much
of the current interest in few-body and many-body processes that are
deeply quantum mechanical in nature.  And indeed, some of the progress
can be traced to the advances that have been made in our understanding
of few-body collisions and resonances in the low-temperature limit.
Some of the most important advances were the development of accurate
theoretical models for atom-atom collisions at sub-millikelvin
temperatures.\cite{WeinerBagnatoZilioJulienne1999,GaoTiesingaWilliamsJulienne2005,Gao2008,RaoultMies2004,MiesRaoult2000,burkejr1999tic,BurkeGreeneBohn98PRL}
Ab initio theory was not sufficiently advanced to predict the
atom-atom interaction potentials to sufficient accuracy, so
refinements and adjustments of a small number of parameters (the
singlet and triplet scattering lengths and in some cases the van der
Waals coefficient and the total numbers of singlet and triplet bound
levels) were needed to specify the two-body models.  Once the two-body
interactions were well understood, the next challenge became
three-body collisions.  In most degenerate quantum gases created
during the past decade or longer, the lifetimes have been controlled
by three-body recombination, i.e. in a single-component BEC, this is
the process $A+A+A\rightarrow A_2+A$.  Advances in understanding and
in the ability to carry out nonperturbative three-body recombination
calculations resulted, by the late 1990s, in some of the first survey
studies of the dependence of the three-body recombination rate $K_3$
on the two-body scattering length.  Two independent treatments
utilizing the adiabatic hyperspherical
representation\cite{nielsen1999ler,EsryGreeneBurke1999} led to the
prediction that destructive interference minima should exist at
positive atom-atom scattering lengths $a$, with universal scaling
behavior connected intimately with the Efimov effect.  Such minima
have apparently been observed recently in
experiments.\cite{Zaccanti2009}  Ref.\cite{EsryGreeneBurke1999}
additionally predicted that three-body shape resonances, also
connected intimately with the Efimov effect, should arise periodically
in $a$, and the first such Efimov resonance was observed
experimentally in 2006 by the Innsbruck group of
Grimm.\cite{kraemer2006eeq}

Not long after the dependence of $K_3$ on $a$ had been identified by
the aforementioned theoretical treatments in hyperspherical
coordinates, alternative treatments provided different ways to
understand many of these results.  Effective field
theory\cite{braaten2006ufb}, functional
renormalization\cite{moroz:042705}, Faddeev treatments in momentum
space\cite{shepard:062713,massignan:030701}, a transition matrix
approach based on the three-body Green's
function\cite{LeeKohlerJulienne}, and an analytically-solvable model
treatment of the Efimov problem\cite{gogolin:140404}  This large
number of independent theoretical formulations, which by and large
reproduce and in some cases extend the 1999 predictions, is an
encouraging confluence that suggests our understanding of the
three-body problem with short-range forces is nicely on track.

In contrast, the description of many four-body scattering processes,
especially those with a final or initial state having four free
particles such as the recombination process $A+A+A+A\rightarrow A_3+A$
or $A_2+A_2$ or $A_2+A+A$, is a field in its infancy by comparison
with the state of the art for the three-body problem.  Most previous
attention to date has concentrated on either four-body bound states
such as the alpha particle ground or excited
states\cite{Feenberg1936}, or else simple exchange reactions with
two-body entrance and exit channels, such as $H+H_2O\rightarrow
H_2+OH$ \cite{HuSchatz2006,ClaryScience1998}.  Some theoretical
results of this class have been derived in the context of ultracold
fermi or bose gases.  One of the most important was the prediction by
Petrov, Salomon, and Shlyapnikov\cite{fedichev_etal} that the rate of
inelastic collisions, between weakly bound dimers composed of two
equal mass but opposite spin fermions, should decay at large
fermion-fermion scattering lengths as $a^{-2.55}$. Two experiments are
consistent with this prediction.\cite{regal2004lma,bourdel2004esb}
This result has been confirmed at a qualitative level in a separate
hyperspherical coordinate treatment discussed below in
Sec.~\ref{Subsec:results:relax}, but with some quantitative,
temperature-dependent differences.\cite{dincao2009ddc}  The real part
of the scattering length, associated with purely elastic scattering,
is also important in the BEC-BCS crossover problem, and its value has
been predicted in a number of independent studies to equal 0.6$a$.

For four identical bosons with large scattering lengths, an insightful
theoretical conjecture by Hammer and Platter\cite{HammerPlatter2007}
suggested that two four-body bound levels should exist that are
attached to every 3-body Efimov state.  When this problem was tackled
using the toolkit of hyperspherical coordinates, the resulting
potential curves and their bound and quasi-bound levels provided
strong numerical evidence in support of this
conjecture.\cite{vonStecher2009NatPhys}  Moreover, once the major
technical challenge of computing the adiabatic hyperspherical
potential curves had been overcome, through the use of a correlated
Gaussian basis set expansion\cite{vonStecherGreene2009pra}, it was
possible to calculate four-body recombination rates and demonstrate
that signatures of four-body physics had in fact already been present
and observed in the 2006 Efimov paper by the experimental Innsbruck
group.\cite{kraemer2006eeq}  The four-body resonance features had not
been interpreted as such in that study, but a subsequent experiment by
the same group\cite{Ferlaino2009PRL} provided strong confirmation of
this point.  This theoretical development was also aided by a general
derivation of the N-body recombination rate\cite{mehta2009general} in
terms of a scattering matrix determined within the adiabatic
hyperspherical representation.  Further extensions have permitted an
understanding of dimer-dimer collisions involving four bosonic
atoms.\cite{DIncaoStecherGreene2009prl}  
Another positive advance during the last few years has been a treatment of
atom-trimer scattering that has determined the lifetime of universal
bosonic tetramer states \cite{Deltuva2010} and the analysis of the
Efimov trimer formation via four-body recombination. \cite{WangEsryPRL}

Our aims in this review are to present some of the technical
developments that have recently enabled an extension of the adiabatic
hyperspherical framework that can handle four or more particles.  The
most technically-challenging aspect of this is the solution of the
fixed-hyperradius Schr\"odinger equation to determine the adiabatic
potential curves and their couplings that drive inelastic,
nonadiabatic processes.  Once those couplings and potential curves are
known, it is comparatively simple and intuitive to understand at a
glance the competing reaction pathways that can contribute to any
given process.  In many cases, those pathways are sufficiently small
in number, and sufficiently localized in the hyperradius, to permit
semiclassical WKB and Landau-Zener-Stueckelberg-type
theories\cite{NikitinBook1974,ZhuNakamura,Child1991,MillerGeorge} to
give a semiquantitative description.  Such approximate treatments are
especially useful for interpreting the results of quantitatively
accurate coupled channel solutions to the coupled equations.

\section{General Form of the Adiabatic Hyperspherical Representation}
\label{Sec:Adiabatic_Rep}

One of the greatest advantages of using the hyperspherical adiabatic representation is that it 
offers a simple, yet quantitative, picture of the bound and quasi-bound spectrum
as well as scattering processes. It reduces the problem to the study
of the hyperradial collective motion of the few-body system in terms
of effective potentials and where inelastic transitions are driven by
non-adiabatic couplings. The effective potentials, and the couplings between different channels,
offer a unified, conceptually clear, picture of all properties of the system. 
Below, we give a general description of the adiabatic hyperspherical
representation for a general $N$-body problem.
Details regarding the Coordinate transformations that accomplish the conversion from Cartesian to angular 
variables (along with $R$) are given in Appendix~\ref{App:Hyper_angs}.

\subsection{Channel Functions and Effective Adiabatic Potentials}
In the adiabatic hyperspherical representation, the $N$-body Schr\"odinger equation can be
written in terms of the rescaled wave function $\psi=R^{(3(N-1)-1)/2}\Psi$ (in atomic units),
\begin{eqnarray} 
\left[-\frac{1}{2\mu}\frac{\partial^2}{\partial R^2}
+\hat{H}_{\rm ad}(R,\Omega)\right]\psi(R,\Omega)=E\psi(R,\Omega),\label{schr} 
\end{eqnarray} 
\noindent
where $\mu$ is the arbitrary, reduced, mass and $E$ is the total energy. 
It is interesting to notice that the above form of the Schr\"odinger equation is
the same irrespective of the system in question, leaving all the
details of the interactions in the adiabatic Hamiltonian $\hat{H}_{\rm ad}(R,\Omega)$,
where $\Omega$ denotes the set of all hyperangles.

The few-body effective potentials are eigenvalues of the adiabatic Hamiltonian $\hat{H}_{\rm ad}$,
obtained for fixed values of $R$, i.e., with all radial derivatives omitted from the operator:
\begin{equation}
\hat{H}_{\rm ad}(R,\Omega)\Phi_\nu(R;\Omega)=U_\nu(R) \Phi_\nu(R;\Omega).
\label{adeq}
\end{equation}
\noindent
where $\Phi_\nu (R;\Omega)$, the eigenstates, are the channel
functions, $U_{\nu}(R)$ the few-body potentials, 
and the adiabatic Hamiltonian given by
\begin{eqnarray}
\hat{H}_{\rm ad}(R,\Omega)=\frac{\hat{\Lambda}^{2}(\Omega)+{(3N-4)(3N-6)}/{4}}{2\mu
  R^2}+\hat V(R,\Omega).
\label{had}
\end{eqnarray} 
\noindent
The operator $\hat\Lambda^{2}$ is the squared grand angular momentum
defined in Eq.~(\ref{Eq:hyperangmoment_def}), 
and $\hat V$ contains {\em all} the interparticle interactions. In the
above equations, $\nu$ is a collective
index that represents all quantum numbers necessary to label each
channel.

From the analysis above, since the channel functions $\Phi_{\nu}(R;\Omega)$ form a
complete set of orthonormal functions at each $R$, they are a natural base to expand the
total rescaled wavefunction,
\begin{equation}
\psi(R,\Omega)=\sum_{\nu}F_{\nu}(R)\Phi_{\nu}(R;\Omega),\label{chfun}
\end{equation}
\noindent
where the expansion coefficient $F_{\nu}(R)$ is the hyperradial wave
function. In this representation, the total wave function is, in
principle, exact.
Upon substituting Eq.~(\ref{chfun}) into the Schr\"odinger equation
(\ref{schr}) and projecting out $\Phi_{\nu}$, the hyperradial motion  
is described by a system of coupled ordinary differential equations  
\begin{widetext}
\begin{eqnarray}
\left[-\frac{1}{2\mu}\frac{d^2}{dR^2}+U_{\nu}(R)\right]F_{\nu}(R) 
-\frac{1}{2\mu}\sum_{\nu'}
\left[2P_{\nu\nu'}(R)\frac{d}{dR}+Q_{\nu\nu'}(R)\right]F_{\nu'}(R)
=EF_{\nu}(R),\label{radeq}
\end{eqnarray}
\end{widetext}
\noindent
where $P_{\nu\nu'}(R)$ and $Q_{\nu\nu'}(R)$ are the nonadiabatic
coupling terms responsible for the inelastic transitions in
$N$-body scattering processes. They are defined as   
\begin{eqnarray} 
P_{\nu\nu'}(R) &=&
\Big\langle\hspace{-0.15cm}\Big\langle\Phi_{\nu}(R,\Omega)\Big|
\frac{\partial}{\partial R}\Big|\Phi_{\nu'}(R,\Omega)\Big\rangle\hspace{-0.15cm}\Big\rangle
\label{puv}
\end{eqnarray}
\noindent
and
\begin{eqnarray} 
Q_{\nu\nu'}(R) &=&
\Big\langle\hspace{-0.15cm}\Big\langle\Phi_{\nu}(R,\Omega)\Big|
\frac{\partial^2}{\partial R^2}\Big|\Phi_{\nu'}(R,\Omega)\Big\rangle\hspace{-0.15cm}\Big\rangle,
\label{quv}
\end{eqnarray} 
\noindent
where the double brackets denote integration over the angular
coordinates $\Omega$ only.

Although in the adiabatic hyperspherical representation the major
effort is usually in solving the adiabatic equation (\ref{adeq}), 
the hyperradial Schr\"odinger equation (\ref{radeq})
is central to the simplicity of this representation.
Since $R$ represents the overall size of the system, the 
hyperradial equation (\ref{radeq})
describes the collective radial motion under the influence of the 
effective potentials $W_{\nu}$, defined by   
\begin{equation}
W_{\nu}(R)=U_{\nu}(R)-\frac{1}{2\mu}Q_{\nu\nu}(R),
\label{effpot}
\end{equation}
\noindent
while the inelastic transitions are driven by the nonadiabatic
couplings $P_{\nu\nu'}$ and $Q_{\nu\nu'}$.  
Scattering observables, as well as bound and quasi-bound spectrum, 
can then be extracted by solving Eq.~(\ref{radeq}).
As it stands, Eq.~(\ref{radeq}) is exact. In practice, of course, the
sum over channels must be truncated, and the accuracy of the
solutions can be monitored with successively larger truncations.
Therefore, in the adiabatic hyperspherical representation the usual
complexity due to the large number of degrees of freedom for few-body
systems is conveniently described by a one-dimensional radial
Shcr\"odinger equation, reducing the problem to a ``standard''
multichannel process.   

The hyperspherical adiabatic representation has been shown to offer a
simple and unifying picture for describing few-body ultracold
collisions in the regime where the short-range two-body interactions
are strongly modified due to a presence of a Fano-Feshbach resonance 
\cite{FRReview}.
In this regime, the long-range properties of the few-body effective
potentials $W_{\nu}$ become very important and other analytical
 few-body collision properties can be derived. For
instance, the asymptotic behavior of the few-body effective 
potentials $W_{\nu}$ determine the generalized Wigner threshold laws
for few-body  collisions \cite{esry2001tlt}, i.e., the energy dependence
of the ultracold collisions rates in the near-threshold
limit. Moreover, when the two-body
interactions are resonant, few-body effective potentials are
modified accordingly to universal physics
\cite{dincao2005sls,dincao2009over}, 
as we will show in the following sections.  From this analysis, a
simple picture describing both elastic and inelastic
transitions emerges. We also discuss the validity of our results
 in the context of numerical calculations carried out through the
 solution of Eqs.(\ref{adeq}) and (\ref{radeq}) for a model two-body interaction.

\subsection{Generalized Cross Sections}
\label{GenCrossSec}

Here we derive a formula for the generalized cross-section describing
the scattering of N-particles. Our formulation is based on the solutions
of the hyperadial equation (\ref{radeq}) but is sufficiently general
to describe any scattering process with particles of either
permutation symmetry. The only information required in our derivation is that 
at large $R$, the solutions to the angular portion of the Schr\"odinger equation yield the 
fragmentation channels of the $N$-body system, i.e., the same
asyptotic form of the adiabatic effective potentials (\ref{effpot}), 
and the quantum numbers labeling those solutions index the $S$-matrix.  

This derivation begins by considering the scattering by a
purely hyperradial finite-range potential in $d$-dimensions, 
then the resulting cross section is generalized to the case of anisotropic finite range potentials in $d$-dimensions 
\textquotedblleft by inspection \textquotedblright, which we interpret in the adiabatic hyperspherical picture.
For clarity, we adopt a notation that resembles the usual derivation in three dimensions.

In $d$-dimensions, the wavefunction at large $R$
behaves as: 
\begin{equation}
\Psi ^{I}\rightarrow e^{i\bm{k}\cdot \bm{R}}+f(\hat{\bm{k}},\hat{\bm{k}}^{\prime })%
\frac{e^{ikR}}{R^{(d-1)/2}}
\end{equation}%
Equivalently, an expansion in hyperspherical harmonics is written in terms of
unknown coefficients $A_{\lambda \mu}$:
\begin{equation}
\Psi ^{II}=\sum_{\lambda ,\mu }{A_{\lambda \mu }Y_{\lambda \mu }(\hat{\bm{R}}%
)(j_{\lambda }^{d}(kR)\cos \delta _{\lambda }-n_{\lambda }^{d}(kR)\sin
\delta _{\lambda })}
\end{equation}%
Here, $Y_{\lambda \mu}$ are hyperspherical harmonics and $j_{\lambda}^d$ ($n_{\lambda}^d$) are hyperspherical Bessel (Neumann) functions~\cite{Avery}.
\begin{equation}
j_{\lambda }^{d}(kR)=\frac{\Gamma (\alpha )2^{\alpha -1}}{(d-4)!!}\frac{%
J_{\alpha +\lambda }(kR)}{(kR)^{\alpha }},
\end{equation}%
 where $\alpha=d/2-1$.  We will make use of the asymptotic expansion,

\begin{equation}
j_{\lambda }^{d}(kR)\;
\stackrel{kR\rightarrow \infty }{\approx}
\;\frac{\Gamma
(\alpha )2^{\alpha -1}}{(d-4)!!}\sqrt{\frac{2}{\pi }}\frac{\cos {(kR-\frac{%
\alpha +\lambda }{2}\pi -\frac{\pi }{4})}}{(kR)^{\alpha +1/2}}
\end{equation}

and the plane wave expansion in $d$-dimensions \cite{Avery}:
\begin{equation}
e^{i\bm{k}\cdot \bm{R}}=(d-2)!!\frac{2\pi ^{(d/2)}}{\Gamma (d/2)}%
\sum_{\lambda ,\mu }{i^{\lambda }j_{\lambda }^{d}(kR)Y_{\lambda \mu }^{\ast
}(\hat{\bm{k}})Y_{\lambda \mu }(\hat{\bm{R}})}.
\end{equation}%
Identifying the incoming wave parts of $\Psi ^{I}$ and $\Psi ^{II}$ yields the coefficients $A_{\lambda \mu }$:
\begin{equation}
A_{\lambda \nu }=e^{i\delta _{\lambda }}(d-2)!!\frac{2\pi ^{d/2}}{\Gamma (d/2)}%
i^{\lambda }Y_{\lambda \mu }^{\ast }(\hat{\bm{k}}).
\end{equation}%
Inserting the coefficients $A_{\lambda\nu}$ back into the expression for $\Psi ^{II}$ gives 
the expression for the scattering amplitude: 
\begin{align}
f(\hat{\bm{k}},\hat{\bm{k}}') = \left(\frac{2 \pi}{i k}\right)^{\frac{d-1}{2}}\sum_{\lambda \mu}{ Y_{\lambda \mu}^*(\hat{\bm{k}})Y_{\lambda\mu}(\hat{\bm{k}}')(e^{2i\delta_\lambda}-1)}.
\end{align}
The immediate generalization of this elastic scattering amplitude to an anisotropic short-range potential is of course:
\begin{align}
f(\hat{\bm{k}},\hat{\bm{k}}') = \left(\frac{2 \pi}{i k}\right)^{\frac{d-1}{2}}\sum_{\lambda \mu \lambda' \mu'}{ Y_{\lambda \mu}^*(\hat{\bm{k}})Y_{\lambda^{\prime}\mu^{\prime}}(\hat{\bm{k}}^{\prime}) 
 (S_{\lambda\mu,\lambda^{\prime}\mu^{\prime}}-\delta_{\lambda\lambda^{\prime}}\delta_{\mu\mu^{\prime}})}  .
\end{align}
Upon integrating  $\vert f(\hat{\bm{k}},\hat{\bm{k}}^{\prime }) \vert^2$ over all final hyperangles $\hat{\bm{k}}$, and {\it averaging} over all initial hyperangles $\hat{\bm{k}}^{\prime}$ as would be appropriate to a gas phase experiment, we obtain the average integrated elastic
scattering cross section by a short-range potential:
\begin{equation}
\sigma^{dist}=\left( \frac{2\pi }{k}\right) ^{d-1} %
\frac{1}{\Omega(d)}
\sum_{\lambda \mu \lambda^{\prime} \mu^{\prime}}
{ \left\vert S_{\lambda\mu,\lambda^{\prime}\mu^{\prime}}-\delta_{\lambda\lambda^{\prime}}\delta_{\mu\mu^{\prime}}%
\right\vert {^{2}}}
\end{equation}%
where $\Omega(d)=2\pi^{d/2}/\Gamma(d/2)$ is the total solid angle in $d$-dimensions~\cite{Avery}.  This last expression is immediately interpreted as the average generalized cross section resulting from a scattering event that takes an initial channel into a final channel, $i \equiv \lambda^{\prime}\mu^{\prime} \rightarrow \lambda\mu \equiv f $.  Since this S-matrix is manifestly unitary in this representation, it immediately applies to inelastic collisions as well, including $N$-body recombination, in the form:
\begin{equation}
\label{multichansigma}
\sigma^{dist}_{i \rightarrow f}=\left( \frac{2\pi }{k_i}\right) ^{d-1}\frac{1}{\Omega(d)}{%
|S_{f i}-\delta _{f i}|^{2}},
\end{equation}%
It is worth noting that this expression needs to be simply summed up
for all initial and final channels contributing to a given process of
interest, including degeneracies.  For instance, we note that in the
case 
of a purely hyperradial potential, each $\lambda$ has 
$M(d,\lambda)= \dfrac{\left( 2\lambda +d-2\right) \Gamma \left( \lambda
+d-2\right) }{\Gamma \left( \lambda +1\right) \Gamma \left( d-1\right) }$ degenerate values of $\mu$.

In this form, we can readily \emph{interpret} the generalized cross section derived above in terms of the unitary S-matrix computed by solving the exact coupled-channels reformulation of the few-body problem in the adiabatic hyperspherical representation~\cite{MacekJPB1968}.
In principle this can describe collisions of an arbitrary number of particles.  
Identical particle symmetry is handled by summing over all indistinguishable amplitudes before taking the square, averaging over solid angle, then integrating over distinguishable final states to obtain the total cross section:
\begin{equation}
\sigma^{indist}=\int{\frac{d \hat{\bm{k}}}{N_p}}\int{\frac{d \hat{\bm{k}}^{\prime}}{\Omega(d)}|N_p f(\hat{\bm{k}},\hat{\bm{k}}')|^2} \notag \\
= N_p \sigma^{dist}.
\end{equation}
Here $N_{p}$ is the number of terms in the permutation
symmetry projection operator (e.g. for $N$ identical particles, $N_{p}=N!$.) 

The cross section for total angular momentum $J$ and parity $\Pi$ includes an explicit $2J+1$ factor.
Hence, the cross section from incoming channel $i$ to the final state $f$, 
properly normalized for identical particle symmetry, is given in terms of general $S$-matrix elements as~\cite{mehta2009general}
\begin{equation}
\label{NBodySigma}
\sigma^{indist}_{f i}({J^{\Pi}})=N_{p}  \left(\frac{2\pi }{k_i}\right) ^{d-1} \frac{1}{\Omega(d)} 
\left( 2J+1\right) \left\vert {S^{J^{\Pi}}_{fi }-\delta _{fi}}%
\right\vert {^{2}}.
\end{equation}%

For the process of \emph{N-body recombination} in an ultracold trapped gas that is not quantum degenerate, the experimental quantity of interest is the recombination event rate constant $K_N$ which determines the rate at which atoms are ejected from the trapping potential:
\begin{equation}
\der{}{t}n(t)=\sum_{N=2}^{N_{max}}\frac{-K_N}{(N-1)!}n^{N}(t),
\end{equation}
where $n$ is the number density.  The above relation assumes that the energy released in the recombination process is sufficient to eject all collision partners from the trap.  
The event rate constant [recombination probability per second for each distinguishable $N$-group within 
a (unit volume)$^{(N-1)}$] is the generalized cross section Eq.~(\ref{NBodySigma}) multiplied by a factor of the 
$N$-body hyperradial \textquotedblleft velocity\textquotedblright\ (including factors of $\hbar$ to explicitly show the units of $K_N$):
\begin{equation}
K^{J^{\Pi}}_{N}=\sum_{i,f}{\frac{\hbar k_i}{\mu _{N}}\sigma^{indist}_{f i}({J^{\Pi}}).}
\end{equation}%
Here, the sum is over all initial and final channels that contribute
to atom-loss.

The relevant $S$-matrix element appearing in Eq.~(\ref{NBodySigma}) from an adiabatic hyperspherical viewpoint is 
$S_{fi}^{J^{\Pi}}$
where $i$ and $f$ are the initial and final channels (i.e. solutions to Eq.~(\ref{adeq}) in the limit $R\rightarrow \infty$).  In the ultracold limit, the energy dependence of the recombination process is controlled by the long-range potential Eq.~(\ref{effpot}) in the entrance channel $i\rightarrow \lambda_{min}$, where $\lambda_{min}$ is the lowest hyperangular momentum quantum number allowed by the permutation symmetry of the $N$-particle system.  For any combination of bosons and distinguishible particles, $\lambda_{min}=0$, while for fermions, the permutation antisymmetry adds nodes to to the hyperangular wavefunction leading to $\lambda_{min}>0$.  

As a concrete example, consider the recombination formula for the
four-fermion process $F+F'+F+F' \rightarrow FF'+FF'$. In applying the
permutation symmetry operator, it is convenient to employ the H-type
coordinates given in Eq.~(\ref{symm_coords}). Expressing the
hyperangular momentum operator in these coordinates, it is possible to
show (see Section~\ref{Sec:var_bas_sym}) that
$\lambda=(l_1+l_2+2n_1)+l_3+2n_2$, 
where $l_1, l_2$ and $l_3$ are the angular momentum quantum numbers
associated with the three 
Jacobi vectors in Eq.~(\ref{symm_coords}) and $n_1, n_2$ are both
non-negative integers.  
Antisymmetry under exchange of identical particles in these
coordinates implies that $l_1$ 
and $l_2$ must be odd.  The lowest allowed values are then
$l_1=l_2=1$, $l_3=n_1=n_2=0$ 
such that $\lambda_{min}=2$.  

The preceding arguments enable us to calculate 
the generalized Wigner threshold law for strictly four-body recombination processes where the four particles undergo an inelastic transition at $R\sim a$;  any nonadiabatic couplings Eqs.~(\ref{puv}) and~(\ref{quv}) at $R\gg |a|$ can be viewed as three-body processes with the fourth particle acting as a spectator.
The asymptotic ($R\gg |a|$) form of the effective potential can be written in terms of an \emph{effective} angular momentum quantum number $l_e$
\begin{equation}
W_{\lambda_{min}}(R)  \;  \begin{array}{c} {\underrightarrow{\ \ \ \ \ \ \ \ \ \ \  } }\\
{{R \gg |a|}} \end{array}
\frac{l_e(l_e+1)}{2\mu_{N}R^2} \;\;\text{with} \;\; l_e=(2\lambda_{min}+d-3)/2.
\end{equation}
It was shown in~\cite{mehta2009general} following the treatement of
Berry~\cite{BerryProcPhysSoc1966}, the WKB tunneling integral gives
the threshold behavior of the $S$-matrix element $S_{fi}\propto
e^{(-2\gamma)}$ with
\begin{equation}
\gamma = \text{Im} {\int_{R^*}^{(3N-5+2\lambda_{min})/2k}{dR\;\sqrt{2\mu_N(E-W'(R))}}}
\end{equation}
and where $W'(R) = W(R)+\frac{1/4}{2\mu_N R^2}$ is the effective potential with the Langer correction~\cite{langer1937cfa}.  The lower limit of the integral $R^*$ coincides with the maximum of the nonadiabatic coupling strength $P^2_{fi}/|U_i(R) - U_f(R) |$ defined in Eq.~(\ref{puv}).  
For recombination into weakly bound dimers or trimers (of size $|a|$), $R^*\approx |a|$ so that in the threshold limit $E\rightarrow 0$~\cite{mehta2009general}
\begin{equation}
1-|S_{ii}|^2\propto e^{-2\gamma} = (k |a|)^{2\lambda_{min}+3N-5}.
\end{equation}
Unitarity of the $S$-matrix implies that $1-|S_{ii}|^2=\sum_{f\ne i}{|S_{fi}|^2}$, which is related to the total inelastic cross section through Eq.~(\ref{NBodySigma}).  If inelastic transitions are dominated by recombination then the scaling law for the recombination event rate constant is:
\begin{equation}
K_N\propto k^{2\lambda_{min}} |a|^{2\lambda_{min}+3N-5}
\end{equation}
We stress that the above expression gives the \emph{overall} scaling
of the event rate, and in cases where the coupling to lower channels
occurs in the region $r_0 \ll R^*\ll |a|$, 
one must include the additional WKB phase leading to a modified
scaling with respect to $a$. The $k$ dependence arises through the
outer turning point limit in the WKB tunneling phase integral.
This occurs, for example, in the case of four-identical
bosons treated in~\cite{mehta2009general}.  For the four-fermion
problem, the effective angular momentum quantum numbers for the
universal potentials in the region $r_0\ll R \ll |a|$ are calculated
in Ref.~\cite{vonstech08a}.  Table~\ref{RecombScaling} gives the value
of $\lambda_{min}$ along with the overall recombination rate scaling with $|a|$ for a few select cases.

\begin{table}
  \centering
  \begin{tabular}{|c|c|c|c|}
    \hline
    Process & $\lambda_{min}$ & Energy-dependence of $K_4$ & $a$-dependence of $K_4$ \\
    \hline
    $B+B+B+B \rightarrow BBB+B$ & $0$ & constant & $|a|^7$ \\
    $F+F+F'+F' \rightarrow FF'+FF'$ & $2$ & $E^2$ & $|a|^{11}$ \\
   $F+F+X+Y \rightarrow FFX+Y$ & $1$ & $E$ & $-$\\
\hline
  \end{tabular}
  \caption{The generalized Wigner threshold laws are given for a limited set of four-body recombination processes.  Here, $B$ denotes a boson, $F$ and $F'$ are fermions in different ``spin'' states, and $X$ and $Y$ are distinguishible atoms.  Note that since in general the scattering lengths for the $F-X$, $F-Y$ and $X-Y$ interactions are different, the scaling with respect to $a$ is not given for this 3-component case.}
\label{RecombScaling}
\end{table}

\section{Variational Basis Methods for the Four-Fermion Problem}
\label{Sec:Var_bas} 

Solution of the four-body hyperangular equation, Eq.~(\ref{adeq}), poses
significant challenges, since the difficulty grows exponentially
with the number of particles. For four particles with zero total
angular momentum, Eq.~(\ref{adeq}) 
consist of a 5-dimensional partial differential equation.  
Some state-of-the-art methods for three-particle systems often employ
B-splines or finite elements. In fact, if $40-100$
B-splines were used in each dimension to solve Eq.~(\ref{adeq}) (a common number in three-body calculations
\cite{EsryGreeneBurke1999,suno2002tbr,dincao2005sls}), there would
be $10^{8}-10^{9}$ basis functions resulting in $10^{11}-10^{13}$ non-zero
matrix elements in a sparse matrix. The computational power required for such
a calculation is currently beyond reach. Therefore, in order to
proceed numerically, a different strategy must be developed. 
In this review, we describe two of our current numerical techniques.

The method of this section for a two-component system of four fermions
uses a non-orthogonal variational basis set
consisting of some basis functions that accurately describe the system at
very large hyperradii, $R\gg\left\vert a\right\vert $, and other functions that
describe the system at very small hyperradii, $R\ll\left\vert a\right\vert $.
If both possible limiting behaviors are accurately described within the
basis, then a linear combination of these two behaviors might be expected to describe the
intermediate behavior of the system.\cite{Lin80sRef,SadeghpourGreene90}

\begin{figure}[htbp]
\begin{center}
\includegraphics[width=2in]{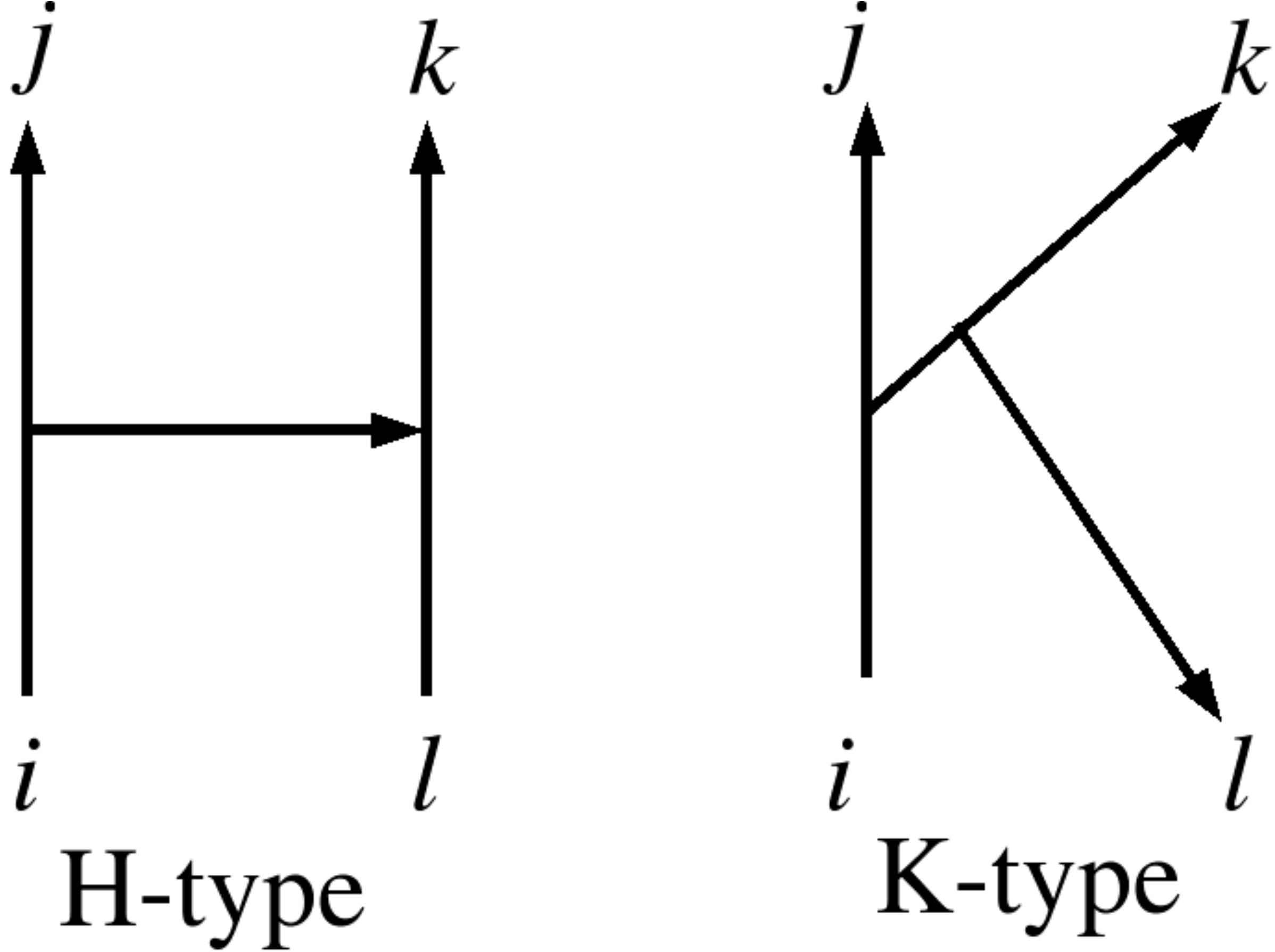}
\end{center}
\caption{The two possible configurations Jacobi coordinates in the four-body
problem are shown schematically.}%
\label{Htype_Ktype}%
\end{figure}

As with the correlated Gaussian method of Section~\ref{CGSec}, the use of
different Jacobi coordinates plays a central role in the variational basis
method. Depending on the symmetries, interactions, and fragmentation channels
inherent in the problem, different coordinates may significantly affect the
ease with which the problem can be described. For example, in the four fermion
problem, the fermionic symmetry of the system can be used to significantly
reduce the size of the basis set needed to describe the possible scattering
processes. Describing this symmetry in a poorly-chosen coordinate system can create
considerable difficulty. The two main types of Jacobi coordinate
systems are called H-type and K-type, shown schematically in Fig.
\ref{Htype_Ktype}. We discuss some of the relevant properties of the different
coordinate systems here. Appendix \ref{App:Jac_coords} gives a detailed
account of the Jacobi coordinate systems used in this review and of the
transformations between them.

H-type Jacobi coordinates are constructed by considering the separation vector
for a pair of two-body subsystems, and the separation vector between the centers of
mass of those two subsystems. Physically, H-type coordinates are useful
for describing correlations between two particles, for example a two-body
bound state or a symmetry between two particles, or two separate two-body
correlations. K-type Jacobi coordinates are constructed in an iterative way by
first constructing a three-body coordinate set as in Eq.~(\ref{Jac_coord_Ktype}), and
then taking the separation vector between the fourth particle and the center
of mass of the three particle sub-system. When two particles coalesce (e.g.
when $\bm{r}_{i}=\bm{r}_{j}$ in Eq. \ref{Jac_coord_Htype}), the H-type
coordinate system reduces to a three-body system with two of the four
particles acting like a single particle with the combined mass of its
constituents. Locating these \textquotedblleft coalescence
points\textquotedblright\ on the surface of the hypersphere is crucial for an
accurate description of the interactions between particles, and this coordinate
reduction will prove useful for the construction of a variational basis set.

Examination of Fig. \ref{Htype_Ktype} shows that K-type Jacobi coordinate systems
are useful for describing correlations between three particles within the four-particle system. 
In the four-fermion system, there are no weakly-bound trimer
states, whereby K-type Jacobi coordinates will not be used here, but the
methods described in this section can be readily generalized to
include such states. 
Unless explicitly stated, all Jacobi coordinates from here on
will be of the H-type.

The task of parametrizing the 3 Jacobi vectors in hyperspherical coordinates
remains. There is no unique way of choosing this
parameterization. The simplest method comes in the form of Delves
coordinates. Construction of these hyperangular coordinates is outlined in
Appendix \ref{App:Hyper_angs} and is described in detail in a number of
references (see Refs. \cite{SmirnovShitikova,Avery} for example). This
construction method also allows for a physically meaningful grouping of the
cartesian coordinates. For example a hyperangular coordinate system that
treats the dimer-atom-atom system as a separate three-body subsystem can
be created. This type of physically meaningful coordinate system plays a
crucial role in the construction of the variational basis set that follows.

After adoption of the Jacobi vectors, the center of mass of the four-body system is
removed, which leaves a 9-dimensional partial differential equation to solve. By
applying hyperspherical coordinates, this becomes an 8-dimensional hyperangular PDE that
must be solved at each hyperradius, a daunting task. A further simplification is achieved by initially considering only zero total angular momentum states of the system. This
implies that there is no dependence on the three Euler angles in the final
wavefunction, and in a body-fixed coordinate system these three degrees
of freedom can be removed. The body-fixed coordinates adopted here are called
democratic coordinates, adequately described in several references (see Refs.
\cite{aquilanti1997qmh,kuppermann1997rsr,littlejohn1999qdk}). The parameterization of Aquilanti and Cavalli is convenient for our purposes (for more detail see
their work in 
Ref.~\cite{aquilanti1997qmh}).

At the heart of democratic coordinates is a rotation from a space-fixed frame
to a body-fixed frame:%
\begin{equation}
\bm{\varrho}=\bm{D}^T\left(  \alpha,\beta,\gamma\right)  \bm{\varrho}_{bf}
\label{BF_rot1}%
\end{equation}
where $\bm{\varrho}$ is the matrix of "lab frame" Jacobi vectors defined in Eq.
\ref{coord_mats}, $\bm{\varrho}_{bf}$ is the matrix of body-fixed Jacobi
coordinates, and $\bm{D}\left(  \alpha,\beta,\gamma\right)  $ is an Euler rotation
matrix defined in the standard way as%
\begin{equation}
\bm{D}=\left[
\begin{array}
[c]{ccc}%
\cos\alpha & -\sin\alpha & 0\\
\sin\alpha & \cos\alpha & 0\\
0 & 0 & 1
\end{array}
\right]  \left[
\begin{array}
[c]{ccc}%
\cos\beta & 0 & \sin\beta\\
0 & 1 & 0\\
-\sin\beta & 0 & \cos\beta
\end{array}
\right]  \left[
\begin{array}
[c]{ccc}%
\cos\gamma & -\sin\gamma & 0\\
\sin\gamma & \cos\gamma & 0\\
0 & 0 & 1
\end{array}
\right]  .
\end{equation}
This parameterization is described in detail in
Appendix \ref{App:Hyper_angs}. After removing the Euler angles, the body fixed
Jacobi vectors are then described by a set of $5$ angles $\left\{  \Theta
_{1},\Theta_{2},\Phi_{1},\Phi_{2},\Phi_{3}\right\}  $ and the hyperradius $R$. The
angles $\Theta_{1}$ and $\Theta_{2}$ parameterize the overall $x,$ $y,$ and
$z$ spatial extent of the four-body system in the body-fixed frame, while the
angles $\Phi_{1},$ $\Phi_{2},$ and $\Phi_{3}$ describe the internal
configuration of the four particles.

\begin{figure}[htbp]
\begin{center}
\includegraphics[width=6in]{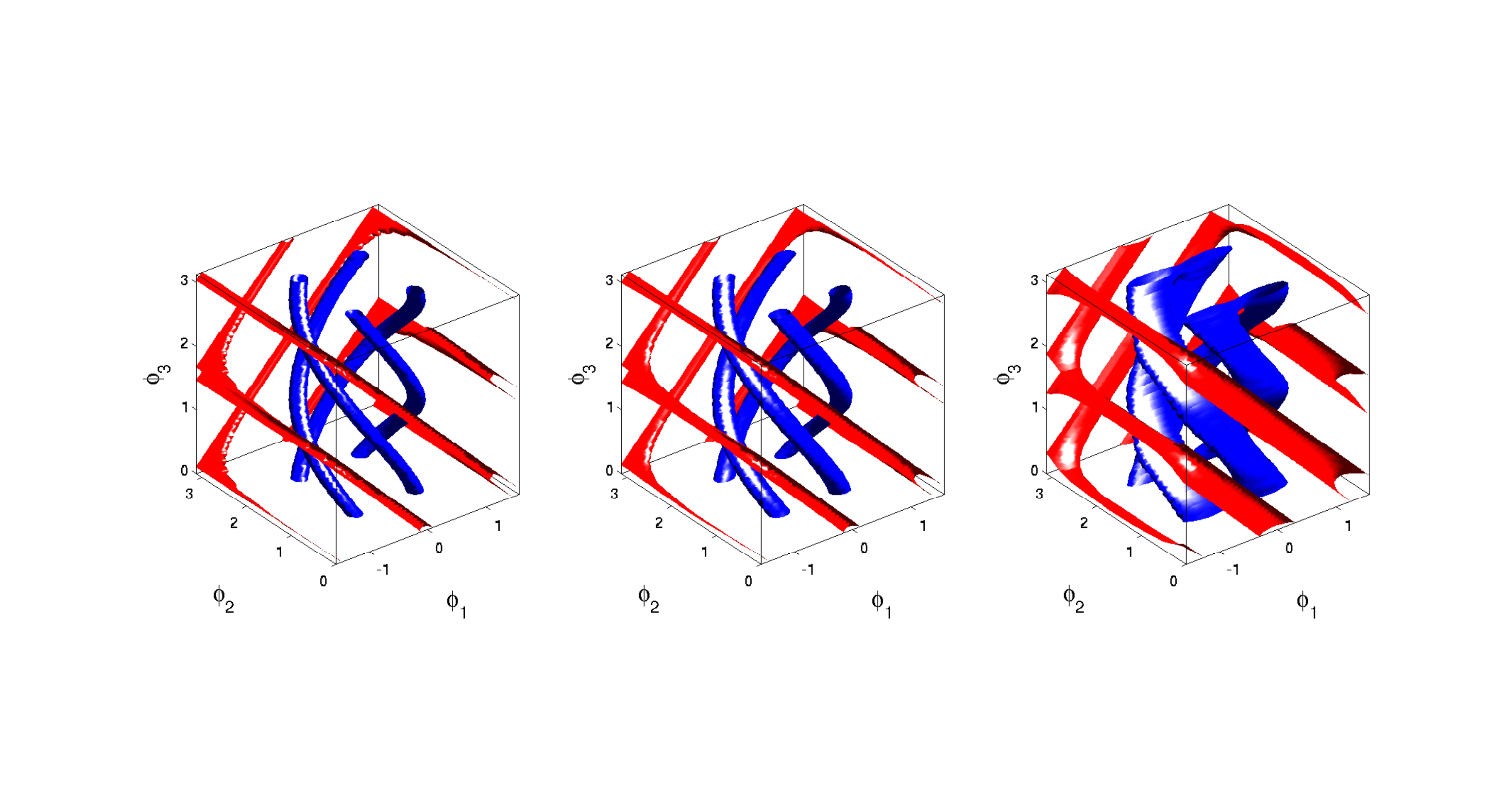}
\end{center}
\caption{Surfaces surrounding the coalescence points in the body-fixed
democratic coordinates are shown for $\theta_{1}=\pi/2$ and $\theta_{2}%
=\dfrac{\pi}{4}$ (a)$,\dfrac{\pi}{6}$ (b)$,$ and $\dfrac{\pi}{12}$ (c)
respectively. Blue surfaces surround interaction coalescence points while red
surfaces surround Pauli exclusion nodes.}%
\label{coalplots}%
\end{figure}

The description of coalescence points in democratic coordinates is especially
important. These are the points at which interactions occur and also where
nodes must be enforced for symmetry. Figure \ref{coalplots} shows these
points for $\Theta_{1}=\pi/2$, which enforces planar configurations, and 
for several values of $\Theta_{2}$. The body fixed coordinates in question are
H-type Jacobi coordintes that connect identical fermions, so symmetry is
is easily described. The $\Phi_{3}$ axis in Fig. \ref{coalplots} is shown from
$\pi/2$ to $\pi$ and then $0$ to $\pi/2$ to emphasize symmetry. The red
surfaces surround Pauli exclusion nodes while the blue surfaces surround
interaction points. It is clear that using a symmetry-based
coordinate system leaves a simple description of the Pauli exclusion
nodes.

\subsection{Unsymmetrized basis functions}

With the Jacobi vectors and democratic coordinates in hand, the 12-dimensional
four-body problem is reduced to a 6-dimensional problem for total orbital
angular momentum $J=0$. After the hyperradius is treated adiabatically, the
remaining 5-dimensional hyperangular partial differential equation Eq.~(\ref{adeq}) 
must be solved at each $R$ to obtain the adiabatic channel functions and potentials used
in the adiabatic hyperspherical representation. In Eq.~(\ref{had}),
$\hat{V}(R;\Omega)$ is chosen as 
a sum of short-range pairwise interactions, which to an excellent 
approximation affects only the $s$-wave for each pair:
$\hat{V}(R;\Omega)=\sum_{i,j}{\hat{V}\left(  r_{ij}\right)}  $, where the sum runs over all possible pairs of
distinguishable fermions. This section only considers a potential whose
zero energy s-wave scattering length $a$ is positive and large compared with
the range $r_{0}$ of the interaction. Further, unless otherwise stated, we
assume that the potential can support only a single weakly-bound dimer.

The strategy used here is not unknown \cite{sadeghpour1990vhe}. It involves
using a variational basis that diagonalizes the adiabatic Hamiltonian
in two limits asymptotically 
($R\gg a$) and at small distances 
($R\ll r_{0}$).
It is thought that linear combinations of these basis elements will provide a
variationally accurate description of the wavefunction at intermediate
$R$-values. 

Next we describe the unsymmetrized basis functions that exactly
diagonalize Eq.~(\ref{adeq}) in the small-$R$ and large-$R$ regimes.
At large $R$, three scattering thresholds arise: a threshold energy corresponding to
weakly-bound dimers 
at twice the dimer binding energy, another threshold consisting of a
single weakly bound dimer and two free particles, and finally a threshold
associated with four free particles. In general, it would be necessary to
consider another set of thresholds associated with trimer states plus a free atom (for instance, a set of
Efimov states for bosons). But for equal mass fermions, such
considerations are irrelevant since no weakly bound trimers occur in the $a\gg
r_{0}$ regime. 
At small $R$, the physics is dominated by the kinetic energy, and the eigenstates 
of the adiabatic Hamiltonian are simply the 4-body hyperspherical
harmonics which also describe four free particles at large $R$.
For a detailed description of hyperspherical harmonics, see Appendix~\ref{Sec:Hyperharms}.
Identification of these threshold regimes gives a simple
interpretation of the corresponding channel functions and provides a
starting point for the 
construction of our variational basis.

\subsubsection{Dimer-Atom-Atom Three-Body Basis Functions ($2+1+1$)}

One fragmentation possibility that must be incorporated into the
asymptotic behavior of the four-fermion system is that of an s-wave dimer with
two free particles. The dimer wave function $\phi_{d}$ is best incorporated 
using a hyperangular parameterization that treats the dimer-atom-atom system
with a set of three-body hyperangles, described by
\begin{equation}
\Psi^{\lambda_{3B}\mu_{3B}}\left(  R,\Omega\right)  =\phi_{d}\left(
r_{12}\right)  Y_{\lambda_{3B}\mu_{3B}}\left(  \Omega_{3B}^{12}\right)
,\label{d_3bod_arb}%
\end{equation}
where $Y_{\lambda_{3B}\mu_{3B}}$ is a three-body hyperspherical harmonic defined in Eq.~(\ref{3bod_harm}), 
$\lambda_{3B}$ is the three-body hyperangular momentum, and $\mu_{3B}$ indexes the degenerate states
for each value of $\lambda_{3B}$.
The dimer wave function $\phi_{d}$ is chosen as the bound state solution to
the two-body Schr\"{o}dinger equation:%
\begin{equation}
\left[  -\dfrac{\hbar^{2}}{2\mu_{2b}}\dfrac{\partial^{2}}{\partial r^{2}%
}+V\left(  r\right)  \right]  r\phi_{d}\left(  r\right)  =-E_{b}r\phi
_{d}\left(  r\right)  .\label{dd_SE}%
\end{equation}
Here the superscript $12$ in
$\Omega_{3B}^{12}$ indicates that the third particle in the three-body
subsystem is a dimer of particles 1 and 2. Further, for notational
simplicity, $\mu_{3B}$ has been used to denote the set of quantum numbers,
$\left\{  l_{2},l_{3},m_{2},m_{3}\right\}  $, which enumerate the degenerate
states for each $\lambda_{3B}$.

So far the basis function defined by Eq.~(\ref{d_3bod_arb}) is easily
written in Delves coordinates. However, in order to ensure that $L$ is
a good quantum number, one must couple the angular momenta corresponding to
the interaction Jacobi coordinates $i1$ (defined in
Eq.~\ref{interact_coords1}) to 
total angular momentum $L=0$. The
angular momentum of the (s-wave) dimer is by definition zero and all that remains is to
restrict the angular momentum of the three-body sub-system to zero. This can
be achieved by recognizing that the angular momentum associated with the individual Jacobi
vectors are good quantum numbers for the hyperspherical harmonics defined by
Eq.~(\ref{3bod_harm}), meaning that we can proceed using normal
angular momentum coupling, i.e.%
\begin{equation}
\Psi_{2+1+1}^{\lambda_{3B}l_{1}l_{2}}\left(  R,\Omega\right)  =\phi_{d}\left(
r_{12}\right)  \sum_{m_{2}=-l_{2}}^{l_{2}}\sum_{m_{3}=-l_{3}}^{l_{3}%
}\left\langle l_{2}m_{2}l_{3}m_{3}|00\right\rangle Y_{\lambda_{3B}\mu_{3B}%
}\left(  \Omega_{3B}^{12}\right)  , \label{d_3bod_unsymm}%
\end{equation}
where $\left\langle l_{2}m_{2}l_{3}m_{3}|LM\right\rangle $ is a Clebsch-Gordan
coefficient, and $l_{2}$ ($l_{3}$) is the angular momentum quantum number
associated with $\bm{\rho}_{2}^{i1}$ ($\bm{\rho}_{3}^{i1}$) from the
interaction Jacobi coordinates defined in Eqs. (\ref{interact_coords1}). Now
with the total angular momentum set to $L=0$, there must be no Euler angle
dependence in the total wavefunction. The Delves coordinates can then be
defined for this system in the body fixed frame using
Eq. (\ref{symm_hyp_angs}). 
The Delves hyperangles
are accordingly rewritten in terms of the democratic coordinates without including
the Euler angle dependence.

\subsubsection{Four-Body Basis Functions ($1+1+1+1$)}

Another important asymptotic threshold that must be considered is that of four
free particles. Using Delves coordinates, the free-particle eigenstates are 
four-body hyperspherical harmonics (see Appendix \ref{Sec:Hyperharms}):
\begin{align*}
\Phi_{\lambda\mu}^{\left(  4b\right)  }\left(  \Omega\right)  = &
N_{l_{l}l_{m}\lambda_{l,m}}^{33}N_{\lambda_{l,m}l_{n}}^{63}\sin^{\lambda
_{l,m}}\left(  \alpha_{lm,n}\right)  \cos^{l_{n}}\left(  \alpha_{lm,n}\right)
P_{\left(  \lambda-\lambda_{l,m}-l_{n}\right)  /2}^{\lambda_{l,m}+5/2,l_{n}%
+1}\left(  \cos2\alpha_{lm,n}\right)  \\
&  \times N_{l_{l},l_{m}}^{\lambda_{l,m}}\sin^{l_{l}}\left(  \alpha
_{l,m}\right)  \cos^{l_{m}}\left(  \alpha_{l,m}\right)  P_{\left(
\lambda_{l,m}-l_{l}-l_{m}\right)  /2}^{l_{l}+1,l_{m}+1}\left(  \cos
2\alpha_{l,m}\right)  \\
&  \times Y_{l_{l}m_{l}}\left(  \omega_{l}\right)  Y_{l_{m}m_{m}}\left(
\omega_{m}\right)  Y_{l_{n}m_{n}}\left(  \omega_{n}\right)  ,
\end{align*}
where $\mu$ has again been used to denote the set of quantum numbers $\left\{
\lambda_{12},l_{1},l_{2},l_{3},m_{1},m_{2},m_{3}\right\}  $ that enumerate the
degenerate states for each $\lambda$. Here $l_{i}$ is the spatial angular
momentum quantum number associated with the Jacobi vector $\bm{\rho}%
_{i}^{\sigma}$ with z-projection $m_{i}$, and $\lambda_{l,m}$ is the
sub-hyperangular momentum quantum number associated with the sub-hyperangular
tree in Fig. \ref{delves_4bod} (For example, $\lambda_{12}=l_1+l_2+2n_{12}$ where $n_{12}$ is a non-negative integer.) The hyperangles $\left\{  \alpha_{lm,n}%
,\alpha_{l,m}\right\}  $ are defined here using Delves coordinates as
described in Appendix \ref{App:Hyper_angs} and $\omega_{n}$ refers to the
spherical polar angles associated with the Jacobi vetor $\bm{\rho}_{n}$.

The choice of quantum numbers described above does not give the total orbital angular momentum of
the four particle system as a good quantum number. To accomplish this, the
three angular momenta of the Jacobi vectors must be coupled to a resultant total
, in this case to $L=0$. This gives a variational basis function of the form%
\begin{align}
\Psi_{1+1+1+1}^{\lambda\lambda_{12}l_{1}l_{2}l_{3}}\left(  \Omega\right)   &
=\sum_{M_{12}=-L_{12}}^{L_{12}}\sum_{m_{3}=-l_{3}}^{l_{3}}\sum_{m_{2}=-l_{2}%
}^{l_{2}}\sum_{m_{1}=-l_{1}}^{l_{1}}\left\langle L_{12}M_{12}l_{3}%
m_{3}|00\right\rangle \label{4_bod_unsymm}\\
&  \times\left\langle l_{1}m_{1}l_{2}m_{2}|L_{12}M_{12}\right\rangle
\Phi_{\lambda\mu}^{\left(  4b\right)  }\left(  \Omega\right)  .\nonumber
\end{align}
Now that the total angular momentum is set to $L=0$ the same procedure used
for the $\Psi_{2+1+1}^{\lambda_{3B}l_{1}l_{2}}$ basis functions can be
employed. However, this time the hyperangular parameterization is defined
using the symmetry Jacobi coordinates in Eqs. (\ref{symm_coords}). Since there is
no dependence on the Euler angles, the Jacobi coordinates can then be defined
in the body-fixed frame.

\subsubsection{Dimer-Dimer Basis Functions ($2+2$)}

The asymptotic behavior of the two-component four-fermion system must include a description of
two s-wave dimers separated by a large distance. To incorporate this behavior
the variational basis must include a basis function of the form,%
\begin{equation}
\Psi_{2+2}\left(  R,\Omega\right)  =\phi_{d}\left(  r_{12}\right)  \phi
_{d}\left(  r_{34}\right)  ,\label{dd_unsymm}%
\end{equation}
where the subscript 2+2 indicates the dimer-dimer nature of this function, and
the dimer wavefunction, $\phi_{d}$, is given by the two-body Schr\"{o}dinger
equation.  Here $\mu_{2b}$ is the reduced mass of the two distinguishable
fermions, and $E_{b}\approx\hbar^{2}/2\mu_{2b}a^{2}$ is the binding energy of
the weakly bound dimer. At first glance the right-hand side of Eq.
(\ref{dd_unsymm}) depends only implicitly on the hyperradius and hyperangles. To
make this dependence explicit, Eqs. (\ref{interpart12}) and (\ref{interpart34})
are employed to extract $r_{12}\left(  R,\Omega\right)  $ and $r_{34}\left(
R,\Omega\right)  $. It can also be noted that the basis function, Eq.
\ref{dd_unsymm}, does not respect the symmetry of the identical fermions, i.e.
$P_{13}\Phi_{2+2}\neq-\Phi_{2+2}$. The antisymmetrization of the variational
basis is discussed in 
the next section.

\subsection{ Symmetrizing the Variational Basis}
\label{Sec:var_bas_sym}

The definitions of the basis functions developed in the previous subsection do
not include the fermionic symmetry of the four particle system in question.
Until this point, we have only been concerned 
with
Jacobi coordinate
systems in which the particle exchange symmetry is well described and
with a single set of Jacobi vectors that describe some of the
interactions. In order to impose the $S_{2}\otimes S_{2}$ symmetry of two
sets of two identical fermions, we now incorporate the extra H-type
Jacobi coordinates described in Appendix~\ref{App:Jac_coords}. As a first step 
we define the projection operator,
\begin{equation}
\hat{P}=\dfrac{1}{4}\left(  \hat{I}-\hat{P}_{13}\right)  \left(  \hat{I}%
-\hat{P}_{24}\right)  , \label{Projection}%
\end{equation}
where $\hat{I}$ is the identity operator, and $\hat{P}_{ij}$ is the operator
that permutes the coordinates of particles $i$ and $j.$ This operator will
project any wavefunction onto the Hilbert space of wavefunctions that are
antisymmetric under exchange of identical fermions. Since we are treating the
fermionic species as distinguishable, permutations of members of different
species are ignored. Applying this projection operator to the dimer-dimer basis function yields an unnormalized basis function,%
\begin{equation}
\Psi_{2+2}^{\left(  symm\right)  }\left(  R,\Omega\right)  =\hat{P}\Psi
_{2+2}\left(  R,\Omega\right)  =\dfrac{1}{2}\left(  \phi_{d}\left(
r_{12}\right)  \phi_{d}\left(  r_{34}\right)  -\phi_{d}\left(  r_{14}\right)
\phi_{d}\left(  r_{23}\right)  \right)  , \label{dd_symm}%
\end{equation}
where the inter-particle distances $r_{14}$ and $r_{23}$ are given in Eqs.
(\ref{interpart14}) and (\ref{interpart23}).

Imposition of the antisymmetry constraints on the dimer-atom-atom basis functions in Eq.~(\ref{d_3bod_unsymm}) yields%
\begin{align}
\Psi_{2+1+1}^{\left(  symm\right)  \lambda_{3b}l_{2}l_{3}}\left(
R,\Omega\right)   &  =\hat{P}\Psi_{2+1+1}^{\lambda_{3b}l_{2}l_{3}}\left(
R,\Omega\right) \nonumber\\
=  &  \dfrac{1}{4}\phi_{d}\left(  r_{12}\right)  \sum_{m_{2}=-l_{2}}^{l_{2}%
}\sum_{m_{3}=-l_{3}}^{l_{3}}\left\langle l_{2}m_{2}l_{3}m_{3}|00\right\rangle
Y_{\lambda_{3B}\mu_{3B}}\left(  \Omega_{3B}^{12}\right) \label{d_3bod_symm}\\
&  -\dfrac{1}{4}\phi_{d}\left(  r_{23}\right)  \sum_{m_{2}=-l_{2}}^{l_{2}}%
\sum_{m_{3}=-l_{3}}^{l_{3}}\left\langle l_{2}m_{2}l_{3}m_{3}|00\right\rangle
Y_{\lambda_{3B}\mu_{3B}}\left(  \Omega_{3B}^{23}\right) \nonumber\\
&  -\dfrac{1}{4}\phi_{d}\left(  r_{14}\right)  \sum_{m_{2}=-l_{2}}^{l_{2}}%
\sum_{m_{3}=-l_{3}}^{l_{3}}\left\langle l_{2}m_{2}l_{3}m_{3}|00\right\rangle
Y_{\lambda_{3B}\mu_{3B}}\left(  \Omega_{3B}^{14}\right) \nonumber\\
&  +\dfrac{1}{4}\phi_{d}\left(  r_{34}\right)  \sum_{m_{2}=-l_{2}}^{l_{2}}%
\sum_{m_{3}=-l_{3}}^{l_{3}}\left\langle l_{2}m_{2}l_{3}m_{3}|00\right\rangle
Y_{\lambda_{3B}\mu_{3B}}\left(  \Omega_{3B}^{34}\right)  ,\nonumber
\end{align}
where $\Omega_{3B}^{ij}$ is the set of three-body hyperangles associated with
particles $i$ and $j$ in a dimer and the remaining two particles free. The
democratic parameterizations for the inter-particle distances from 
Eqs.~(\ref{interpart12})-(\ref{interpart34}) can be used in the dimer wavefunction
directly. Through the use of symmetry coordinates, the hyperangles of the four-body
system can be divided into a dimer subsystem and a three-body subsystem where
the third particle is the dimer itself. Using the three-body hyperangles in
the three-body harmonic in each term of Eq.~(\ref{d_3bod_symm}), combined with
the kinematic rotations from Eqs.~(\ref{Kin_rot_s_i1}) and (\ref{Kin_rot_s_i2}),
the three-body harmonics are then fully described in the hyperangles defined using
symmetry Jacobi coordinates. Since $\Psi_{2+1+1}^{\left(  symm\right)
\lambda_{3b}l_{2}l_{3}}$ has been constrained to zero total spatial angular
momentum, $L=0$, the body-fixed parameterization of the Jacobi vectors can be
inserted directly without worrying about the Euler angles $\alpha,\beta$ and
$\gamma$.

The final set of basis functions that must be antisymmetrized with respect to
identical fermion exchange are the hyperspherical harmonics
representing 
four free particles.
Permutation of the identical fermions is accomplished in the symmetry
coordinates using Eqs. (\ref{Eq:perm13symm1})-(\ref{Eq:per24symm3}). Using these
permutations gives  
\begin{align*}
\hat{P}_{13}\Psi_{1+1+1+1}^{\lambda\lambda_{12}l_{1}l_{2}l_{3}}\left(
\Omega\right)   &  =\left(  -1\right)  ^{l_{1}}\Psi_{1+1+1+1}^{\lambda
\lambda_{12}l_{1}l_{2}l_{3}}\left(  \Omega\right)  ,\\
\hat{P}_{24}\Psi_{1+1+1+1}^{\lambda\lambda_{12}l_{1}l_{2}l_{3}}\left(
\Omega\right)   &  =\left(  -1\right)  ^{l_{2}}\Psi_{1+1+1+1}^{\lambda
\lambda_{12}l_{1}l_{2}l_{3}}\left(  \Omega\right),
\end{align*}
where antisymmetry of the four free particle basis functions is 
enforced simply by choosing $l_{1}$ and $l_{2}$ to be odd.

Another symmetry in this system is that of inversion (parity), in which all Jacobi
coordinates are sent to their negatives,%
\[
\bm{\rho}_{j}^{\sigma}\rightarrow-\bm{\rho}_{j}^{\sigma},
\]
where $\sigma=s,i1,i2$ and $j=1,2,3$. Following the definitions of the Jacobi
coordinates, positive inversion symmetry in the $1+1+1+1$ basis functions,
$\Psi_{1+1+1+1}^{\lambda\lambda_{12}l_{1}l_{2}l_{3}}\left(  \Omega\right)  $,
is imposed by choosing $\lambda$ to be even. The $2+1+1$ basis functions,$\Psi
_{2+1+1}^{\left(  symm\right)  \lambda_{3b}l_{2}l_{3}}\left(  R,\Omega\right)
$, must already have positive inversion symmetry since $\phi_{d}\left(
r\right)  $ is an s-wave dimer wavefunction and $l_{2}=l_{3}$ for zero total
spatial angular momentum, $L=0$. The dimer-dimer basis function, $\Psi
_{2+2}^{\left(  symm\right)  }\left(  R,\Omega\right)  $, is already symmetric
under inversion and does not need further restrictions placed on it.

The final symmetry to be imposed is not quite as obvious as the symmetries
discussed so far. By performing a \textquotedblleft spin-flip" operation in
which the distinguishable species of fermions are exchanged, i.e. 
$\hat{P}_{12}\hat{P}_{34}$, the Hamiltonian in Eq. \ref{had} (with $N=4$) remains unchanged.
This operation is identical to inverting the two dimers in the
dimer-dimer basis function. One can see that $\Psi_{2+2}^{\left(  symm\right)
}$ is unchanged under this operation. We will limit ourselves to
dimer-dimer collisions in this section and will only be concerned with basis
functions that have this symmetry. This symmetry is imposed on both
 $\Psi_{2+1+1}^{\left(  symm\right)  \lambda_{3b}l_{2}l_{3}}$ and 
$\Psi_{1+1+1+1}^{\lambda\lambda_{12}l_{1}l_{2}l_{3}}$ 
by demanding $l_3$ to be \emph{even}.

Recalling that $\lambda=(l_1+l_2+2n_1)+l_3+2n_2$ where $n_1$ and $n_2$ are both non-negative integers, the combination of these symmetries implies that the minimum $\lambda$ for $\Psi_{1+1+1+1}^{\lambda\lambda_{12}l_{1}l_{2}l_{3}}$ must be $\lambda_{min}=2$.  This argument plays a pivotal role in determining the overall threshold scaling law for four-body recombination, as is discussed in Section~\ref{GenCrossSec}.

\section{Correlated Gaussian and Correlated Gaussian Hyperspherical Method}
\label{Sec:CorrGaussian}

\subsection{Correlated Gaussian method}
\label{CGSec}

In this Section, we discuss alternative numerical techniques to study the four-body problem.
First, we present a powerful technique to describe
few-body trapped systems where the solutions are expanded in
correlated Gaussian (CG) basis set. Additional details regarding the CG basis set, 
including the evaluation of matrix elements, symmetrization, and basis set selection 
are discussed in Appendix~\ref{Sec:Impl}.  We then present an
innovative method which combines the adiabatic hyperspherical representation with
the CG basis set and Stochastic Variational method (SVM).  For additional information on 
the methods described in this section, see~(\cite{vonstecherthesis,von2009correlated}).

\subsubsection{General procedure}

Different types of Gaussian basis functions have
long been used in many different areas of physics.  In particular,
the usage of Gaussian basis functions is one of the key elements of
the success of {\it ab initio} calculations in quantum chemistry.
The idea of using an explicitly correlated Gaussian to solve quantum
chemistry problems was introduced in 1960 by Boys~\cite{boys1960ifv}
and Singer~\cite{singer1960uge}. The combination of a Gaussian basis
and the stochastical variational method SVM was first introduced by
Kukulin and Krasnopol'sky~\cite{kukulin1977svm} in nuclear physics
and was extensively used by Suzuki and
Varga~\cite{varga1996svm,varga1995psf,varga1997sfb,varga1994mmd}.
These methods were also used to treat ultracold many-body Bose
systems by Sorensen, Fedorov and Jensen~\cite{sorensen2005cgm}. A
detailed discussion of both the SVM and CG methods can be found in
a thesis by Sorensen \cite{sorensen2005cmb} and, in
particular, in Suzuki and Varga's book~\cite{suzuki1998sva}. In the
following, we present the CG method and its application to few-body
trapped systems.

Consider a set of coordinate vectors that describe the system
$\{\bm{x}_1,...,\bm{x}_N\}$. In this method, the eigenstates
are expanded in a set of basis functions,
\begin{equation}
\Psi (\bm{x}_1,\cdots,\bm{x}_N) = \sum_A C_A\,\Phi
_A(\bm{x}_1,\cdots,\bm{x}_N)=\sum_A
C_A\,\braket{\bm{x}_1,\cdots,\bm{x}_N|A}.
  \label{TotalWF}
\end{equation}
Here $A$ is a matrix with a set of parameters that characterize the
basis function. In the second equality we have introduced a
convenient ket notation. Solving the time-independent Schr\"odinger
equation in this basis set reduces the problem to a diagonalization
of the Hamiltonian matrix:
\begin{equation}
  \bm{\mathcal{H}}\bm{C}_i =E_i \bm{\bm{\mathcal{O}}}\bm{C}_i
  \label{HamMatrix}
\end{equation}
Here, $E_i$ are the energies of the eigenstates, $\bm{C}_i$ is a
vector form with the coefficients $C_A$ and $\bm{\mathcal{H}}$ and
$\bm{\mathcal{O}}$ are matrices whose elements are
$\mathcal{H}_{BA}=\braket{B|\mathcal{H}|A}$ and
$\mathcal{O}_{BA}=\braket{B|A}$. For a 3D system, the evaluation of
these matrix elements involves $3N$-dimensional integrations which
are in general very expensive to compute.
Therefore, the effectiveness of the basis set expansion method
relies mainly on the appropriate selection of the basis functions.
As we will see, the CG basis functions permit a fast evaluation of
overlap and Hamiltonian matrix elements; they are flexible enough to
correctly describe physical states.

To reduce the dimensionality of the problem we take advantage of
its symmetry properties. Since the interactions considered are
spherically symmetric, the total angular momentum, $J$, is a good
quantum number, and here we restrict ourselves to $J=0$.  Observe that if
the basis functions only depend on the interparticle distances, then
Eq.~(\ref{TotalWF}) only describes states with zero angular
momentum and positive parity ($J^\Pi=0^+$).
 Furthermore, in the problems we consider,
the center-of-mass motion decouples from the system. Thus
the CG basis functions take the form
\begin{equation}
\Phi _{\{\alpha_{ij}\}}(\bm{x}_1,\cdots,\bm{x}_N)
=\psi_0(\bm{R}_{CM})\mathcal{S}\left\{\exp\left(-\sum_{j>i=1}^N
\alpha_{ij} r_{ij}^2/2\right)\right\},
  \label{BF}
\end{equation}
where $\mathcal{S}$ is a symmetrization operator and $r_{ij}$ is the
interparticle distance between particles $i$ and $j$. Here, $\psi_0$ is the ground state of the center-of-mass
motion. For trapped systems, $\psi_0$ takes the form,
$\psi_0(\bm{R}_{CM})=e^{-R_{CM}^2/2a_{ho}^{M}}$. Because of its
simple Gaussian form, $\psi_0$ can be absorbed into the
exponential factor. Thus, in a more general way, the basis function
can be written in terms of a matrix $A$ that characterizes them,
\begin{equation}
\label{Basis1}
\Phi_A(\bm{x}_{1},\bm{x}_{2},...,\bm{x}_{N})=
\mathcal{S}\left\{\exp(-\frac{1}{2}
\bm{x}^T.\bm{A}.\bm{x})\right\}=\mathcal{S}\left\{\exp(-\frac{1}{2}
\sum_{j>i=1}^N A_{ij} \bm{x}_i.\bm{x}_j)\right\},
\end{equation}
where $\bm{x}=\{\bm{x}_1,\bm{x}_2,...,\bm{x}_N\}$
and $A$ is a symmetric matrix. The matrix elements $A_{ij}$ are
determined by the $\alpha_{ij}$ (see Appendix~\ref{AJac}).
Because of the simplicity of the basis functions, Eq.~(\ref{BF}),
the matrix elements of the Hamiltonian can be calculated
analytically.

Analytical evaluation of the matrix
elements is enabled by selecting the set of coordinates that simplifies the
integrals.  For basis functions of the form of Eq.~(\ref{Basis1}),
the matrix elements are characterized by a matrix $M$ in the
exponential. Hence the matrix element integrand can be
greatly simplified if we write it in terms of the coordinate eigenvectors
that diagonalize that matrix $M$. This change of coordinates
permits, in many cases, the analytical evaluation of the matrix
elements. The matrix elements are explicitly evaluated
in Appendices~\ref{ASymm} and~\ref{AEvBasisFunc}.

Two properties of the CG method are worth mentioning.
First, the CG basis set is numerically linearly-dependent and over-complete, so
a systematic increase in the number of basis functions will in principle converge
to the exact eigenvalues~\cite{sorensen2005cmb}. Secondly, the basis
functions $\Phi_A$ are square-integrable only if the matrix $A$ is
positive definite. We can further restrict the basis function by
introducing real widths $d_{ij}$ such that
$\alpha_{ij}=1/d_{ij}^2$ which ensures
that $A$ is positive definite. Furthermore, these widths are
proportional to the mean interparticle distances in each basis
function. Thus, it is easy to select them after considering the
physical length scales relevant to the problem. Even though we have
restricted the Hilbert space with this transformation, we have
numerical evidence that that the results converge to the exact
eigenvalues.

The linear dependence in the basis set causes problems in the
numerical diagonalization of the Hamiltonian matrix
Eq.~(\ref{HamMatrix}).
Different ways to reduce or eliminate such problems
are explained in the Appendix~\ref{ALinDep}.

Finally, we stress the importance of selecting an appropriate
interaction potential. For the problems considered
in this review, the interactions are expected to be
characterized primarily by the scattering length, i.e., to be independent of the
shape of the potential.  We capitalize on that flexibility by choosing a model
potential that permits rapid evaluation of the matrix elements. A Gaussian form,
\begin{eqnarray}
\label{potNumChap} V_0(r)=-d \exp \left(- \frac{r^2}{2r_0^2}
\right),
\end{eqnarray}
is particularly suitable for this basis set choice.  If the range
$r_0$ is much smaller than the scattering length, then the
interactions are effectively characterized only  by the scattering
length. The scattering length is tuned by changing the strength of
the interaction potential, $d$, while the range, $r_0$, of the
interaction potential remains unchanged. This is particularly
convenient in this method since it implies that we only need to
evaluate the matrix elements once and we can use them to solve the
Schr\"odinger equation at any given potential strength ( or
scattering length). Of course, this procedure will give accurate
results only if the basis set is complete enough to describe the
different configurations that appear at different scattering
lengths.

In general, a simple version of this method includes four basic
steps: generation of the basis set, evaluation of the matrix
elements, elimination of the linear dependence, and
evaluation of the spectrum. The stochastical variational method (SVM), briefly discussed in
Appendix~\ref{SVM}, combines the first three of these steps in an
optimization procedure where the basis functions are selected
randomly.

\subsection{Correlated Gaussian Hyperspherical method}
\label{CGHSSec}

Several techniques have been developed to solve few-body systems in
the last few decades
\cite{faddeev1960mat,suzuki1998sva,malfliet1969sfe,yaku67,MacekJPB1968}.
Among these methods, the Correlated Gaussian (CG) technique
presented in the previous Section has proven to be capable of
describing trapped few-body systems with short-range interactions.
Because of the simplicity of the matrix element calculation, the CG
method provides an accurate description of the ground and excited
states up to $N=6$ particles~\cite{blumePRL07}. However, CG can only describe
bound states. For this reason, it is numerically convenient to treat
trapped systems where all the states are quantized. The CG cannot
(without substantial modifications) describe states above the
continuum nor the rich behavior of atomic collisions such as
dissociation and recombination.

The hyperspherical representation, on the other hand, provides an
appropriate framework to treat the continuum. In the adiabatic
hyperspherical representation (see Sec.~\ref{Sec:Adiabatic_Rep}), the Hamiltonian is solved as a
function of the hyperradius $R$, reducing the many-body
Schr\"odinger equation to a single variable form with a set of
coupled effective potentials. The asymptotic behavior of
the potentials and the channels describe different dissociation or
fragmentation pathways, providing a suitable framework for analyzing
collision physics. However, the standard hyperspherical methods
expand the channel functions in B-splines or finite element basis
functions~\cite{pack1987qrs,zhou1993hac,esry1996ahs,suno2002tbr},
and the calculations become very computationally demanding for $N>3$
systems.

Ideally, we would like to combine the fast matrix element evaluation
of the CG basis set with the capability of the hyperspherical
framework to treat the continuum.
Here, we explore how the CG basis set can be used within the
adiabatic hyperspherical representation. We call the use of CG basis function to
expand the channel functions in the hyperspherical framework the CG
hyperspherical method (CGHS).

In the hyperspherical framework, matrix elements
of the Hamiltonian must be evaluated at fixed $R$.
To proceed, consider first how the matrix
element evaluation is carried out in the standard CG approach

In the CG method, we select, for each matrix element evaluation, a
set of coordinate vectors that simplifies the integration, i.e., the
set of coordinate vectors that diagonalize the basis matrix $M$
which characterizes the matrix element (see Appendix~\ref{AJac}). The flexibility to choose
the best set of coordinate vectors for each matrix element
evaluation is key to the success of the CG method.

The optimal set of coordinate vectors are formally
selected by making an orthogonal transformation from an initial set of
vectors $\bm{x}=\{\bm{x}_1,...,\bm{x}_N\}$ to a final
set of vectors $\bm{y}=\{\bm{y}_1,...,\bm{y}_N\}$:
$\bm{x}\bm{T}=\bm{y}$, where $\bm{T}$ is the $N \times N$ orthogonal transformation matrix.
The hyperspherical method is particularly suitable for such
orthogonal transformations because the hyperradius $R$ is an
invariant under them. Consider the hyperradius defined in terms of a
set of \emph{mass-scaled} Jacobi
vectors~\cite{delves1959tag,delves1960tag,suno2002tbr,mehta2007haf},
$\bm{x}=\{\bm{x}_1,...,\bm{x}_N\}$,
\begin{equation}
R^2=\sum_i\bm{x_i}^2,
\end{equation}
If we apply an orthogonal transformation to a new set of vectors
$\bm{y}$, then
\begin{equation}
R^2=\sum_i\bm{x_i}^2=\bm{y}\bm{T}^T\bm{T}\bm{y}=\sum_i\bm{y_i}^2
\end{equation}
where we have used that $\bm{T}^T\bm{T}=\bm{I}$, and $\bm{I}$ is the identity.
Therefore, in the hyperspherical framework we can also select the
most convenient set of coordinate vectors for each matrix element
evaluation.
This is the key to reducing the dimensionality of
the matrix element integrals. One can view the flexibility afforded by such orthogonal transformations
of the Jacobi vectors instead in terms of the hyperangles $\Omega$ that best simplify the evaluation
of matrix elements.

As an example of how the dimensionality of matrix-elements is reduced, consider a three dimensional $N$-particle
system in the center of mass frame and with zero orbital angular momentum ($J=0$). We will
show that this technique reduces a $(3N-7)$-dimensional numerical
integral~\cite{Javfootnote}
to a sum over ($N-3$)-dimensional numerical integrals (see Subsec.~\ref{Exp}). Hence, for $N=3$ the
matrix elements can be evaluated analytically, and the
$N=4$ matrix elements require a sum of one-dimensional numerical integrations.

The next three subsections discuss the implementation of the CGHS.
Many of the techniques used in the standard CG method can be
directly used in the CGHS approach. For example, the selection and
symmetrization of the basis function can be directly applied in the
CGHS method. Also, the SVM method can be used to
optimize the basis set at different values of the hyperradius $R$.
Subsection~\ref{Exp} describes how the hyperangular Schr\"odinger
equation (Eq.~\ref{LamHRnew}) can be solved using a CG basis set
expansion and shows, as an example, how the
unsymmetrized matrix elements can be calculated analytically for a
four particle system.
Finally, subsection~\ref{GenCon} discusses the general implementation of this method.

\subsubsection{Expansion of the channel functions in a CG basis set and calculation of matrix elements}
\label{Exp}

In the hyperspherical method (see Sec.~\ref{Sec:Adiabatic_Rep}), channel
functions are eigenfunctions of the adiabatic Hamiltonian
$\mathcal{H}_A(R;\Omega)$,
\begin{equation}
\label{LamHRnew} \mathcal{H}_A(R;\Omega)\Phi_\nu(R;\Omega)=U_\nu
(R)\Phi_\nu(R;\Omega).
\end{equation}
The eigenvalues of this equation are the hyperspherical potential
curves $U_\nu (R)$. The adiabatic Hamiltonian has the form,
\begin{equation}
\mathcal{H}_A(R;\Omega)=\frac{\hbar^2\Lambda^2}{2\mu
R^2}+\frac{(d-1)(d-3)\hbar^2}{8\mu R^2}+V(R,\Omega).
\end{equation}
Here, $d=3N_J$ where $N_J$ is the number of Jacobi vectors.

A standard way to solve Eq.~(\ref{LamHRnew}) is to expand the
channel functions in a basis,
\begin{equation}
 \label{Phiexp}
\ket{\Phi_\mu(R;\Omega)}=\sum_i c_\mu^i(R)\ket{B_i(R;\Omega)}.
\end{equation}
Here $\mu$ labels the channel functions and $\ket{B_i(R;\Omega)}$
are the CG basis functions [Eq.~(\ref{Basis1})] written in hyperspherical coordinates. With this expansion, Eq.~(\ref{LamHRnew})
reduces to the generalized eigenvalue equation
\begin{equation}
\bm{\mathcal{H}}_A(R)\bm{c}_\mu =U_\mu
(R)\bm{\mathcal{O}}(R)\bm{c}_\mu.
\end{equation}
The vectors $\bm{c}_\mu=\{c_\mu^1,...,c_\mu^D\}$, where $D$ is the
dimension of the basis set. $\bm{\mathcal{H}}_A$ and $\bm{\mathcal{O}}$ are
the Hamiltonian and overlap matrices whose matrix elements are given
by
\begin{gather}
\mathcal{H}_A(R)_{ij}=\braket{\braket{B_i|\mathcal{H}_A(R;\Omega)|B_j}},\label{CGHSME1}\\
\mathcal{O}(R)_{ij}=\braket{\braket{B_i|B_j}}.\label{CGHSME2}
\end{gather}

Efficient evaluation of the matrix elements, e.g. Eqs.~\eqref{CGHSME1} and \eqref{CGHSME2}, is essential for the
optimization of the basis functions and the overall feasibility of the
four-body calculations. Here, we demonstrate how to speed up the
calculation by reducing the dimensionality of the numerical
integrations involved in the matrix element evaluation.

Consider a four-body system described by three Jacobi vectors,
$\bm{x}\equiv\{\bm{x}_1,\bm{x}_2,\bm{x}_3\}$, once
the center-of-mass motion is decoupled. The overlap matrix elements
between two unsymmetrized basis functions $\Phi_A$ and $\Phi_B$
(characterized by matrices $A$ and $B$ in the respective exponents) is significantly simplified if
we change variables to the set of coordinates
that diagonalize $A+B$. We call $\beta_1$, $\beta_2$ and
$\beta_3$ the eigenvalues and
$y\equiv\{\bm{y}_1,\bm{y}_2,\bm{y}_3\}$ are the
eigenvectors of $A+B$. In this new coordinate basis set the overlap integrand
takes the form
\begin{equation}
\label{O4pI}
\Phi_A(\bm{x}_{1},\bm{x}_{2},\bm{x}_{3})\Phi_B(\bm{x}_{1},\bm{x}_{2},\bm{x}_{3})
=\exp\left(-\frac{\beta_1y_1^2+\beta_2y_2^2+\beta_3y_3^2}{2}\right).
\end{equation}

In this set of eigencoordinates, the integration over the polar angles of
$\bm{y}_i$, vectors is easily carried out.  To fix the
hyperradius, we express the magnitude of the $\bm{y}_i$ vectors in
spherical coordinates, i.e. $y_1=R\sin\theta\cos\phi$,
$y_2=R\sin\theta\sin(\phi)$and $y_3=R\cos\theta$. In these coordinates
the overlap matrix elements reads
\begin{multline}
\label{O4pI3}
\braket{B|A}\Big|_R=(4 \pi)^3\int
\exp\left(-\frac{R^2(\beta_1\sin^2\theta\cos^2\phi+\beta_2\sin^2\theta\sin^2\phi+\beta_3\cos^2\theta)}{2}\right)\\
\sin^5\theta\cos^2\theta\cos^2\phi\sin^2\phi d\theta d\phi.
\end{multline}
The integration over one of the angles can be carried out
analytically. Introducing a variable dummy $y$, the overlap matrix
element takes the form
\begin{multline}
 \label{O4pI4}
 \braket{B|A}\Big|_R=\frac{(4\pi)^3 \pi  }{2R^2(\beta_1-\beta_2)}\int_0^1  \exp\left(-\frac{R^2}{4}
 [(\beta_1+\beta_2)(1-y^2)+2\beta_3y^2]\right)\\
 I_1\left[R^2\frac{(\beta_1-\beta_2)(1-y^2)}{4}\right]y^2(1-y^2)
 dy,
\end{multline}
where $I_1$ is the modified Bessel function of the first kind.

To simplify the interaction matrix element evaluation, it is advantageous to use a
Gaussian model potential as was used in the CG method. In this
case, the interaction term can be evaluated in the same way as the overlap term since the interaction is also a
Gaussian. Each pairwise interaction can be written as
$V_{ij}=V_0\exp(-\frac{r_{ij}^2}{2d_0^2})=V_0
\exp(-x^T.M^{(ij)}.x/(2d_0^2))$ (see Subsec.~\ref{AJac} for the
definition of $M^{(ij)}$). Therefore the interaction
matrix element has the structure
\begin{equation}
 \label{V1}
 \braket{B|V_{ij}|A}=V_0\int d\Omega
 \exp(-\frac{x^T.(A+B+M^{(ij)}/d_0^2).x}{2}).
\end{equation}
This integration can be performed following the same steps used for the overlap
matrix element. Equation~(\ref{O4pI4}) can be used directly if we
multiply it by $V_0$, and $\beta_1$, $\beta_2$  and $\beta_3$ are replaced by
the eigenvalues of $A+B+M^{(ij)}/d_0^2$. Note that for each pairwise
interaction, the matrix $M^{(ij)}$ changes and requires a new
evaluation of the eigenvalues.

The third term we need to evaluate is the hyperangular kinetic term
at fixed $R$. This kinetic term is proportional to the grand angular
momentum operator $\Lambda$, defined for the $N=3$ case as
\begin{equation}
\label{kinHR} \frac{\Lambda^2\hbar^2}{2\mu
R^2}=-\sum_i\frac{\hbar^2\nabla_i^2}{2
\mu}+\frac{\hbar^2}{2\mu}\frac{1}{R^{5}} \frac{\partial}{
\partial R} R^{5}\frac{\partial}{
\partial R}.
\end{equation}
The expression can be formally written as
\begin{equation}
\label{To} \mathcal{T}_\Omega=\mathcal{T}_T-\mathcal{T}_R,
\end{equation}
where
\begin{equation}
\label{Tto} \mathcal{T}_\Omega=\frac{\Lambda^2\hbar^2}{2\mu
R^2},\;\;\;\;\mathcal{T}_T=-\sum_i\frac{\hbar^2\nabla_i^2}{2
\mu},\;\;\mbox{and}\;\;\mathcal{T}_R=-\frac{\hbar^2}{2\mu}\frac{1}{R^{5}}
\frac{\partial}{
\partial R} R^{5}\frac{\partial}{
\partial R}.
\end{equation}
In typical calculations, $\mathcal{T}_\Omega$ is evaluated by
directly applying the corresponding derivatives in the hyperangles
$\Omega$. However, in this case, it is convenient to evaluate
$\mathcal{T}_T$ and $\mathcal{T}_R$ separately, since it is easier to differentiate over the Jacobi vectors and the hyperradius.
 These two matrix elements are not separately symmetric, but the angular
kinetic energy matrix, i.e., the total kinetic energy minus the hyperradial
kinetic energy, is symmetric. To obtain an explicitly symmetric
operator, we symmetrize the operation
$\braket{B|\mathcal{T}_\Omega|A}|_R=(\braket{B|\mathcal{T}_T-\mathcal{T}_R|A}|_R+\braket{A|\mathcal{T}_T-\mathcal{T}_R|B}|_R)/2$
and obtain
\begin{equation}
 \label{Tang1}
\braket{B|\mathcal{T}_\Omega|A}\Big|_R= \frac{(4 \pi)^3}{R^8}\int
 \exp\left(-\frac{\beta_1y_1^2+\beta_2y_2^2+\beta_3y_3^2}{2}\right)\Upsilon(y_1,y_2,y_3)y_1^2y_2^2y_3^2dy_1dy_2dy_3\Big|_R,
\end{equation}
where
\begin{multline}
 \label{Tang2}
 \Upsilon(y_1,y_2,y_3)=\frac{1}{2}\left\{\sum_{i=1}^3 \left[-3\beta_i+\left(\beta_i^2-2(A.B)_{ii}+\frac{
 \beta_i}{R^2}\right)y_i^2\right]\right.\\
\left. -\left(\sum_{i=1}^3\frac{
 \beta_i y_i^2}{R^2}
 \right)^2+\frac{\overline{(\bm{y}.\bm{A}.\bm{y})(\bm{y}.B.\bm{y})}}{R^2}\right\}.
\end{multline}
It is easy to show that $(A.B)_{ii}=\sum_{j=1}^3 a_{ij}b_{ij}$ since
$A$ and $B$ are symmetric matrices. Here the bar sign indicates the
integration over the angular degrees of freedom of $\bm{y}_1$,
$\bm{y}_2$, and $\bm{y}_3$. We then divide the total result
by $(4\pi)^3$. Making these integrations analytically we obtain
\begin{equation}
 \label{Tang3x}
 \overline{(\bm{y}.\bm{A}.\bm{y})(\bm{y}.B.\bm{y})}=\sum_{i=1}^3 a_{ii}
 b_{ii}
 y_i^4+\sum_{i>j}^3\left(a_{ii}b_{jj}+b_{ii}a_{jj}+\frac{4}{3}a_{ij}b_{ij}\right)y_i^2y_j^2.
\end{equation}
Rewriting the $y_i$ variables in spherical coordinates, we separate the hyperradial dependence in Eq.~\eqref{Tang1}. As in Eq.~\eqref{O4pI3}, one of the angular integrations
can be evaluated analytically and the final expression reduces to a one dimensional integral involving modified Bessel function of the first kind (see Ref.~\cite{vonstecherthesis} for more details).

The matrix elements involved in the $P$ and $Q$ couplings can be evaluated by following the above strategy, and it also reduces to a one dimensional numerical integration. The symmetrization of the matrix elements is handled just as in the standard CG method and is described in Appendix~\ref{ASymm}.

\subsubsection{General considerations}
\label{GenCon}

Many of the procedures of the standard CG method can be easily
extended to the CGHS. The selection, symmetrization, and
optimization of the basis set follow the the standard CG
method (see Appendices \ref{ASymm}, \ref{AJac},
\ref{SelectBasisSet}, \ref{ALinDep} and \ref{SVM}). However, the evaluation of the unsymmetrized matrix elements at fixed $R$ is
clearly different. Furthermore, the hyperangular Hamiltonian
[Eq.~\ref{LamHRnew}] needs to solved at different hyperradii $R$.

There are several properties that make the CGHS method particularly
efficient. For the model potential used, the scattering length is
tuned by varying the potential depths of the two-body interaction.
Therefore, as in the CG case, the matrix elements need only be
calculated once; then they can be used for a wide range of
scattering lengths. Of course, the basis set should be sufficiently complete
to describe the relevant potential curves at all desired
scattering length values.

The selection of the basis function generally depends on $R$. To
avoid numerical problems, the mean hyperradius of each basis
function $\braket{R}_B$ should be of the same order of the
hyperradius $R$ in which the matrix elements are evaluated. We can
ensure that $\braket{R}_B\sim R$ by selecting some (or all) the
weights $d_{ij}$ to be of the order of $R$.

We consider two different optimization procedures. The first
possible optimization procedure is the following: First, we select a
few basis functions and optimized them to describe the lowest few
hyperspherical harmonics.  The widths of these basis functions are
rescaled by $R$ at each hyperradius so that they represent the
hyperspherical harmonics equally well at different hyperradii. These basis functions are
used at all $R$, while the remaining are optimized at each $R$.
Starting from small $R$ (of the order of the range of the
potential), we optimize a set of basis functions. As $R$ is
increased, the basis set is increased and reoptimized. At every $R$
step, only a fraction of the basis set is optimized, and those basis
functions are selected randomly. After several $R$-steps, the
basis set is increased.

Instead of optimizing the basis set at each $R$, one can
alternatively try to create a complete basis set at large $R_{max}$.
In this case, the basis functions should be complete enough to
describe the lowest channel functions with interparticle distances
varying from interaction range $r_0$ up to the hyperradius
$R_{max}$.  Such a basis set can be rescaled to any $R<R_{max}$ and
should efficiently describe the channel functions at that $R$.  The
rescaling procedure is simply $d_{ij}/R=d_{ij}^{max}/R_{max}$. This
procedure avoids the optimization at each $R$. Furthermore, the
kinetic, overlap, and couplings matrix elements at $R$ are
straightforwardly related to the ones at $R_{max}$. The
interaction potential is the only matrix element that needs to be
recalculated at each $R$. This property can be understood by
dimensional analysis. The kinetic, overlap and couplings matrix
elements only have a single length scale $R$, so a rescaling of the widths is simply
related to a rescaling of the matrix elements. In
contrast, the interaction potential introduces a new length scale,
so the matrix elements depend on both $R$ and $d_0$, and the
rescaling does not work.

These two choices, the complete basis set or the small optimized
basis set, can be appropriate in different circumstances. If a large
number of channels are needed, the complete basis method is
often the best choice. But, if only a small number of channels are
needed, then the optimized basis set might be more efficient.

The most convenient method we have found to optimize the basis functions
in the four-boson and four-fermion problem is the following: First
we select a hyperradius $R_m$ that is $R_m\approx 300\, d_0$ where
the basis function will be initially optimized. The basis set is
increased and optimized until the relevant potential curves are
converged and, in that sense, the basis is complete. This basis is
then rescaled, as proposed in the second optimization method, to all
$R<R_m$. For $R>R_m$, it is too expensive to have a ``complete"
basis set. For that reason, we use the first optimization method
to find a reliable description of the lowest potential curves.

Note that for standard correlated Gaussian calculations, the
matrices $A$ and $B$ need to be positive definite. This condition
restricts the Hilbert space to exponentially decaying functions. In
the hyperspherical treatment, this is not necessary since the matrix
elements are always calculated at fixed $R$, even for
exponentially growing functions. This gives more flexibility in
the choice of optimal basis functions.

\section{Application to the Four-Fermion Problem}
\label{Sec:results}

This section presents our results for the four-body fermionic
problem using the methods discussed in Sections \ref{Sec:Var_bas}
and \ref{Sec:CorrGaussian}. Our finite-energy calculations for elastic and inelastic processes are compared to established zero-energy results and are seen to exhibit significant qualitative and quantitative differences. 
Several properties of trapped
four-fermion systems are also discussed, along with the connections between this few-body system and the many-body BEC-BCS crossover physics.

\subsection{Four-fermion potentials and the dimer-dimer wavefunction}

Calculation of the hyperradial potentials and channel functions using the
variational basis method of Sec.~\ref{Sec:Var_bas} is conceptually
simple. Matrix
elements of the hyperangular part of the full Hamiltonian are required,%
\[
H_{ad}=\dfrac{\hbar^{2}}{2\mu}\dfrac{\Lambda^{2}}{R^{2}}+\sum_{i,j}V\left(
r_{ij}\right)  ,
\]
where the sum runs over all interacting pairs of distinguishable fermions. 
Sec. \ref{Sec:Var_bas}, considered the specific two-body interaction to be general, but required the two-body potential to support 
a weakly bound dimer state (and hence a positive scattering
length much larger than the range of the interaction). At this point we adopt
the so-called P\"oschl-Teller potential,%
\begin{equation}
V\left(  r\right)  =-\dfrac{U_{0}}{\cosh^{2}\left(  r/r_{0}\right)
},\label{Eq:sechpot}%
\end{equation}
where $r_{0}$ is the range of the interaction. Unless otherwise stated $U_{0}$
is tuned so that $V\left(  r\right)$ gives the appropriate scattering length
with only a single shallow bound state. This potential is adopted because the bound
state wavefunctions and binding energies are known analytically
\cite{landau1981qmn}, but any two-body interaction could be used here, provided that one obtains the
wavefunctions and energies numerically or analytically.

Application of the variational basis results in a generalized eigenvalue problem,%
\begin{equation}
\bm{H}\left(  R\right)  \bm{x}_{\nu}\left(  R\right)  =U_{\nu}\left(
R\right)  \bm{S}\left(  R\right)  \bm{x}_{\nu}\left(  R\right)
\label{Eq:4Bod_ME}%
\end{equation}
where $U_{\nu}\left(  R\right)  $ is the $\nu$-th adiabatic hyperradial
potential, and $\bm{x}_{\nu}$ is the channel function expansion in the
variational basis. The matrix elements of $\bm{H}$ are given by matrix
elements of the adiabatic Hamiltonian at fixed hyperradius,%
\[
{H}_{nm}=\left\langle \Psi_{n}\left\vert H_{ad}\right\vert \Psi
_{m}\right\rangle .
\]
Because the variational basis is not orthogonal, a real, symmetric overlap matrix, $\bm{S}$,
appears in this matrix equation. While the method employed here is
conceptually simple, the actual calculation of the matrix elements is
numerically demanding because the interaction valleys in the hyperangular potential
surface, $\sum_{i,j}V\left(  r_{ij}\right)  $, become localized into narrow cuts of the
hyperangular space at large hyperradii. Further, examination of
Fig.~\ref{coalplots} shows that the locus of coalescence points where the interatomic potential is appreciable has a complicated structure
in the five dimensional body-fixed hyperangular space. To accurately calculate
the matrix elements in Eq.~(\ref{Eq:4Bod_ME}) numerically, a large number of
integration points must be placed within the interaction valleys.

\begin{figure}[htbp]
\begin{center}
\includegraphics[width=3in,angle=90]{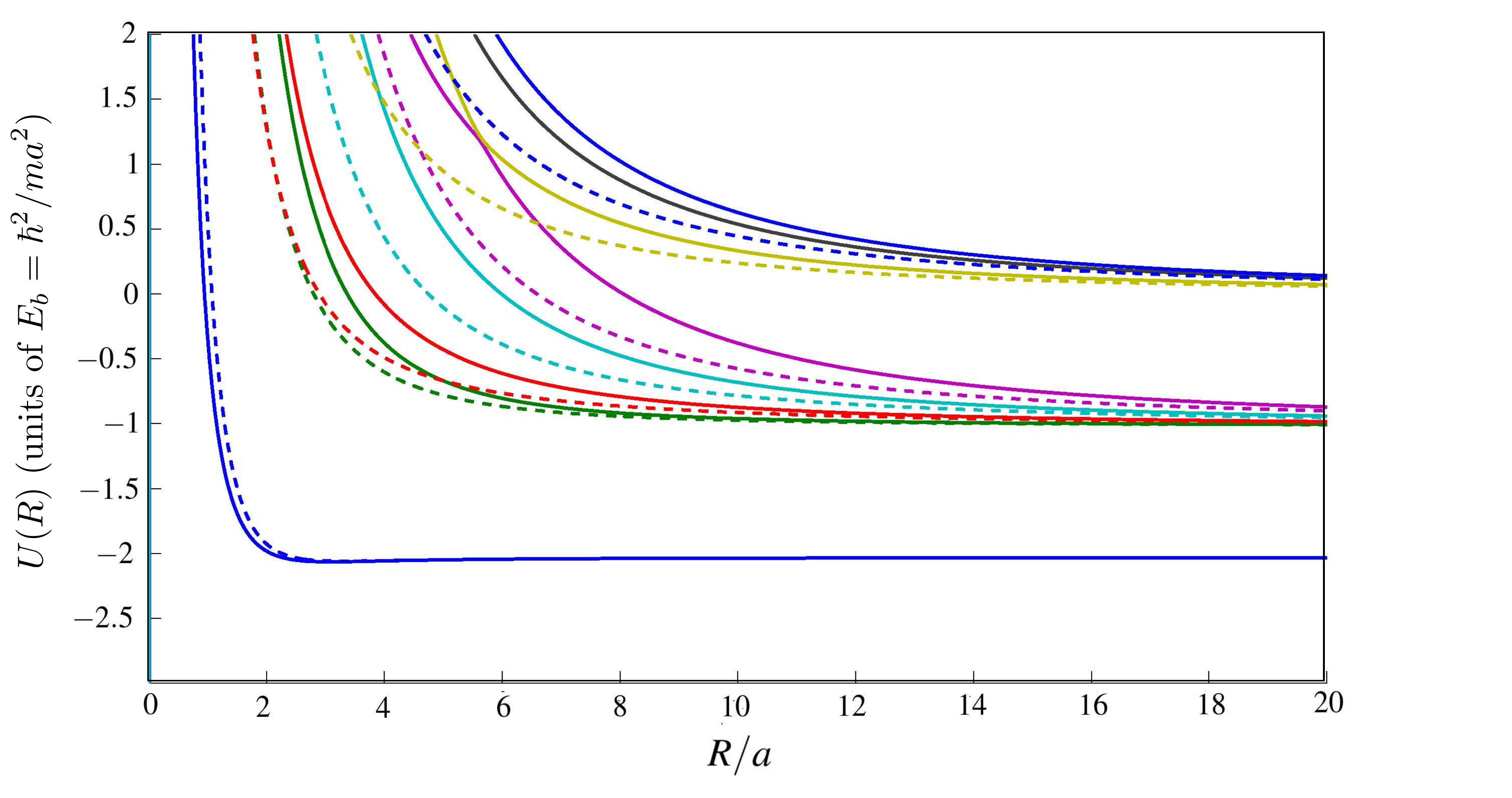}
\end{center}
\caption{The calculated hyperradial potentials (solid lines) for
$a=100 r_0$ are shown as a function of $R/a$. Also shown are the
expectation values of the fixed-$R$ Hamiltonian for the individual
variational basis functions (dashed curves).}%
\label{4bodpots}%
\end{figure}

Despite all of these complications, the adiabatic potential can be found
approximately. Figure \ref{4bodpots} shows the full set of hyperradial
potentials including the diagonal non-adiabatic correction (solid curves)
calculated using $8$ variational basis elements: one $2+2$ element, four
$2+1+1$ elements, and three $1+1+1+1$ elements. Also shown are the expectation
values of the basis elements themselves, i.e. the diagonal of $H(R)$
from
Eq.~(\ref{Eq:4Bod_ME}) (dashed curves). All calculations shown here
are performed for $a=100r_0$. It is clear that the lowest potential curve converges very quickly
with respect to the number of variational basis elements used. The lowest
potentials converge well when only a few variational basis functions are
included, while the higher potentials are somewhat suspect.
According to the universal theory of zero-range interactions, the hyperspherical potential curves should only depend on $a$ in the regime where $a$ is the dominant length scale in the problem.
Thus, in our finite range interaction calculations, the adiabatic
potentials should become universal in the $R \gg r_0$ regime for large scattering lengths, i.e. $a \gg r_0$. In other words, the
potentials should look the same when scaled by the scattering length and the binding
energy, $U_{\nu}(R \gg r_0) = (\hbar^2 / m a^2)u_{\nu}(R/a)$ where $u_{\nu}(x)$ is a universal function for the $\nu$th effective potential. Comparison with the potential curves computed in the correlated Gaussian method shows excellent agreement in the lowest dimer-dimer potential, and
reasonable agreement for the lowest few dimer-atom-atom potentials
\cite{vonstecherthesis}.

At large $R$, the lowest hyperradial adiabatic potential curve (see Fig.~\ref{4bodpots}) approaches the bound-state energy of two dimers that are approximately separated by a distance $R$.  It is natural to interpret processes for which flux enters and leaves this channel as "dimer-dimer" collisions.
Examining this potential further,
one can see that at hyperradii less than the scattering length, $R<a$, the adiabatic
dimer-dimer potential becomes strongly repulsive. This can be
visualized qualitatively as hard wall scattering, which would give
a dimer-dimer scattering length comparable to the two-body scattering length $a_{dd}\sim a$.
Higher potential curves approach the single dimer binding energy at large $R$,
indicating that these potentials correspond to a dimer with two free particles
in the large $R$ limit. Note that the variational basis functions described in Section~\ref{Sec:Var_bas} give the correct large $R$ adiabatic energies \emph{by construction}.
As the scattering length becomes much larger than the
range of the two-body potential, the effective four-fermion hyperradial potential becomes universal and independent of $a$. In the range of $r_{0}\ll
R\ll a$:%
\begin{equation}
U\left(  R\right)  \rightarrow\dfrac{\hbar^{2}}{2\mu}\dfrac{p_{0}^{2}%
-1/4}{R^{2}},
\end{equation}
where $p_{0}=2.55$. This universal potential was extracted in Refs. \cite{vonstecher2007bbc,vonstecher2008eas} by examining the
behavior of the ground state energy of four fermions in a trap in the
unitarity limit.

\begin{figure}[htbp]
\begin{center}
\includegraphics[width=3in]{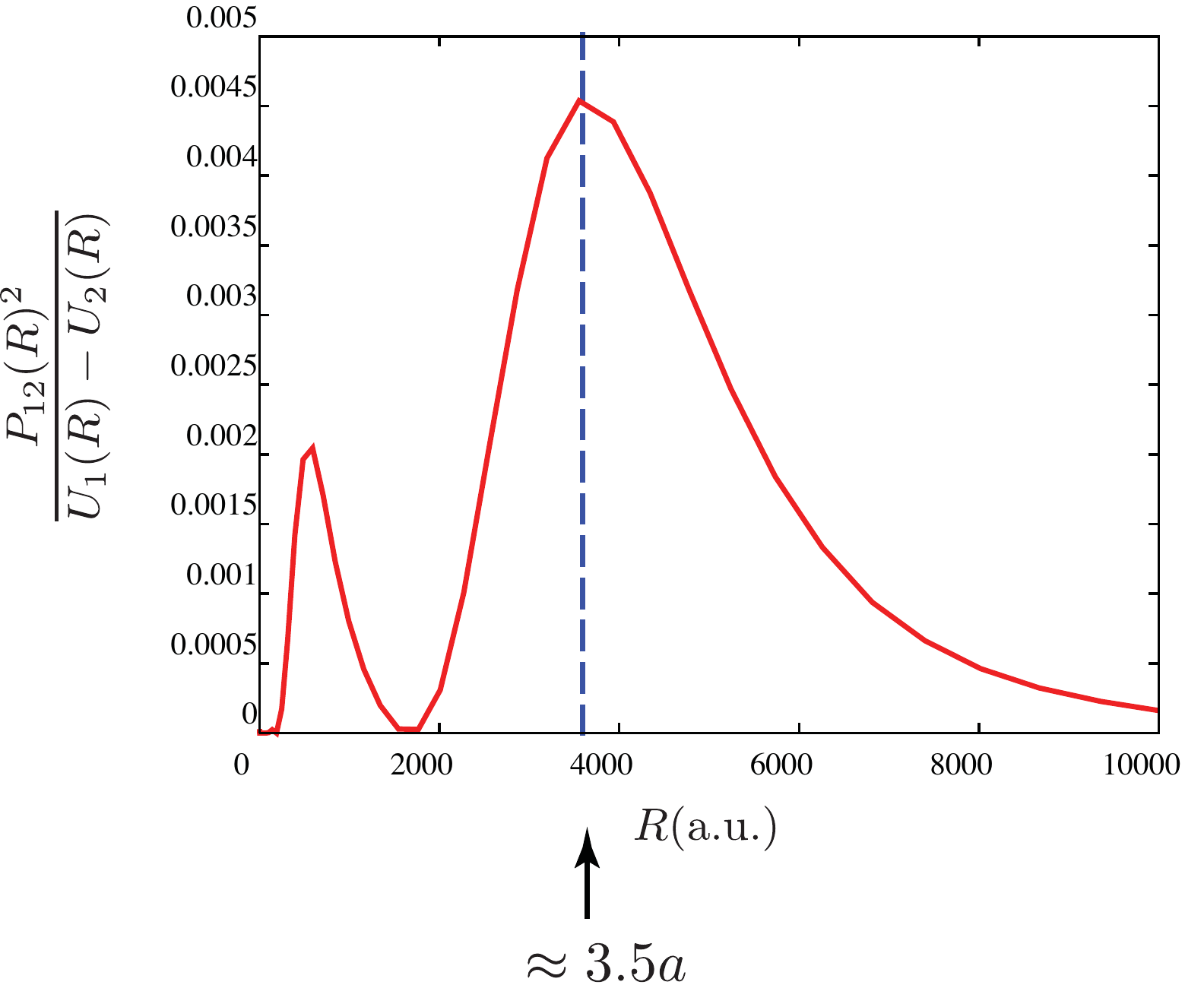}
\end{center}
\caption{The nonadiabatic coupling strength between the dimer-dimer potential and the
lowest dimer-atom-atom potential is shown as a function of $R$ ($r_{0}=100$ a.u. was choosen to be the van der Waals length of $^40$K). The blue
dashed line shows the position of the coupling peak at $R/a\approx3.5$.}%
\label{4bcoupling}%
\end{figure}

Figure \ref{4bcoupling} shows the coupling strengths, $\hbar^{2}P_{nm}%
^{2}/\left\{  2\mu\left[  \left(  U_{m}\left(  R\right)  -U_{n}\left(
R\right)  \right)  \right]  \right\}  $, between the dimer-dimer potential and
the lowest three dimer-atom-atom adiabatic potentials for a two-body
scattering length of $a=100r_{0}$. In each case the coupling strength peaks
strongly near the short range region, $R\sim r_{0}$, and near the scattering
length, $R\sim a$, and then falls off quickly in the large $R$ limit. This
behavior indicates that recombination -- from a state consisting of a dimer and two free particles to the dimer-dimer state -- occurs mainly at
hyperradii of the order of $a$. Looking at Fig.~\ref{4bcoupling} one might think that a
recombination path which occurs at small $R$, $R\sim r_{0}$, could also
contribute, but the strong repulsion in
the dimer-atom-atom potentials between $R\sim r_{0}$ and $R\sim a$, shown in Fig.~\ref{4bodpots}, suppresses this pathway.

\begin{figure}[htbp]
\begin{center}
\includegraphics[width=2.5in]{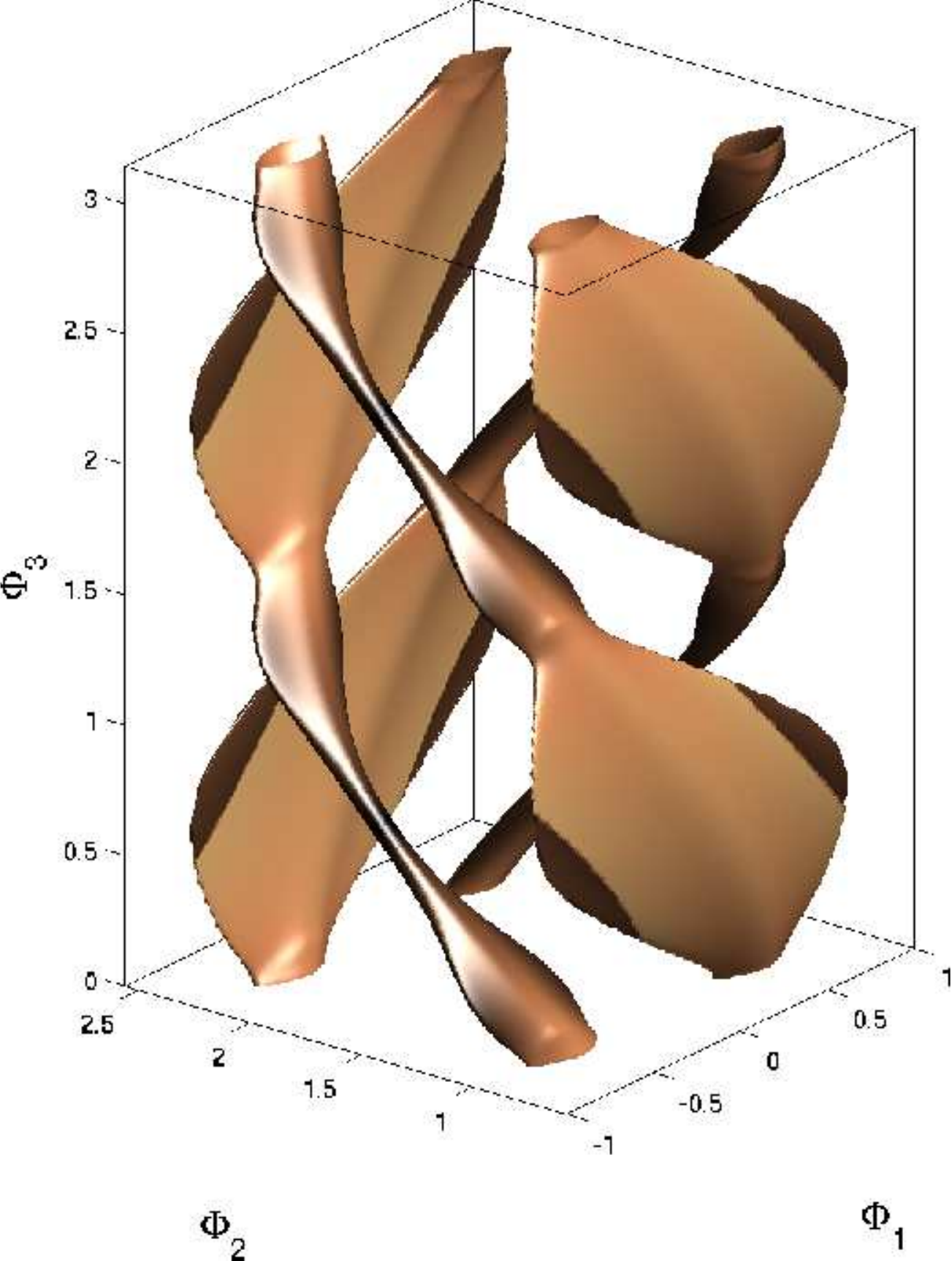}
\end{center}
\caption{An isosurface of the dimer-dimer probability density is shown. The
surfaces are found by integrating the total probability over $\theta_{1}$ and
$\theta_{2}$ and plotting with respect to the remaining democratic angles
$\left(  \phi_{1},\phi_{2},\phi_{3}\right)  $. The peak probability always
occurs in planar configurations, $\theta_{1}=\pi/2$. The coloring (light to dark)
indicates the value of $\theta_{2}$ at the peak.}%
\label{Cobra_plot}%
\end{figure}

\begin{figure}[htbp]
 \begin{center}
 \includegraphics[width=2.5in]{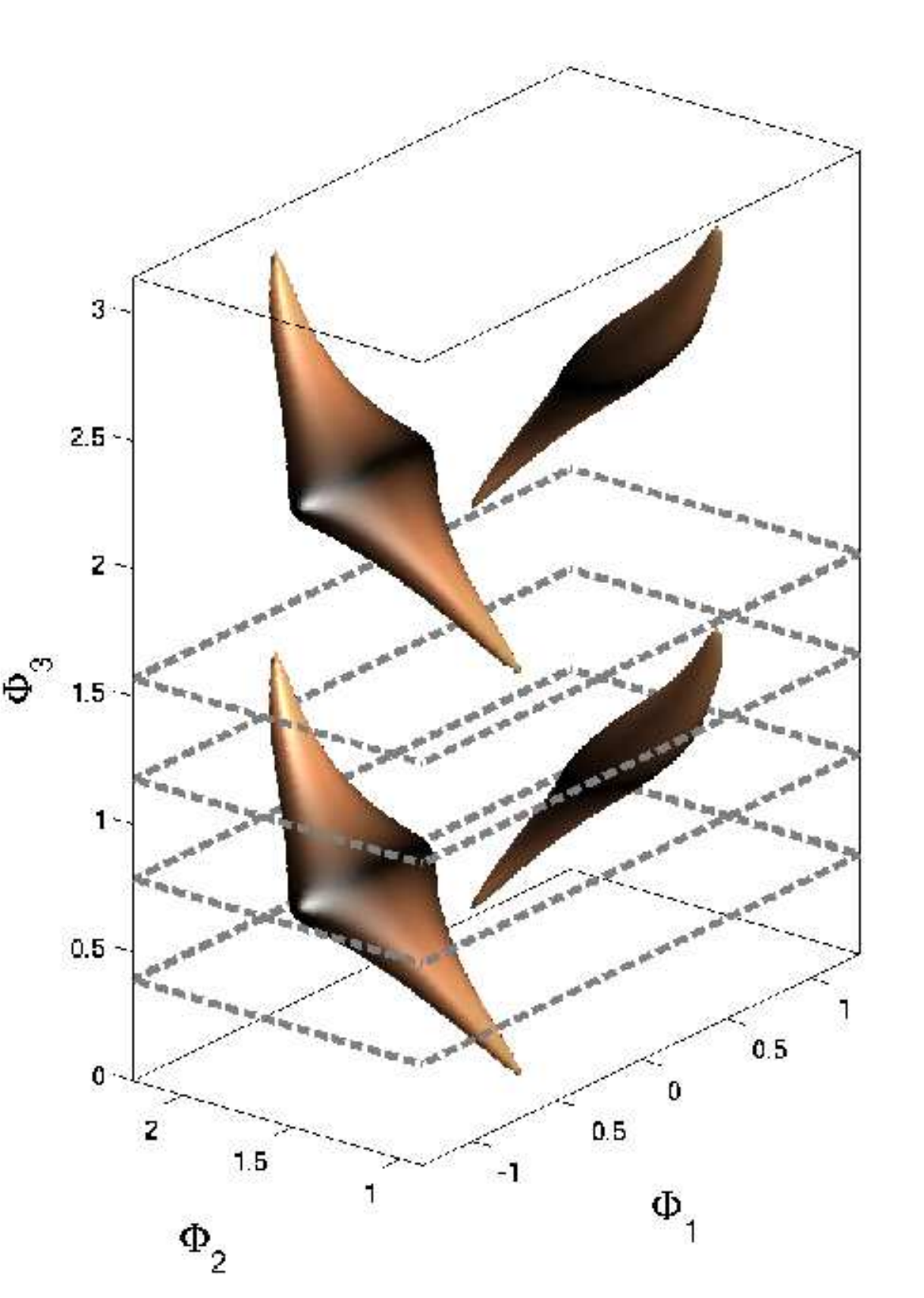}
 \end{center}
 \caption{The same as Fig.~\ref{Cobra_plot}, but using only the $2+2$ 
 function. The dashed gray lines are purely for perspective.}%
 \label{dd_only}%
 \end{figure}

Figure \ref{Cobra_plot} shows an isosurface of the hyperangular probability
density in the configurational angles $\left\{  \phi_{1},\phi_{2},\phi
_{3}\right\}  $ after integrating out $\Theta_{1}$ and $\Theta_{2}$ at a fixed
hyperradius of $R=0.41a$. The $\phi_{1}$ axis has been modified here by
shifting the region $\pi/2\leq\phi_{1}\leq\pi$ to emphasize the symmetry of
the system. Each cobra-like surface corresponds to a peak in the four-body
probability density. By examining Fig.~\ref{coalplots}, one sees that the
spine of each cobra corresponds to the locus of interaction coalescence points. For each
choice of $\left\{  \phi_{1},\phi_{2},\phi_{3}\right\}  $, the maximum of the
probability density in $\Theta_{1}$ and $\Theta_{2}$ is given in a planar
geometry, $\Theta_{1}=\pi/2$. The coloring of each cobra indicates the value
of $\Theta_{2}$ at which the maximum occurs. Darker colors indicate a more
linear geometry, i.e. $\Theta_{2}$ is closer to $0$. Figure \ref{dd_only}
shows the same plot for the $2+2$ basis function only. A comparison of Figs.~\ref{Cobra_plot} and \ref{dd_only} indicates that the added variational
basis elements are critical for describing the full dimer-dimer channel
function for hyperradii less than the scattering length.

\subsection{Elastic dimer-dimer scattering}

With the hyperradial potentials and non-adiabatic couplings in hand, low
energy dimer-dimer scattering properties can be examined. The zero-energy
dimer-dimer scattering length in the limit of large two-body scattering length
was first calculated by Petrov et. al \cite{petrov2004wbd} and found to be
\begin{equation}
a_{dd}\left(  0\right)  =0.60\left(  2\right)  a,\label{Eq:zeroEadd}%
\end{equation}
where the number in the parentheses indicates $\pm0.02$, the $2\%$ error
stated in Ref.~\cite{petrov2004wbd}. This result has been confirmed using several
different theoretical approaches
\cite{levinsen2006psp,vonstecher2007bbc,vonstecher2008eas,dincao2009ddc}.

Using the adiabatic potentials shown in Fig.~\ref{4bodpots} and the resulting
non-adiabatic couplings, the energy-dependent dimer-dimer scattering length
defined by%
\begin{equation}
a_{dd}\left(  E_{\operatorname{col}}\right)  =\dfrac{-\tan\delta_{dd}}{k_{dd}}
\label{Eq:addvsEdef}%
\end{equation}
can be calculated. Here $E_{\operatorname{col}}$ is the collision energy of
the two dimers with respect to the dimer-dimer threshold, and $\delta_{dd}$ is
the s-wave dimer-dimer phase shift. When the
collision energy becomes greater than the dimer binding energy,
the two dimers collide with enough energy to potentially dissociate one of them. When this
happens, the four fermion system can fragment in an excited hyperspherical channel causing a loss
of flux from the dimer-dimer channel. This process is parameterized by the
imaginary part of the dimer-dimer scattering length becomes non-zero
when $E_{\operatorname{col}}>E_{b}$.

\begin{figure}[htbp]
\begin{center}
\includegraphics[width=3in]{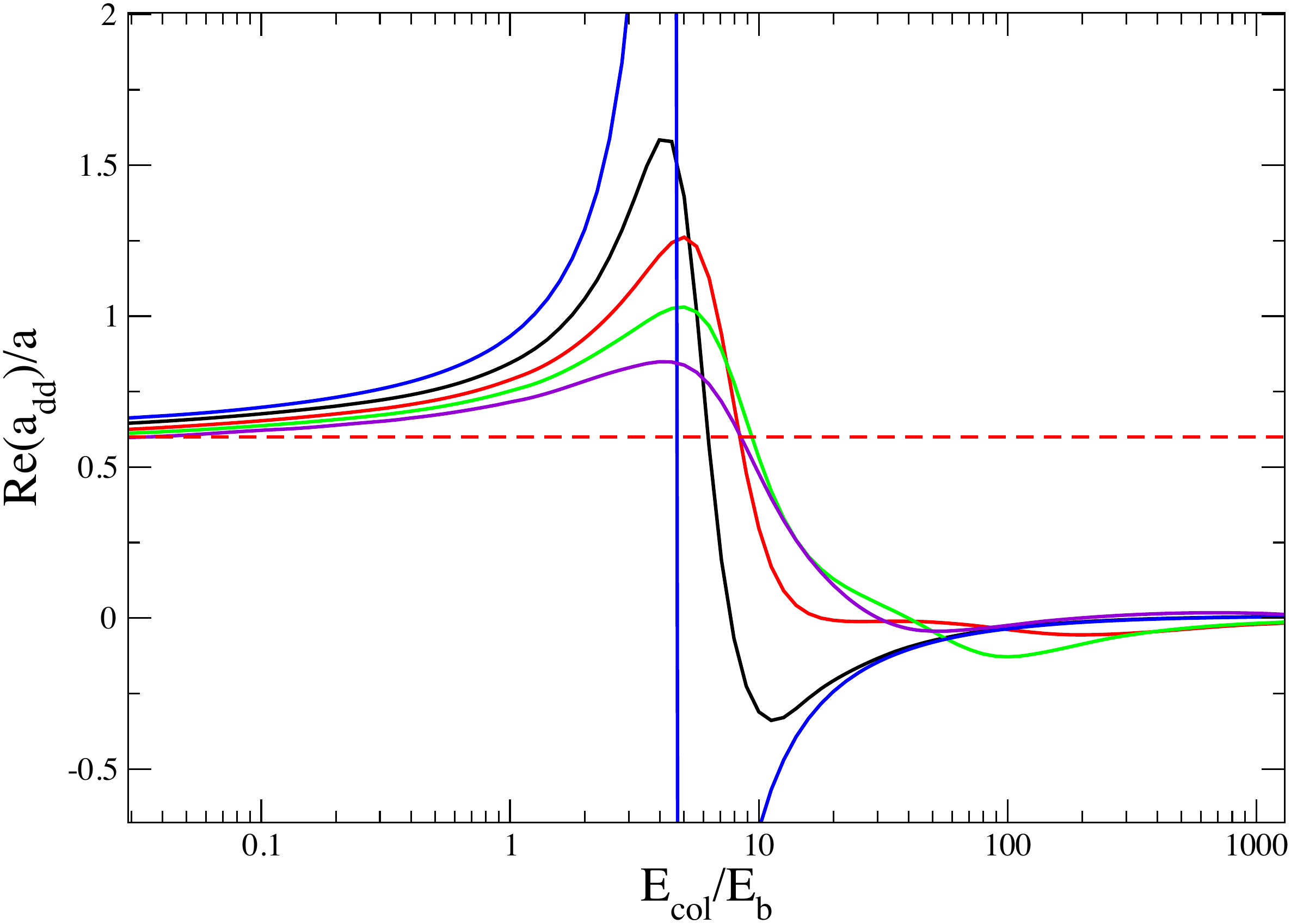}
\end{center}
\caption{The real part of the energy dependent dimer-dimer scattering length
is shown in units of the atom-atom scattering length $a$, plotted versus the collision energy in units of
the dimer binding energy. The calculation is carried out with one, two, three, four, and
five adiabatic channels (blue, black, red, green, and purple curves
respectively) from the adiabatic potential curves and couplings computed using $8$ basis functions. The red dashed line
shows $a_{dd}=0.6a$, the prediction of Ref. \cite{petrov2005spw}.}%
\label{Reddscatt}%
\end{figure}

\begin{figure}[htbp]
\begin{center}
\includegraphics[width=3in]{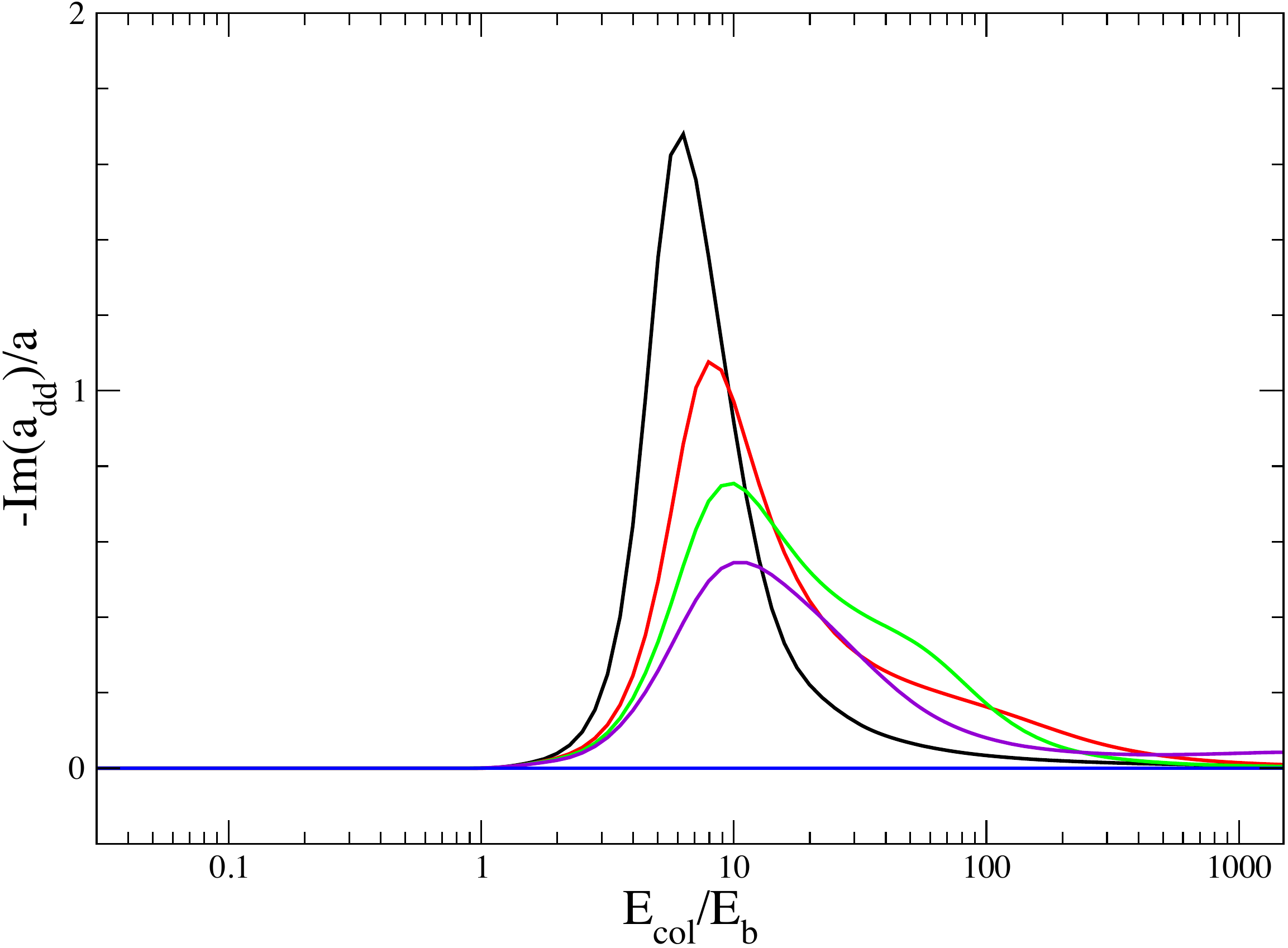}
\end{center}
\caption{The imaginary part of the energy dependent dimer-dimer scattering
length is shown in dimensions of $a$ plotted versus the collision energy in
units of the binding energy.
The calculation is carried out with one, two, three, four, and
five adiabatic channels (blue, black, red, green, and purple curves
respectively) from the adiabatic potential curves and couplings computed using $8$ basis functions.} %
\label{Imddscatt}%
\end{figure}

Figures \ref{Reddscatt} and \ref{Imddscatt} respectively show the real and
imaginary parts of the dimer-dimer scattering length calculated with different
numbers of adiabatic channels plotted as functions of $E_{\operatorname{col}%
}$ in units of the dimer binding energy. Also shown in Fig.~\ref{Reddscatt} is
the dimer-dimer scattering length calculated from the variational potential
that results from using a single variational basis element. It is important to
note that the single adiabatic channel calculation and the single basis
function calculation are not the same. In the former, the single potential used
is the lowest potential resulting from a calculation using multiple basis
functions, while the latter is the result of using only the $2+2$ variational
basis function and is guaranteed to be less accurate. Not surprisingly, the scattering length at collision energies
comparable to the binding energy depends strongly on the number of channels used.
With just a single channel in use, there is no decay pathway available for the
system. As more channels are included the system has more pathways into which it can fragment, which modifies the high energy behavior.

What is more surprising is the low energy behavior seen in Fig.~\ref{Reddscatt}. For a single variational basis element, the dimer-dimer zero
energy scattering length is found to be $a_{dd}=0.72a$, which is already
within 20\% of the result of Ref. \cite{petrov2004wbd}, $a_{dd}\left(
0\right)  =0.6a$. A single channel calculation using the dimer-dimer potential
and channel function that results from using 5 basis elements improves
considerably on this yielding $a_{dd}\left(  0\right)  =0.64a$, showing
that inclusion of correlations characteristic of two free particles at hyperradii less
than $a$ gives a significant contribution to the physics of dimer-dimer
scattering. It is somewhat unexpected that the single channel calculation is
only $8\%$ off of the predicted value. As the scattering energy approaches
zero, the higher fragmentation channels become strongly closed but still
apparently play a small role in the dimer-dimer scattering process. By
including progressively more channels in the scattering calculation the
zero-energy dimer-dimer scattering length can be extracted for large two body
scattering length:%
\begin{equation}
a_{dd}\left(  0\right)  =0.605\left(  5\right)  a.\label{Eq:add0}%
\end{equation}
This result is in agreement with the results of Ref.
\cite{vonstecher2007bbc,vonstecher2008eas} and the results of Section~\ref{extraction} which found the zero-energy
dimer-dimer scattering length to similar accuracy using different methods.

\subsection{Energy Dependent Dimer-Dimer Scattering}

\begin{figure}[htbp]
\begin{center}
\includegraphics[width=3in]{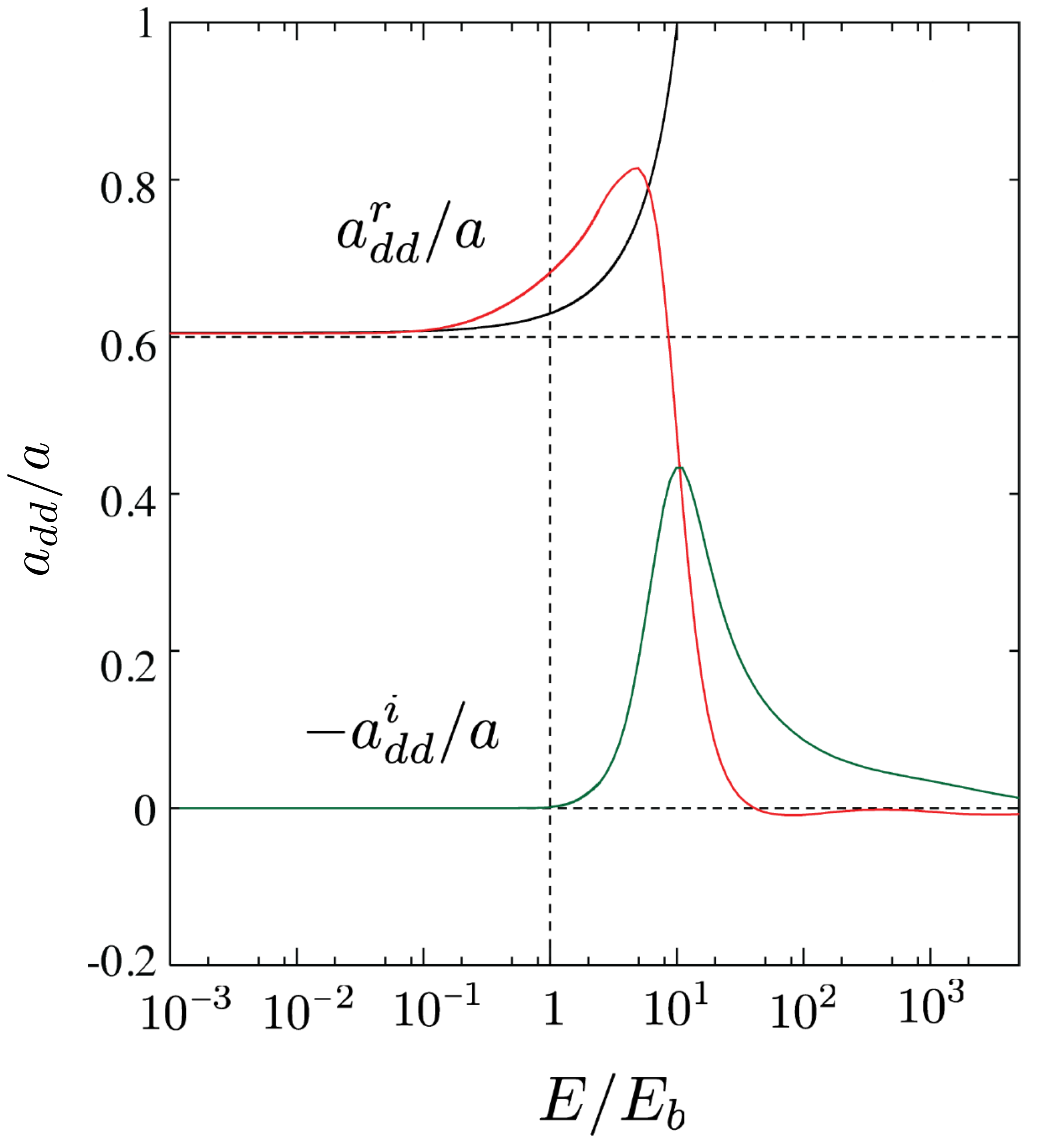}
\end{center}
\caption{The real (red) and imaginary (green) parts of the energy dependent
dimer-dimer scattering length are shown in units of $a$, plotted versus the
collision energy in units of the binding energy. Also shown is the energy
dependent scattering length predicted by the effective range expansion. Figure from
Ref. \cite{dincao2009ddc}.}%
\label{ReandImadd}%
\end{figure}

By examining the low energy behavior of the energy dependent dimer-dimer
scattering length, the effective range can be extracted. The two dimers
\textquotedblleft see" each other when their wavefunctions are overlapping,
i.e. when the hyperradius is approximately equal to the scattering length,
$R\sim a$. If one thinks of the effective range of an interaction as
proportional to the size of the interaction region, then one would expect the
effective range for dimer-dimer scattering to be proportional to the
scattering length. By fitting the low energy scattering phase shift to the
effective range expansion,%
\begin{equation}
\frac{-1}{a_{dd}(E)}=k_{dd}\cot\delta_{dd}=-\dfrac{1}{a_{dd}\left(  0\right)  }+\dfrac{1}{2}r_{dd}{k_{dd}}^{2} + \mathcal{O}({k_{dd}}^4)
\label{Eq:r0_expansion}%
\end{equation}
this intuitive behavior is born out, giving an effective range:%
\begin{equation}
r_{dd}=0.13(1)a, \label{Eq:r_effdd}%
\end{equation}
where $a$ is the two-body scattering length. Figure \ref{ReandImadd} shows
both the real and imaginary parts of the energy dependent dimer-dimer
scattering length as a function of collision energy in units of the binding
energy compared to the effective range expansion, Eq.~(\ref{Eq:r0_expansion}).
This clearly shows that, while the low energy behavior of dimer-dimer
scattering is well described by the effective range expansion, it is only
accurate over a small range of collision energies. In fact, for collision
energies larger than the binding energy, $a_{dd}\left(  E_{\operatorname{col}%
}\right)  $ actually turns over and decreases as dimer breakup channels become
open. Further, when the collision energy exceeds the dimer binding energy,
$E_{\operatorname{col}}=E_{b}$, the dimer-dimer scattering length becomes
complex, with an imaginary part that parameterizes inelastic processes. These
results indicate that both the real and imaginary dimer-dimer scattering
lengths are universal functions of the collision energy.  Specifically, they are insensitive to
the short range nature of the two-body interaction, for scattering lengths
much larger than the two-body interaction length scale, $r_{0}$. Because very
few basis functions were used in these calculations, the results at higher
energies, $E_{b}\ll E_{\operatorname{col}}\ll\hbar^{2}/mr_{0}^{2}$, are not
well converged, though their qualitative nature is expected to persist. Well
above the dissociation threshold, the real part of $a_{dd}$ exhibits an oscillatory behavior. These oscillations are caused by
interference between different scattering pathways. As more basis functions are
included and the high energy results converge, the large number of available
pathways generally washes out the oscillatory behavior and produces incoherence , but the decrease in
the real part of the $a_{dd}$ at higher energies is expected to survive as the calculations become better converged.

\begin{figure}[htbp]
\begin{center}
\includegraphics[width=3in]{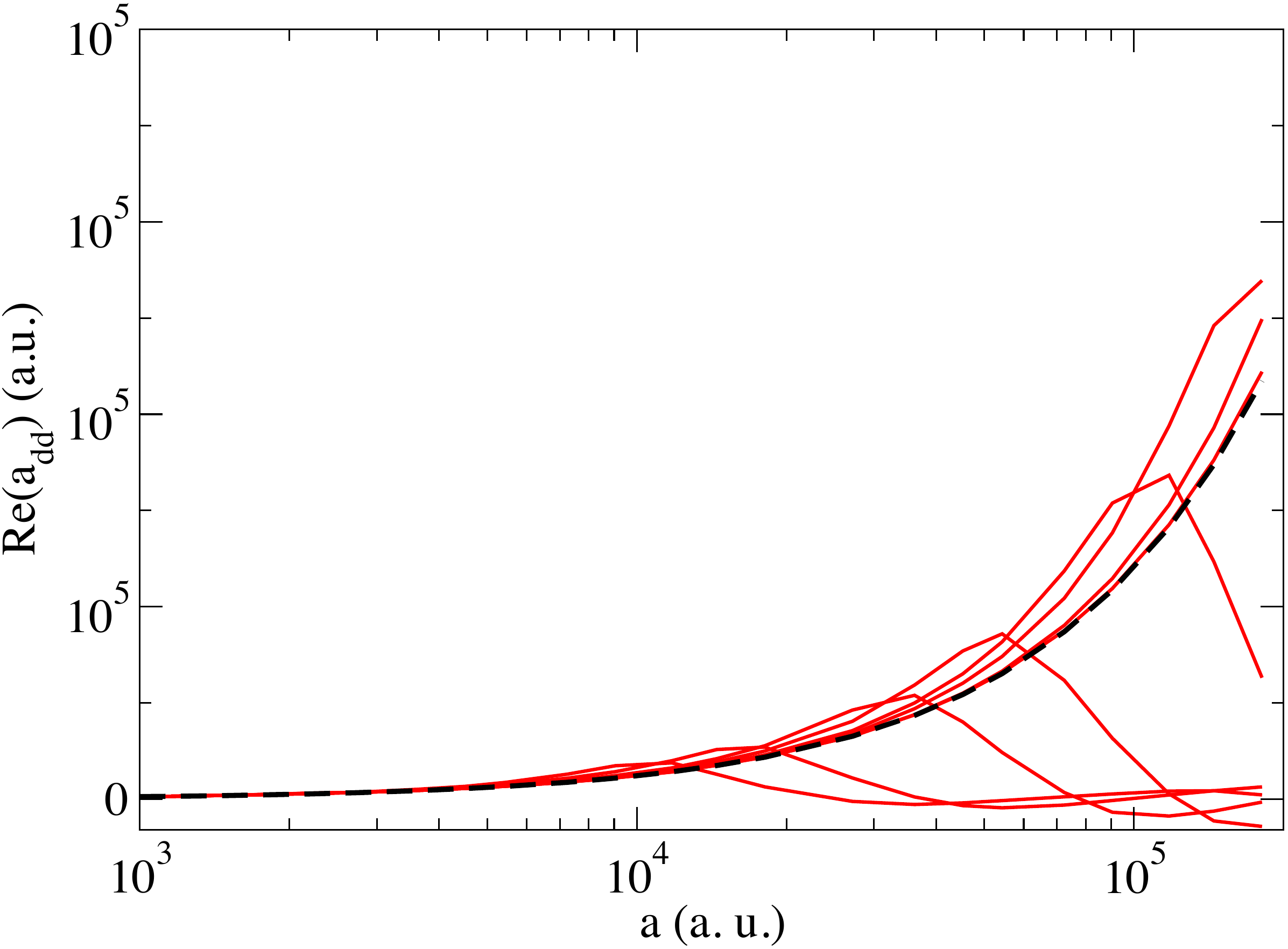}
\end{center}
\caption{The real part of the energy dependent dimer-dimer scattering length
is shown as a function of the two-body scattering length in atomic
units for several collision energies:
$E_{\operatorname{col}}/k_{b}=$250nK,100nK, 25nK, 10nK, 2.5nK, 1nK, $10^{-1}%
$nK, and $10^{-2}$nK. Also shown is the zero energy prediction (black dashed
curve). Figure from Ref. \cite{dincao2009ddc}.}%
\label{addvsa}%
\end{figure}

The dependence of $a_{dd}$ on $a$ at finite collision energy is particularly interesting. In
the large $a$ limit, the dimer binding energy becomes
$E_{b}\approx\hbar^{2}/ma^{2}$, so that as $a$ increases, the
binding energy decreases. At the critical value of the scattering length,%
\[
a_{c}=\dfrac{\hbar}{\sqrt{mE_{\operatorname{col}}}},
\]
the collision energy coincides with the binding energy, and the dimer-atom-atom channel becomes open. As a result (see Fig.~\ref{ReandImadd}), one expects the real part of $a_{dd}$
to turn over and remain finite for all values of $a$. This behavior is
demonstrated in Fig.~\ref{addvsa} which compares the real part of the dimer-dimer
scattering length at several fixed collision energies with the
zero-energy result, $a_{dd}\left(  0\right)  =0.6a$. The scattering length scale
has been fixed by setting the range of the interaction to be approximately the Van
der Waals length of $^{40}$K, $r_{0}\approx100$ a.u. Another aspect of the
finite collision energy behavior is that at large scattering length, the
dimer-atom-atom channels become open, and dimer dissociation is allowed. Thus, near unitarity the Fermi gas might be viewed as a
coherent mixture of atoms and weakly bound dimers.

\subsection{Dimer-dimer relaxation}
\label{Subsec:results:relax}

A significant loss process in an ultracold gas of bosonic dimers is that of
dimer-dimer relaxation, in which two dimers collide and in the process at least one of
the dimers relaxes to a deeply bound two-body state. The extra binding energy
is released as kinetic energy which is sufficient to eject the remaining fragments
from the trap. This process was studied by Petrov, Salomon and Shlyapnikov
\cite{petrov2004wbd,petrov2005spw}, who assumed that the relaxation rate is controlled
by the probability for three particles to be found in close proximity to one another. With
this assumption and the further assumption that the fourth particle is far
away and plays no role in the scattering process, they predict that the
relaxation rate is suppressed at large two-body scattering lengths with a
scaling law, $V_{rel}^{dd}\propto a^{-2.55}$.

Here we introduce a new method for finding the dimer-dimer relaxation rate
based directly on Fermi's golden rule. The key observation in this section is
that the final allowed states appear as an infinite set
of hyperspherical potentials corresponding to a deeply bound dimer with two
free atoms. The transition rate to a single one of these potentials can be
described by the Fermi-popularized golden rule, i.e.%
\begin{equation}
T_{p}^{\lambda}\propto\left\vert \left\langle \Psi_{dd}\left(  R;\Omega
\right)  \left\vert V\left(  R,\Omega\right)  \right\vert \Psi_{\lambda
}\left(  R,\Omega\right)  \right\rangle \right\vert ^{2}%
.\label{Eq:Fermi_golden}%
\end{equation}
Here $\Psi_{\lambda}$ is the final outgoing state, $\Psi_{dd}$ is the
dimer-dimer wavefunction, and $V\left(  R,\Omega\right)  $ is the sum of the
two-body interactions. This matrix element and the sum of probabilities over
final states are evaluated in Appendix~\ref{App:DD_relax}. The final result of this analysis is
expressed as an integral over the hyperradius,%
\begin{equation}
V_{rel}^{dd}\propto\int\dfrac{P_{WKB}\left(  R\right)  \mathcal{F}\left(
R\right)  }{R\kappa\left(  R\right)  }\rho\left(  R\right)
dR\label{Eq:Relaxation_rate}%
\end{equation}
where $P_{WKB}\left(  R\right)  $ is the WKB probability density of the dimer-dimer
wavefunction at hyperradius $R$, $\kappa\left(  R\right)  $ is the WKB
wavenumber:%
\begin{equation}
\kappa\left(  R\right)  =\sqrt{\dfrac{2\mu}{\hbar^{2}}\left(  V_{dd}\left(
R\right)  +\dfrac{\hbar^{2}}{2\mu}\dfrac{1/4}{R^{2}}-E_{\operatorname{col}%
}\right)  }.\label{Eq:WKB_wave_num}%
\end{equation}
In Eq.~(\ref{Eq:Relaxation_rate}) $\rho\left(  R\right)  $ is the nearly
constant density of final states, and $\mathcal{F}\left(  R\right)  $ is the
probability for three particles to be near one another in the
dimer-dimer wavefunction at hyperradius $R$:%
\begin{equation}
\mathcal{F}\left(  R\right)  =\left\langle \Phi_{dd}\left(  R;\Omega\right)
\left\vert f\left(  R,\Omega\right)  \right\vert \Phi_{dd}\left(
R;\Omega\right)  \right\rangle .\label{Eq:proxinity_opp}%
\end{equation}
Here $\Phi_{dd}$ is the hyperangular dimer-dimer channel function, and
$f\left(  R,\Omega\right)  $ is a proximity function that is appreciable only when
three particles are all approximately within the range of the two-body interaction.

\begin{figure}[htbp]
\begin{center}
\includegraphics[width=3in,angle=90]{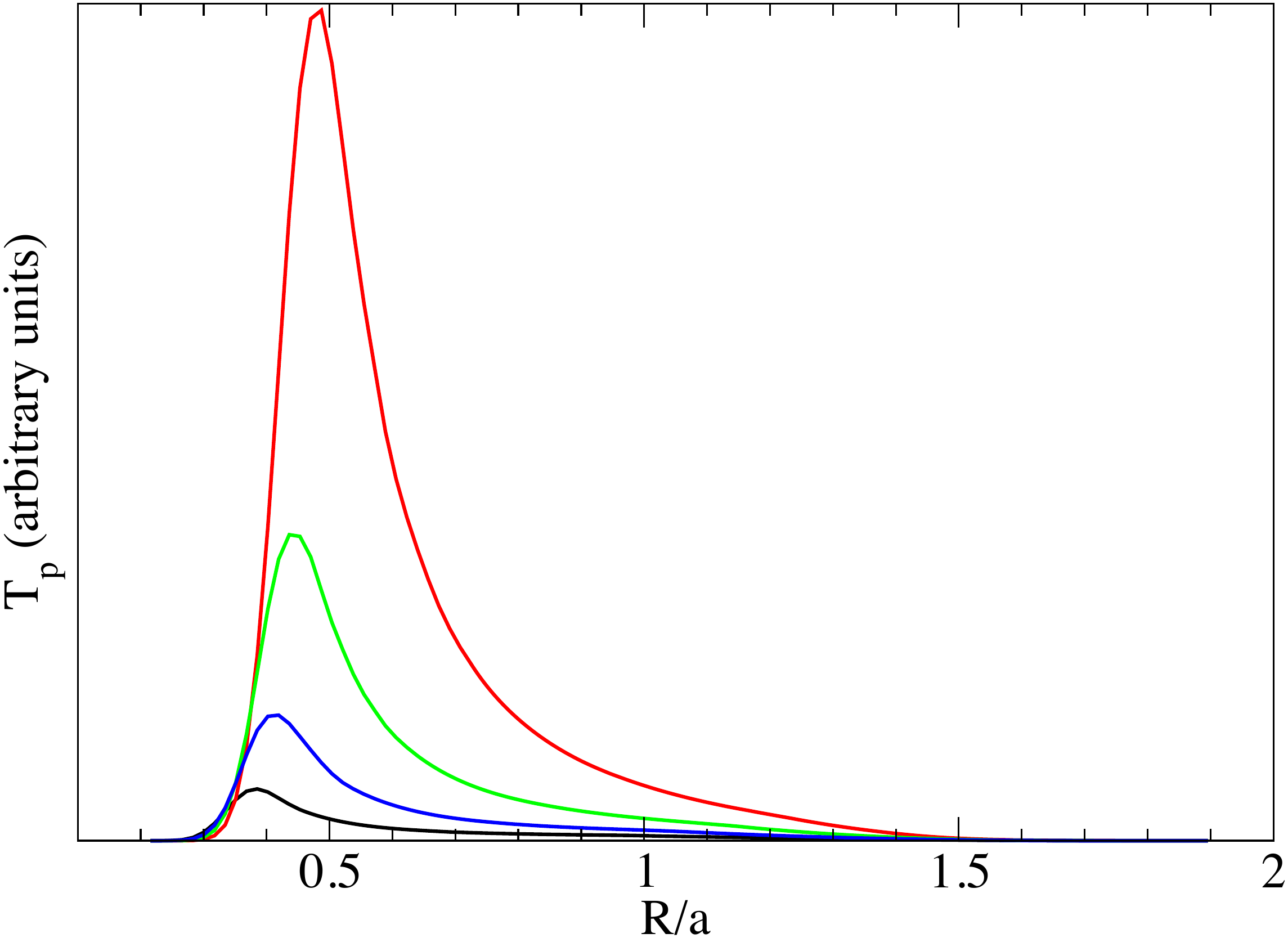}
\end{center}
\caption{The integrand from Eq.~(\ref{Eq:Relaxation_rate}) is shown for
$a=50r_{0}$ (red),$64r_{0}$ (green),$80r_{0}$ (blue), and $100r_{0}$ (black) as a function of $R/a$.}%
\label{TpvsR}%
\end{figure}

Equation (\ref{Eq:Relaxation_rate}) makes physical sense upon closer
examination. It says that the rate at which a dimer relaxes to a deeper state
is determined, with some extra factors, by the probability that three
particles are close enough together so that two of them can fall into a deeply
bound state and release the extra binding energy to the third particle. Figure~\ref{TpvsR} shows the integrand from Eq.~(\ref{Eq:Relaxation_rate}) for several
scattering lengths as a function of the hyperradius in units of the scattering
length. This quantity can be interpreted as being proportional to the
transition rate per unit hyperradius, i.e. the probability that the transition
will occur between $R$ and $R+dR$. The full transition rate is determined
by the nature of the interaction at short range and is not predictable using
this method. By examining the relaxation rate as a function of scattering
length, however, a scaling law can be extracted at each fixed hyperradius.

\begin{figure}[htbp]
\begin{center}
\includegraphics[width=3in]{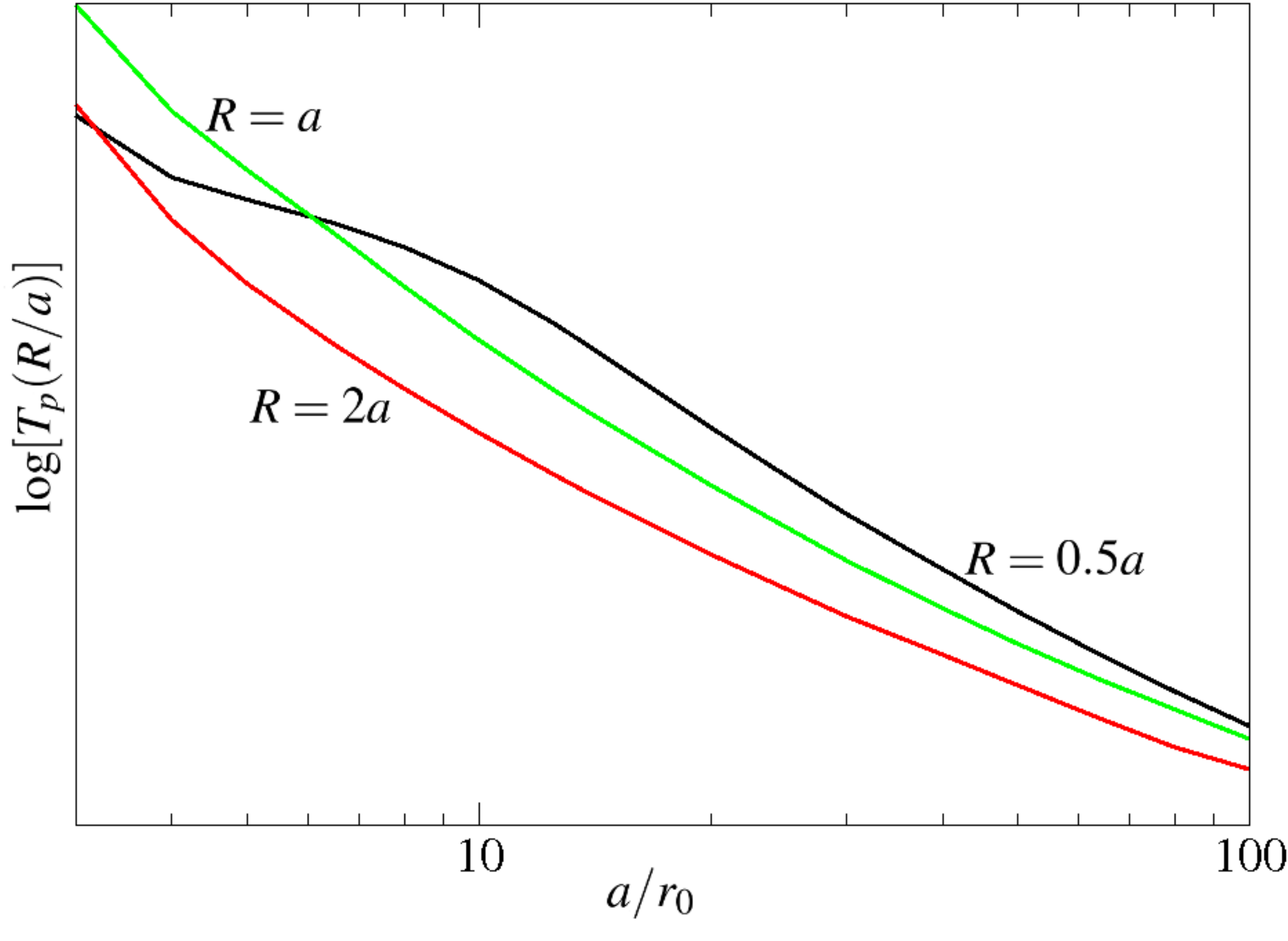}
\end{center}
\caption{The logarithm of the integrand from Eq.~(\ref{Eq:Relaxation_rate}) is shown for
several \emph{fixed} values of $R$ as a function of $a/r_0$}
\label{TpatR}%
\end{figure}

Figure \ref{TpatR} shows the relaxation rate per unit hyperradius for several
fixed values of $R/a$ as a function of the scattering length, $a$. The large
$a$ behavior in each case appears to follow a scaling law, but the scaling law
changes with $R/a$. This behavior indicates that, contrary to the prediction
of Ref.~\cite{petrov2005spw}, when the integral in Eq.~(\ref{Eq:Relaxation_rate}) is evaluated, the relaxation rate will not be
determined by a simple power law. By integrating over different hyperradial
regions, contributions to the transition rate from different processes can be
extracted. For instance, if the integral in Eq.~(\ref{Eq:Relaxation_rate}) is
performed only over small hyperradii, $R\lesssim5r_{0}$, the result is the
transition rate due to processes in which all four particles are in close
proximity. If the integral is evaluated over larger hyperradii, $R>10r_{0}$,
the result is the rate due to three-body processes influenced by the presence of the fourth particle.

\begin{figure}[htbp]
\begin{center}
\includegraphics[width=3in]{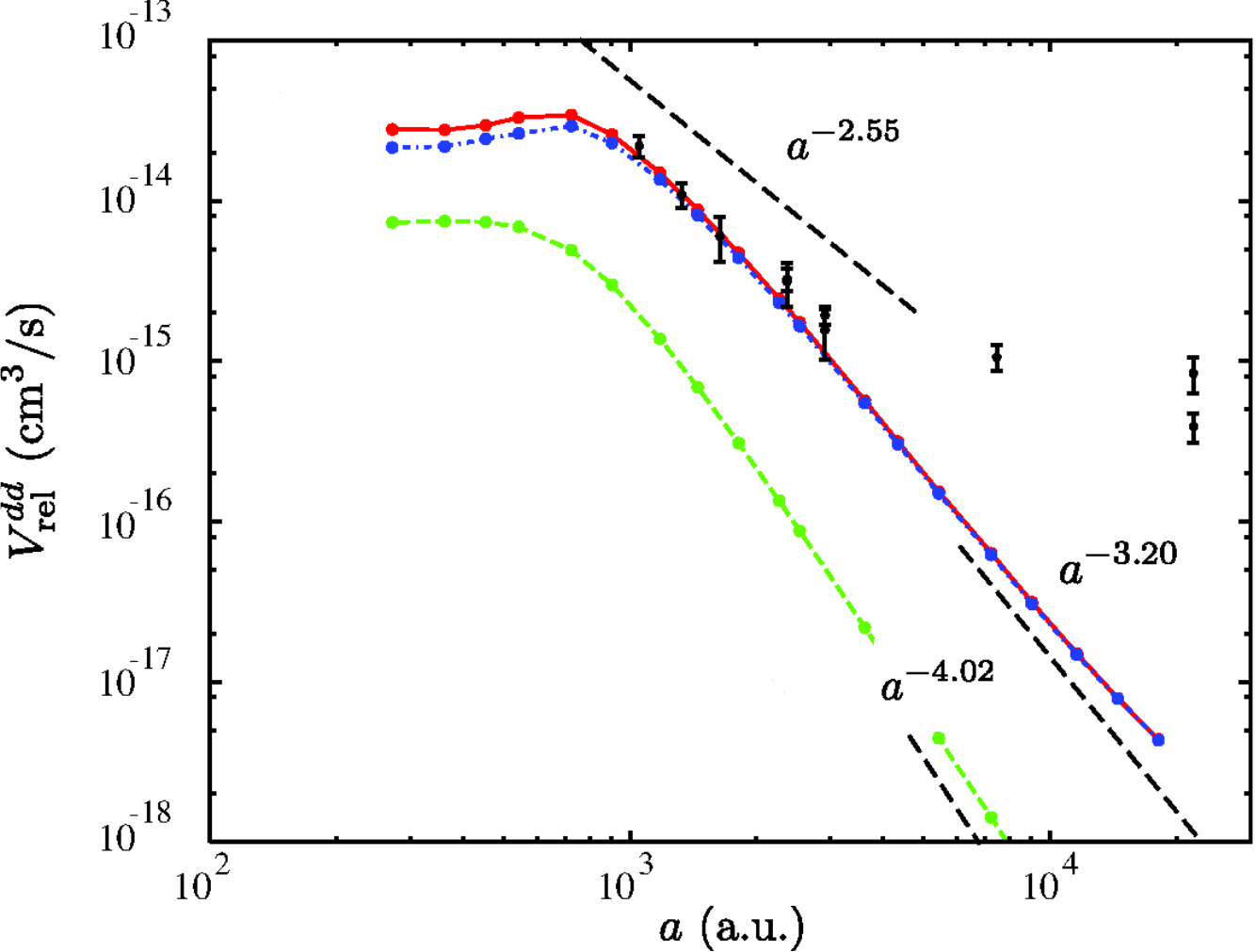}
\end{center}
\caption{The relaxation rate in a dimer-dimer collision is shown as a function of the atom-atom scattering length (see
text). Figure from Ref. \cite{dincao2009ddc}}%
\label{Vrelfig}%
\end{figure}

Figure \ref{Vrelfig} shows the relaxation rate as a function of the
scattering length in atomic units as a solid line. In this result the range of
the interaction is set to the van der Waals length of $^{40}$K, $r_{0}%
\approx100$ a.u. Also shown in Fig.~\ref{Vrelfig} are the contributions to this
relaxation rate due to four-body processes (dashed blue curve), and due to three-body processes (dotted green curve).
Also shown is the expected scaling law for transitions that occur at small
hyperradius, $R=5r_{0}$. Because the hyperradius is small in this regime, the
probability of three particles being in proximity is near unity, meaning that
the transition probability per unit hyperradius is determined by the
probability that the system can tunnel through the repulsive potential seen in
Fig.~\ref{4bodpots} at $R\lesssim a$. The universal repulsive potential in
this regime \cite{vonstecherthesis,dincao2009ddc},%
\begin{align}
U\left(  R\right)   &  =\dfrac{\hbar^{2}}{2\mu}\dfrac{p_{0}^{2}-1/4}{R^{2}},\\
p_{0}  &  =2.55,
\end{align}
leads to a scaling law for transitions in the small $R$ regime that behaves as%
\begin{equation}
V_{rel}^{dd}\propto a^{1-2p_{0}}=a^{-4.20}.
\end{equation}
Figure \ref{Vrelfig} also shows the experimentally determined relaxation rates
from Ref. \cite{regal2004lma}. Both the scaling law predicted in Ref.
\cite{petrov2005spw} of $a^{-2.55}$ and the prediction using
Eq.~(\ref{Eq:Relaxation_rate})
are consistent with the experimental data in the regime for $1000$
a.u.$\lesssim a\lesssim4000$ a.u.. The experimental data for $a>3000$ a.u. are
in the regime where the average dimer separation is less than the
dimer size, where the dimer-dimer scattering picture discussed here no longer applies.

\subsection{Trapped four-body system}

In this section we abandon the hyperspherical methods of the previous sections and examine the case of trapped four-fermion systems. The four-body system in confined geometries has recently become computationally accessible.
In particular, the trapped two-component Fermi system has been
intensely studied in the last
few-years and has become a benchmark for different theories and universal predictions.
One of the challenges present in the study of universal four-body physics in
dilute gases with short range forces is the description of disparate length scales associated with the large interpaticle distances and the short-range two-body interactions.

In this section, we analyze the spectrum, dynamics and universal properties of four-body solutions in confined geometries.
The four-body system is described by a the model Hamiltonian
\begin{eqnarray}
\label{eq_ham} H = \sum_{i=1}^{2} \left(\frac{-\hbar^2}{2m_1}
\nabla_i^2 + \frac{1}{2} m_1 \omega^2 \bm{r}_i^2 \right) +
\sum_{i'=1}^{2} \left( \frac{-\hbar^2}{2m_2} \nabla_{i'}^2 +
\frac{1}{2} m_2 \omega^2 \bm{r}_{i'}^2 \right) +\sum_{i=1}^{2}
\sum_{i'=1}^{2} V(r_{ii'})
\end{eqnarray}
where unprimed indices label the fermionic species with mass $m_1$, primed indices label the species with mass
$m_2$, and $\bm{r}_i$ is the position vector of
the $i$th fermion. The trapping frequency $\omega$ is assumed to be equal for both species.
 In order to facilitate a calculation with the CG method described in Section~\ref{Sec:CorrGaussian}, we take the interaction potential $V$ to be a purely
attractive Gaussian (see Eq.~(\ref{potNumChap})) and tune the depth of $V$ to give the desired (large) scattering length. The mass ratio
$\kappa$ is defined by $m_1/m_2$, and throughout the analysis we
assume $m_1 \ge m_2$. A trap
length $a_{ho}^{(i)}=\sqrt{\hbar/m_i\omega}$ is defined for each species as well as a trap length associated with the pair
$a_{ho}=\sqrt{\hbar/2\mu\omega}$, where $\mu=m_1m_2/(m_1+m_2)$. For
equal mass systems, $a_{ho}=a_{ho}^{(1)}=a_{ho}^{(2)}$.

This section reviews a series of predictions for the two-component
four-fermion system. The spectrum and structural
properties of the four-fermion system are analyzed throughout the BCS-BEC
crossover, followed by an exploration of the system dynamics as the
scattering length is tuned close to a Fano-Feshbach resonance.
Finally, we review a series of numerical studies that confirm
and quantify universal predictions.

\subsubsection{Spectrum in the BCS-BEC crossover}
\label{spect}

To obtain the $J=0$ spectra for
the four-fermion system in the BCS-BEC crossover problem, 
the CG method (Section~\ref{Sec:CorrGaussian}) is utilized to solve the time-independent Schr\"odinger
equation for different values of $a$. Like most numerical methods,
this method provides an adiabatic spectrum (in time), i.e., the energies of
the spectrum are labeled according to their energy values as $a$
changes. The four-body spectra present a series of crossings
or narrow avoided crossings when the scattering length is tuned across
the BCS-BEC crossover. For this reason, it is convenient to use a
representation where these narrow avoided crossings are treated
diabatically, and the spectrum smoothly evolves from the BCS to the
BEC side. The diabatic representation is more relevant from the
physical point of view since the diabatic states are usually
associated with good or ``approximately good" symmetries of the
problem.

\begin{figure}[htbp]
\begin{center}
\includegraphics[scale=0.1]{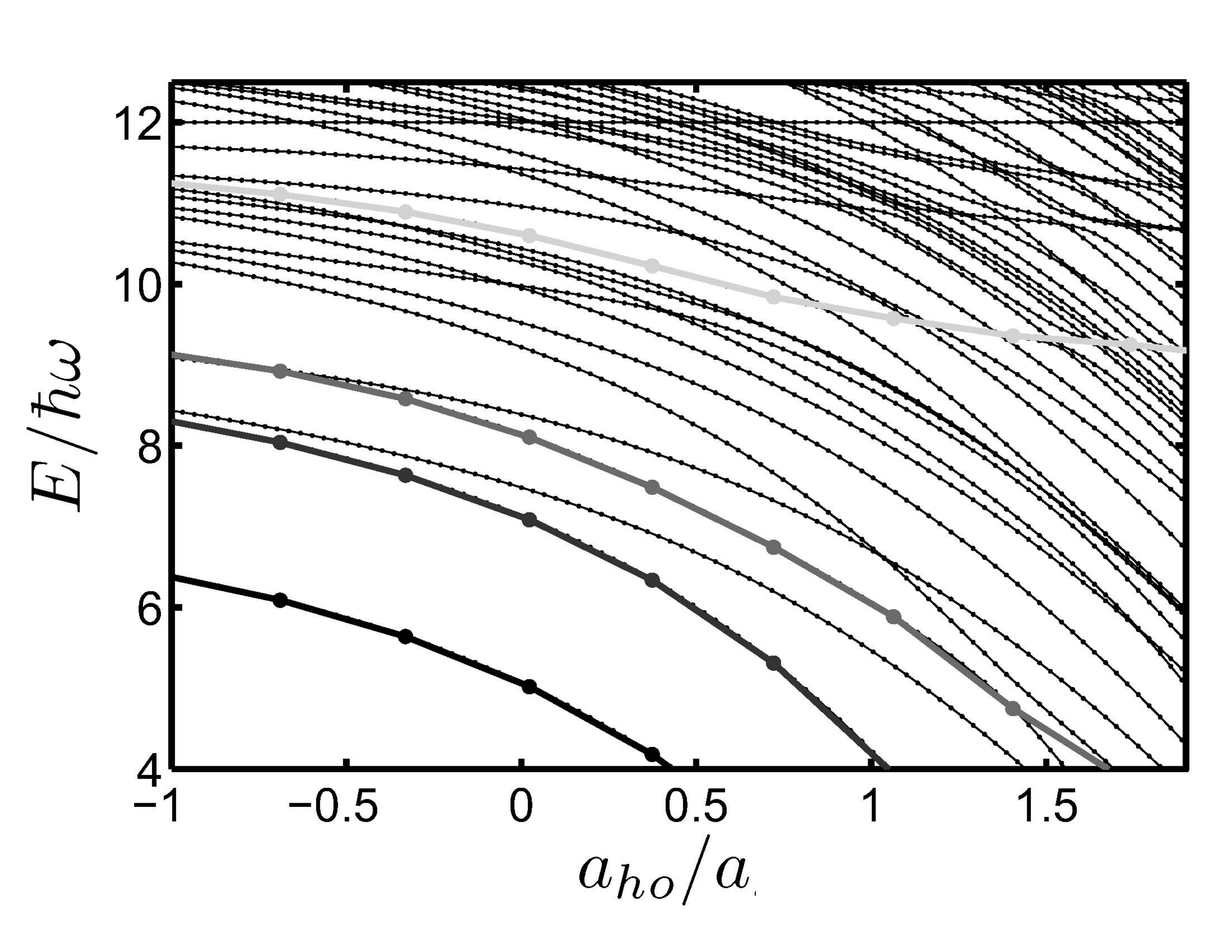}
\caption{  Four-fermion energy spectrum as a function of $\lambda=1/a$ in the
unitarity region with $J=0$. The thin solid black lines
correspond to the adiabatic spectrum. The wide black line with
circles is the diabatic ground state labeled $\Psi_{DD1}$. The blue
curve with circles is the diabatic first-excited state labeled
$\Psi_{DD2}$, the wide red curve with circles is the diabatic state
$\Psi_{DAA}$, and the wide green curve with circles is the diabatic
state $\Psi_{4A}$.} \label{SpectrumDen}
\end{center}
\end{figure}

To illustrate the diabatization procedure, consider the spectrum
of the four-fermion system in the strongly interacting region shown in
Fig.~\ref{SpectrumDen}. A series of crossings and avoided
crossings occurs when the adiabatic parameter $\lambda\equiv1/a$ is varied in the strongly interacting region. The avoided crossings can be
roughly characterized by their width $\Delta\lambda$, the
range where the two adiabatic energy curves are coupled. There are
narrow crossings where $\Delta\lambda\ll 1/a_{ho}$ and there are wide
crossings where $\Delta\lambda\gtrsim 1/a_{ho}$. To obtain smooth
energy values, we use variation of the diabatization procedure
presented in Ref.~\cite{hess04}.

The objective of the diabatization algorithm is to make the
one-to-one connection between states and energies in consecutive
points of the $\lambda$ grid that maximize the sum of the overlaps
between connected states. The diabatization procedure starts from
the BCS ($a<0$) side of the resonance and connects the states (and their
energies) between consecutive values of $\lambda$ for which their
overlap is maximum. When two initial energies connect to the same
final energy, a refinement of the diabatization procedure is
applied.

Diabatization is controlled by the spacing between consecutive
values of $\lambda$ given by $\Delta\lambda_g$. If the width of the
avoided crossing is smaller than $\Delta\lambda_g$, then that
crossing is diabatized. But if the width of the avoided crossing is
larger than $\Delta\lambda_g$, then that crossing is not diabatized.
Thus, $\Delta\lambda_g$ is selected so that narrow crossings are
diabatized and wide crossings remain adiabatic. For example, in
Fig.~\ref{SpectrumDen} we see how this procedure diabatizes the
narrow crossings of $\Psi_{4A}$, however, wide
crossings such as the one between $\Psi_{DAA}$ and $\Psi_{4A}$ are
still adiabatic in this representation.

This structure of avoided crossings permits a global view of the
manner in which states evolve from weakly interacting fermions at
$a<0$ to all the different configurations of a Fermi gas at $a>0$,
i.e., molecular bosonic states, fermionic states, and molecular
Bose-Fermi mixtures. 
Furthermore, it allow us to visualize concretely the alternative
pathways of the time-dependent sweep experiments.

\begin{figure} [htbp]
\begin{center}
\includegraphics[scale=0.25]{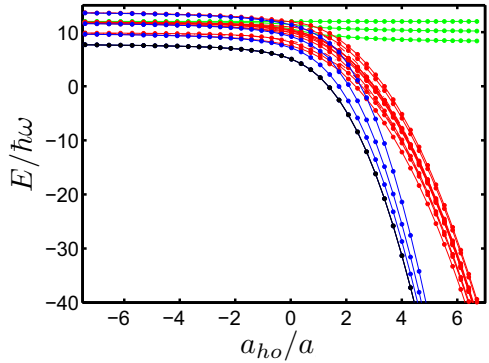}
\caption{  Energy spectrum for four particles with $J=0$ in the
crossover region (lowest 20 diabatic states). The black curve
corresponds to the ground state. The blue curves are the states that
go diabatically to excited dimer-dimer configurations. The red
curves correspond to states that connect diabatically to configurations
of a dimer plus two free atoms, and the green curves correspond to
states that connect diabatically to configurations of four free atoms.
The lowest green curve is the {\it atomic ground state} on the BEC
side of the resonance.  Results from Ref.~\cite{stech07}.}
\label{Spectrum4p}
\end{center}
\end{figure}

The diabatic spectrum of the four-fermion system is presented in
Fig.~\ref{Spectrum4p}. The structure of avoided crossings is
complicated because of two different thresholds
exist, one corresponding to the dimer-atom-atom and one to the dimer-dimer state. We identify three
different families of diabatic states in this spectrum. The
dimer-dimer family, represented by the black and blue curves,
describes the ground and excited dimer-dimer states. These states
are separated by approximately $2\hbar\omega$ on the BEC side. The
dimer--two-atom family, represented by the red energy curves,
follows the dimer binding energy. In the BEC limit, the dimer--two-atom
family reproduces the degeneracies of three distinguishable
particles: a spin up atom, a spin down atom, and a dimer. The third family
describes four-atom bound states, for which the atoms form
no dimers, and whose energy remains positive in the crossover region. In
the BEC limit as $a \rightarrow 0$, the four-atom family reproduces the spectrum of the
noninteracting four-body system.

The evolution of the $N=4$ spectra through the BCS-BEC
crossover region can be understood qualitatively by considering the
important quantum numbers for the description of the dimer.
For each vibrational
excitation of $2\hbar\omega$ in the noninteracting limit, there is
one state that diabatically becomes a dimer-dimer state. These
states correspond qualitatively to states where the relative angular
momentum of two spin-up--spin-down pairs is zero
[$L_{rel}^{\uparrow\downarrow}=0$], and the relative angular
momentum between the pairs is also zero. The spin-up--spin-down
pairs are in the lowest vibrational state. In the weakly interacting
BCS limit, where the degeneracy of the vibrational states is
broken, pair-pair states correspond to the lowest states.

\begin{figure}[htbp]
\begin{center}
\includegraphics[scale=0.2]{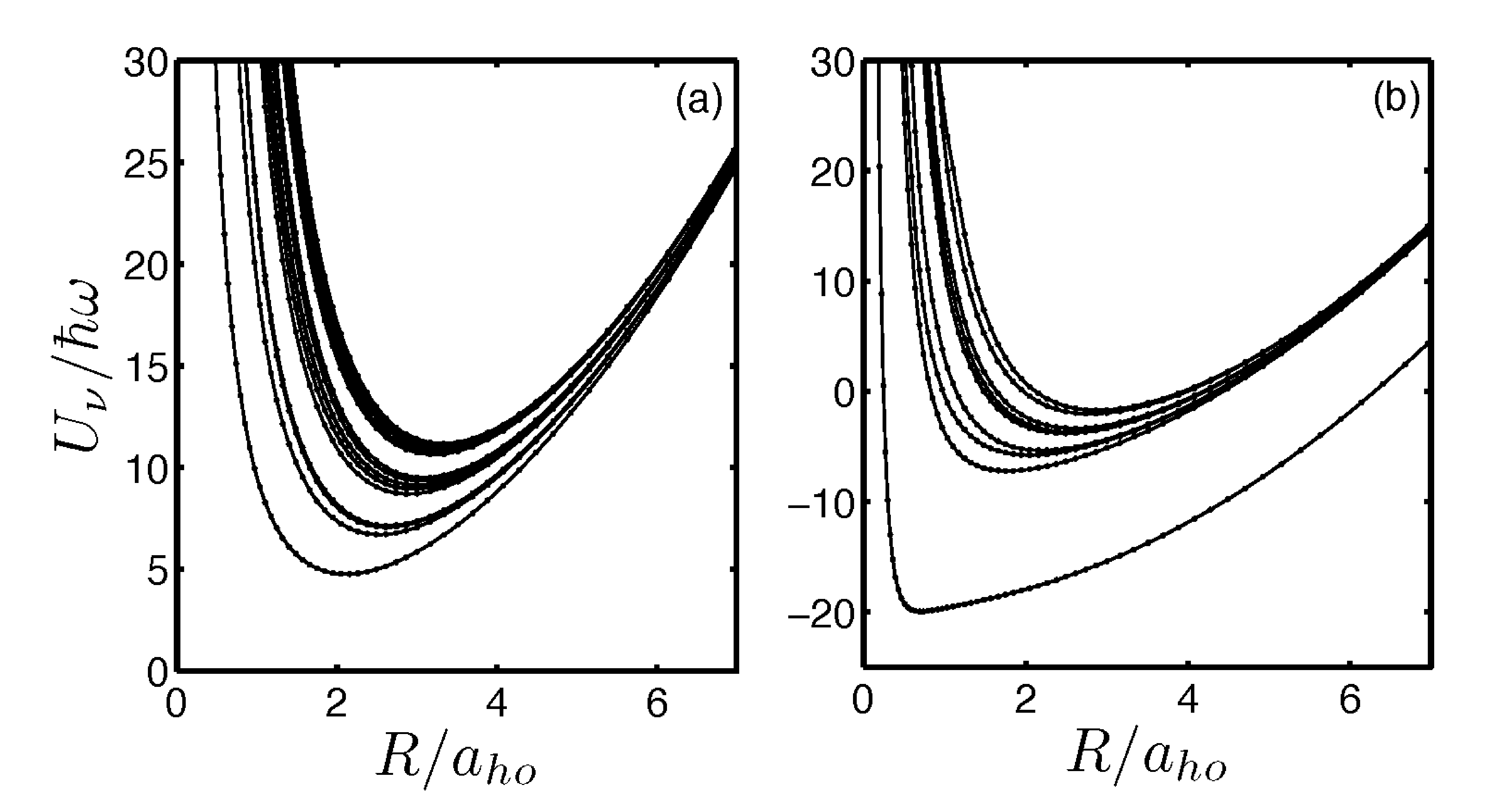}
\caption{ Hyperspherical potential curves in the BCS-BEC crossover
for $N=4$ particles with $J=0$ angular momentum. (a) Potential curves in the BCS regime,
$a\sim-0.3a_{ho}$. (b) Potential curves in the BEC regime,
$a\sim0.3a_{ho}$.} \label{PCBCSBEC}
\end{center}
\end{figure}

A direct and more concrete way to visualize the structure of the
spectrum is to analyze the evolution of the adiabatic hyperspherical
potential curves.
Figure~\ref{PCBCSBEC} presents the four-fermion adiabatic hyperspherical potential
curves $U_\nu(R)$, obtained with
the correlated Gaussian hyperspherical method (CGHS). Panel~(a)
presents the potential curves in the BCS regime, which are clearly grouped into families. Potential curves
belonging to the same family are degenerate in the noninteracting
limit. Thus, the weak interactions in the BCS regime break the degeneracies
of the potential curves forming these families of potential curves. Panel~(b) describes the system in the BEC regime. In this case, the
description of the system is quite clear. The lowest potential curve
is more than twice as deep as the rest of the curves and is
associated with the dimer-dimer threshold.
The family of dimer-dimer states live mainly in the lowest
potential curve. The remainder of the displayed potential curves are associated
with the dimer--two-atom threshold. The dimer--two-atom states are
mainly described by this family of potential curves. A
third family of potential curves, not shown in
Fig.~\ref{PCBCSBEC}~(b), describes four-atom states. This
family of potential curves has a different large-$R$ asymptotic
behavior.

\begin{figure} [htbp]
\begin{center}
\includegraphics[scale=0.8]{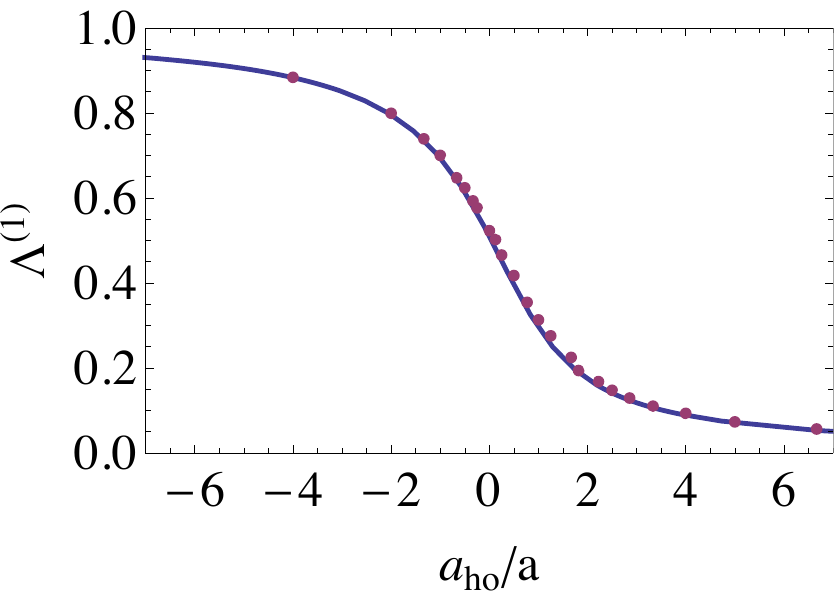}
\caption{ Energy crossover curve $\Lambda_4^{(\kappa)}$ as a
function of $a_{ho}^{(2\mu)}/a$ for $\kappa=1$. Solid lines are calculated by the CG approach, and
 symbols by the FN-DMC method. Adapted from Ref.~\cite{vonstechtbp}. }\label{fig1}
\end{center}
\end{figure}

In order to benchmark the four-body energies, Figure~\ref{fig1} compares CG results with fixed-node diffusion Monte Carlo (FN-DMC) results carried out by Blume.\cite{vonstechtbp}
From the ground-state energy,
the energy crossover curve $\Lambda_4^{(\kappa)}$,
is constructed as in Refs.~\cite{vonstechtbp,vonstech08a}:
\begin{eqnarray}
\label{eq_cross4} \Lambda^{(\kappa)} = \frac{E(4)-2 E(2)}{2
\hbar \omega}.
\end{eqnarray}
Here $E(4)$ is the ground-state energy of the four-particle system
and $E(2)$ is the ground-state energy of the two-particle system.
By construction, the energy crossover curve $\Lambda_4^{(\kappa)}$ varies from 1 in the non-interacting--BCS regime ($a\to0^-$)  to 0 in the BEC limit ($a\to0^+$).
The energy crossover curve is convenient for comparisons
because any effects of finite-range interactions on the two-body
binding energy are significantly reduced by the subtraction in
Eq.~(\ref{eq_cross4}). Therefore, even though both $E(4)$ and $E(2)$
are not completely universal, $\Lambda_4^{(\kappa)}$ is universal to a very good approximation.

The solid lines in Figure~\ref{fig1} correspond to the CG prediction while the symbols correspond to the FN-DMC predictions obtained by Blume, for an equal mass system. To describe the four-fermion ground state with the FN-DMC method, two different guiding wave functions were used: a ``normal'' and a ``paired'' guiding function. The ``normal'' wave function can be written as the non-interacting solution multiplied by a Jastrow term that improves the description of short-range correlation. This guiding function is well suited to describe the ground state of the system in the BCS-unitarity regime. The ``paired'' wave function is constructed as an antisymmetrized product of two-body solutions and leads to a good description of the ground state in the BEC-unitarity regime. 
In order to reduce finite range corrections, the ranges $r_0$ of the
two-body potentials used in Fig.~\ref{fig1} are set to be much smaller than
the oscillator lengths, i.e., $r_0 \approx 0.01a_{ho}^{(2 \mu)}$.
By analyzing the dependence of the energy on the finite range $r_0$, we can extrapolate the zero-range prediction and estimate the deviation in the crossover curve $\Lambda_4^{(\kappa)}$ from the zero range predictions. Such analysis estimates a 1\% deviation in the crossover curve presented in Fig.~\ref{fig1}. For example, at unitarity the CG energies are $E = 5.027 \hbar
\omega$ for $r_0=0.01 a_{ho}^{(2\mu)}$ and $E = 5.099 \hbar \omega$ for $r_0=0.05 a_{ho}^{(2\mu)}$. After a more careful analysis of the range dependence we obtain an extrapolated zero-range value of $E = 5.009 \hbar\omega$.
For comparison, the FN-DMC energy
for the square well potential with $r_0=0.01 a_{ho}^{(2 \mu)}$ is
$E=5.069(9)$, which agrees well with the energy
calculated by the CG approach.

A test of the validity of the guiding function in FN-DMC calculations is fundamental for an accurate description of many-body fermionic systems. Comparisons, such as the one presented in Fig.~\ref{fig1} represent much-needed benchmark tests for the ``normal'' and ``paired'' trial wave-functions used in the FN-DMC approach~\cite{vonstechtbp}
The good agreement between these two numerical methods suggests that, first, $\Lambda_4^{(\kappa)}$ is indeed universal; and  second,
 that both numerical methods accurately describe the BCS-BEC crossover for this four-body system.

\subsection{Extraction of dimer-dimer collisional properties}
\label{extraction}

In the BEC limit, the lowest four body levels describe different
vibrational states of a dimer-dimer configuration. Therefore, the systems
 can be treated effectively in this limit as two-particle systems.  A comparison between the two particle
solutions and the $N=4$ solutions allows us to extract
information on the effective dimer-dimer interactions.

To model the effective dimer-dimer interaction we introduce a zero-range pseudopotential.
Since the size of the
dimers are of the order of $a$, the range of the
effective potentials should also be of order of $a$. Accordingly, effective range effects are discussed using an energy-dependent
scattering length. Inclusion of the scattering length energy dependence is well-known to
extend the validity of the zero-range
pseudopotential when applied to the scattering of
two atoms with finite-range potentials under external
confinement~\cite{blume2002fpa,bolda2002esl}.
This energy dependence is included here through the
effective range expansion 
\begin{eqnarray}
\label{aE}
 -\frac{1}{a_{dd}(E_{col})}\approx-\frac{1}{a_{dd}}+\frac{1}{2} k^2
r_{dd}.
\end{eqnarray}
Here $a_{dd}(E_{col})$ is the energy-dependent dimer-dimer scattering length
parameterized by the (zero-energy) scattering length $a_{dd}$ and the
effective range $r_{dd}$. The momentum $k$ is associated with the
relative kinetic energy of the dimer. Thus, $k^2/2\mu=E_{col}$ where
 $E_{col}=E_{4b}-2E_{2b}$. The appropriate reduced mass $\mu$ is $\mu_{dd}=M/2$, where $M$
is the mass of the bosonic molecules, $M=m_1+m_2$.

Using the effective range expansion [Eq.~(\ref{Eq:r0_expansion})], the
regularized zero-range potential $V(r)$~\cite{huan57} takes the form
$V(r)=g(E) \delta(\bm{r}) (\partial/\partial r) r$. The scattering
strength $g$ is parameterized by the scattering length $a_{dd}$ and
the effective range $r_{dd}$, i.e.,
\begin{eqnarray}
g(E)=\frac{2 \pi \hbar^2 \, a_{dd}} {\mu} \left[1- \frac{\mu E_{col}
r_{dd} a_{dd}}{\hbar^2} \right]^{-1}.
\end{eqnarray}
The $J=0$ spectrum of the
two particle trapped system is given by~\cite{blume2002fpa,bolda2002esl}
\begin{equation}
  \label{Ener2paa}
   \sqrt{2}\frac{\Gamma\left(-\frac{E_{col}}{2\hbar\omega}+\frac{3}{4}\right)}
   {\Gamma\left(-\frac{E_{col}}{2\hbar\omega}+\frac{1}{4}\right)}=\frac{a_{ho}^{(\mu)}}{a_{dd}(E_{col})}.
\end{equation}
Equation (\ref{Ener2paa}) is a transcendental equation that can be
easily solved numerically.  The solutions of Eq.~(\ref{Ener2paa})
are obtained as functions of the $a_{dd}$ and $r_{dd}$ parameters and
fitted to the numerical results. The calculation can be carried out
at different values of the two-body scattering length $a$ and, in
this way, one obtains a reliable estimation of $a_{dd}$ and
$r_{dd}$.

Since the properties of weakly-bound dimers are controlled by the scattering length, it is natural to assume that the properties of dimer-dimer interactions are controlled only by $a$.
This implies that  $a_{dd}$  and
$r_{dd}$ should be proportional to the two-body scattering length
$a$. Therefore, we propose expressions of the form $a_{dd}=c_{dd} a$ and $r_{dd}=d_{dd} a$. The parameters
$c_{dd}$ and $d_{dd}$ are obtained by fitting the zero-range
two-particle solution to the dimer-dimer states of the $N=4$ system.

\begin{figure} [hbtp]
\begin{center}
	 \includegraphics[scale=0.25]{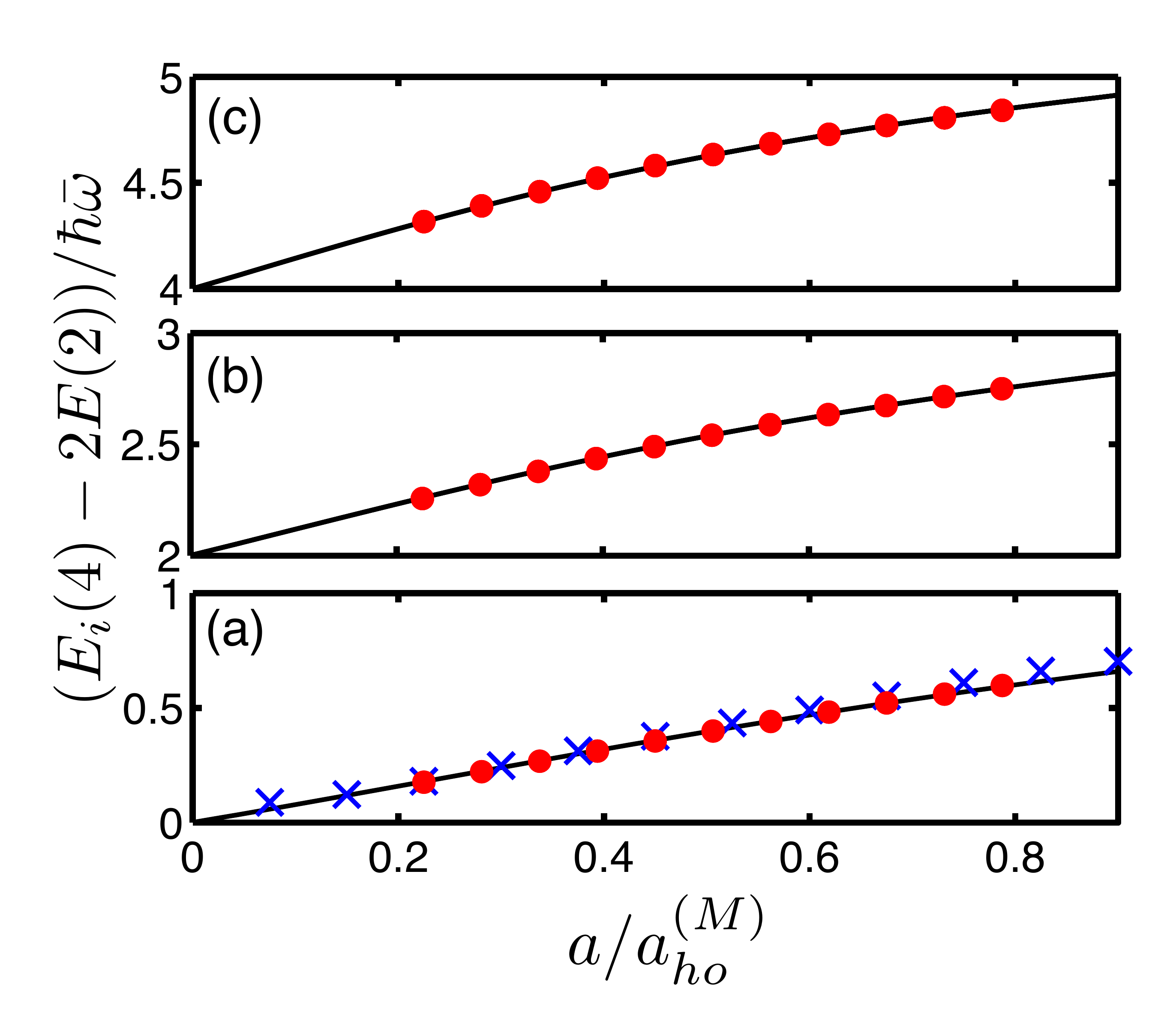}
\caption{ Four-body energies of the three energetically lowest-lying
dimer-dimer states as a function of $a/a_{ho}^{(M)}$ for
$\kappa=8$. Panel~(a) shows the energetically lowest-lying energy
level ($i=0$), panel (b) shows the energetically second-lowest ($i=1$) and
panel (c) shows the energetically third-lowest state ($i=2$). Circles and
crosses show our CG and FN-DMC results, respectively. Solid lines
show the zero-range model results. Figure from
Ref.~\cite{vonstechtbp}.} \label{ddspectrum} \end{center}
\end{figure}

The results in Fig.~\ref{ddspectrum} illustrate the fitting procedure.
The circles in Figs.~\ref{ddspectrum}(a)--(c) show the lowest-lying
dimer-dimer energy levels, referred to as $E_i(4)$, where $i=0$--$2$
with the center-of-mass energy and the dimer-binding energy
subtracted. These energies correspond to a two-heavy two-light four-body system with mass ratio $\kappa=8$.
Solid lines represent the
energy-dependent zero-range pseudopotential predictions obtained by fitting the 
two-boson model [Eq.~(\ref{Ener2paa})] to the four-body energies.
The range of atom-atom scattering lengths, $a$, over which the four-fermion system can be described
by the two-boson model is significantly extended by the inclusion of the effective range $r_{dd}$.
Such inclusion also allows for a more stable and reliable determination of $a_{dd}$.

The crosses in Fig.~\ref{ddspectrum}(a) correspond to FN-DMC energies for the energetically lowest-lying
dimer-dimer state.
The Blume FN-DMC
energies, presented in Ref.~\cite{vonstechtbp}, are found to be slightly higher than the CG energies, the
deviation increasing with $a$.
 This
increasing deviation might be attributed to the variational implications of the fixed-node constraint.
The functional form of the
nodal surface used in the FN-DMC calculations was constructed for noninteracting dimers and should be best
in the very deep BEC regime.
Application of the two-boson model and the fitting procedure to the Blume FN-DMC energies provides an alternative determination of $a_{dd}$ and $r_{dd}$. The increasing deviation between the
FN-DMC and CG energies with increasing $a$ explains why the
effective range predicted by the analysis of the FN-DMC energies is
somewhat larger than that predicted by the analysis of the CG
approach (see discussion of Fig.~\ref{fig2} below).

The formation of few-body states, such as trimers or tetramers, can affect
the scattering properties and limit the applicability of the two-boson model.
For example, the lowest four-body energy for $\kappa=1$,
 constitutes the true ground state of the system,
i.e., no energetically lower-lying bound trimer or tetramer states
with $J^\Pi=0^+$ symmetry exist.
However, for larger mass ratios, bound trimer
and tetramer states exist. These few-body states can in principle be associated with either universal Efimov or nonuniversal physics.
Even in the regime where few-body bound states exist, the
four-body spectrum contains
universal states that are separated by approximately $2
\hbar\omega$ and are best described as two weakly-interacting
composite bosons. For fixed $a$ ($a > 0$), the energy of these
``dimer-dimer states'' changes smoothly as a function of $\kappa$
even in the regime where bound trimer states appear.
Thus, our two-boson model and the fitting procedure can be extended to this regime of mass ratios.

Table~\ref{tablenew} and Fig.~\ref{fig2} summarizes the dimer-dimer scattering length and effective range for for selected values
of the mass ratio $\kappa$.  Circles and crosses in Fig.~\ref{fig2} correspond to the dimer-dimer scattering length, $a_{dd}$, extracted from the energies calculated by the CG and the FN-DMC approach, respectively, as a function of $\kappa$.
The $a_{dd}$ two-boson model predictions compare well with those
calculated by Petrov {\em{et al.}} within a zero-range
framework~\cite{petr05} (solid line in Fig.~\ref{fig2}).
The existence of Efimov states limited Petrov~{\em{et al.}} calculations to $\kappa \lesssim  13.6$ since
beyond this mass ratio a three-body parameter is needed to solve the zero-range four-body
equations. As mentioned above, the four-body solutions for a finite range potential include deeply-bound solutions which correspond to either a trimer-atom or a tetramer. The trimer-atom states affect the dimer-dimer scattering properties only when the dimer-dimer configuration is close in energy to the trimer-atom energy, which usually corresponds to a narrow window in two-body scattering length. Away from this window, the dimer-dimer states are well described by the two-boson mode which predicts a smooth increase of the $a_{dd}$ up to mass ratio $\kappa=20$.

Note that the existence of energetically-open atom-trimer channels at the dimer-dimer collisional energy provides a decay mechanism for the dimers. Clearly, such a decay mechanism cannot be captured within this rather simple two-boson model and a more sophisticated treatment should be used to extract the inelastic scattering properties.
\begin{figure} [htbp]
\begin{center}
\includegraphics[scale=0.25]{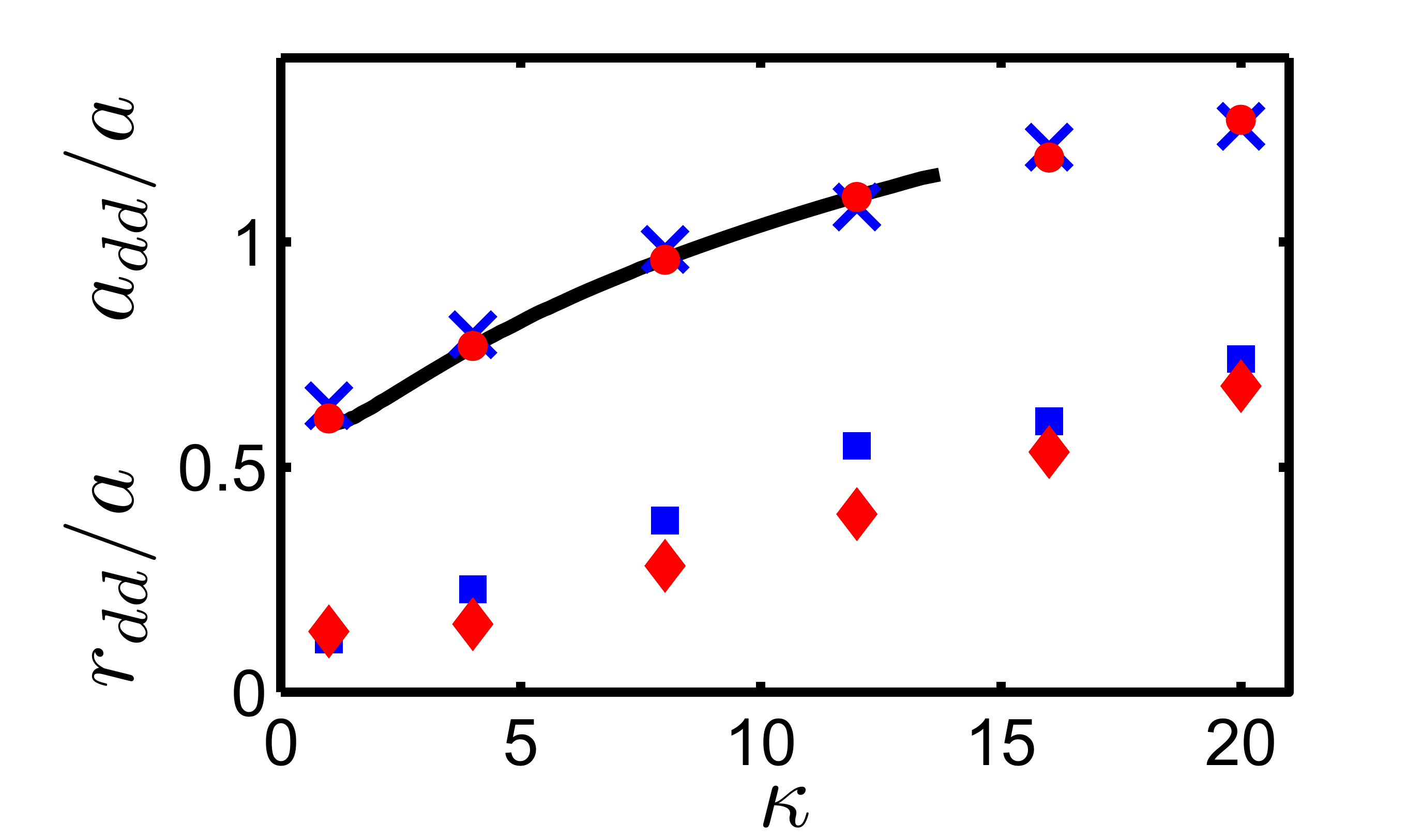}
\caption{ Circles and crosses show $a_{dd}/a$ as a function of
$\kappa$ extracted from the four-fermion CG and FN-DMC energies,
respectively. For comparison, a solid line shows the results from
Fig.~3 of Ref.~\protect\cite{petr05}. Diamonds and squares show
$r_{dd}/a$ extracted from the four-fermion CG and FN-DMC energies,
respectively. Figure from Ref.~\cite{vonstechtbp}.} \label{fig2}
\end{center}
\end{figure}
A recent theoretical study found good agreement with the two-boson model predictions
beyond $13.6$~\cite{marcelPRA08}.

\begin{table}
\caption{\label{tablenew} The dimer-dimer scattering length, $a_{dd}$,
and dimer-dimer effective range, $r_{dd}$, obtained using (a) the CG
spectrum and (b) the FN-DMC energies. The reported uncertainties
reflect the uncertainties due to the fitting procedure; e.g., the
potential limitations of the FN-DMC method to accurately describe
the energetically lowest-lying gas-like stateare not included
here (see Sec.~IIIB of Ref.~\cite{vonstechtbp}).}
\begin{center}
\begin{tabular}{||c|ll|ll||}\hline
  $\kappa$ & $a_{dd}/a$ (a) & $a_{dd}/a$ (b)& $r_{dd}/a$ (a)& $r_{dd}/a$ (b)\\
\hline
1&   0.608(2) &0.64(1) & 0.13(2)&0.12(4)\\
4& 0.77(1) & 0.79(1) & 0.15(1)& 0.23(1)   \\
8& 0.96(1)  & 0.98(1) & 0.28(1) & 0.38(2)  \\
12& 1.10(1) & 1.08(2) & 0.39(2)& 0.55(2)\\
16& 1.20(1) & 1.21(3) & 0.55(2)& 0.60(5) \\
20& 1.27(2) & 1.26(5)& 0.68(2) & 0.74(5) \\ \hline
\end{tabular}
\end{center}
\label{EnU4}
\end{table}

The two-boson model also provides estimates for the dimer-dimer effective range, $r_{dd}$ (shown as 
diamonds and squares in Fig.~\ref{fig2}). 
The uncertainty of $r_{dd}$ obtained from the CG approach is estimated to be
about 10\%; this uncertainty is expected to be larger for the results
extracted from the FN-DMC energies since those calculations only include one energy curve.
As shown in Fig.~\ref{fig2}, the
ratio $r_{dd}/a$ increases from about 0.2 for $\kappa=1$ to
about 0.5 for $\kappa=20$. The existence of a finite effective range can be attributed to the effective broad
soft-core potential that the dimers experience~\cite{petr05}. Ref.~\cite{vonstechtbp} predicts $r_{dd}$ as a function of the mass ratio $\kappa$. The
large value obtained for $r_{dd}$ suggests that effective-range corrections
may need to be considered in order to accurately describe the physics of molecular
Fermi gases.

To conclude, we have shown that studies of few-body trapped systems can be used to
extract information about the collisional properties of free systems.
Dimer-dimer scattering lengths can be extracted by
analyzing the trapped few-body spectrum for different two-body scattering
length values. Furthermore, energy-dependent corrections to
 $a_{dd}$ can also be obtained with this method.

\subsection{Structural properties}
\label{sec_struc1}

The analysis of the BCS-BEC crossover spectrum can be
complemented by an analysis of the wave functions and their
structural properties. The present section determines the one-body densities and pair-distribution functions for two-component Fermi systems in the crossover regime.
The averaged radial densities, $\rho_i(r)$, are normalized
such that $4\pi \int \rho_i(r) r^2 dr=1$; $4\pi r^2 \rho_i(r)$
represents the probability of finding a particle with mass $m_i$ at a
distance $r$ from the center of the trap. Here we focus in the $m_1=m_2$ case, where we find that the radial one-body densities $\rho_1(r)$ and $\rho_2(r)$ coincide and we can omit the species label.
The unequal mass case, in which the radial one-body densities $\rho_1(r)$ and $\rho_2(r)$ differ, will not be considered here but was discussed in Ref.~\cite{vonstecher2008eas}. Also, the averaged radial pair distribution functions, $P_{ij}(r)$, are normalized so that $4\pi \int
P_{ij}(r) r^2 dr=1$; $4\pi r^2 P_{ij}(r)$ represents the probability
of finding a particle of mass $m_i$ and a particle of mass $m_j$ at
a distance $r$ from each other.

These structural properties are computed using the CG method. Since we are only focusing $J=0$ states, all the structural properties presented here are spherically symmetric. The structural properties are calculated using the following general expression:
\begin{equation}
\label{paircoorBC1} 4\pi r^2
F(r)=\braket{\Psi_{n}|\delta(x-r)|\Psi_{n}}=\int
d\bm{r}_1...d\bm{r}_N
\delta(x-r)|\Psi_{n}(\bm{r}_1,\bm{r}_{1'},...,\bm{r}_{N_1},\bm{r}_{N_2}
)|^2.
\end{equation}
Here, $F(r)$ is a generic structural property, e.g. the density
profiles $\rho_1$ or $\rho_2$, the interspecies pair-correlation
function $P_{12}$, or the intraspecies pair-correlation functions
$P_{11}$ or $P_{22}$. $x$ is the length of the coordinate vector
that describes the structural property. For $\rho_1$ and $\rho_2$,
$x=r_1$ and $x=r_{1'}$, respectively. For $P_{12}$, $x$ is the
interparticle distance between opposite-spin or different species,
$x=r_{11'}$. For $P_{11}$ and $P_{22}$, $x$ is the same-spin or
same-species interparticle distance, with $x=r_{12}$ and $x=r_{1'2'}$,
respectively. To evaluate $4\pi r^2 F(r)$, $\Psi_{n}$ is expanded in
the CG basis set, permitting the integral in Eq.~(\ref{paircoorBC1}) to be
carried out analytically.

\begin{figure}[htbp]
\begin{center}
\includegraphics[scale=1]{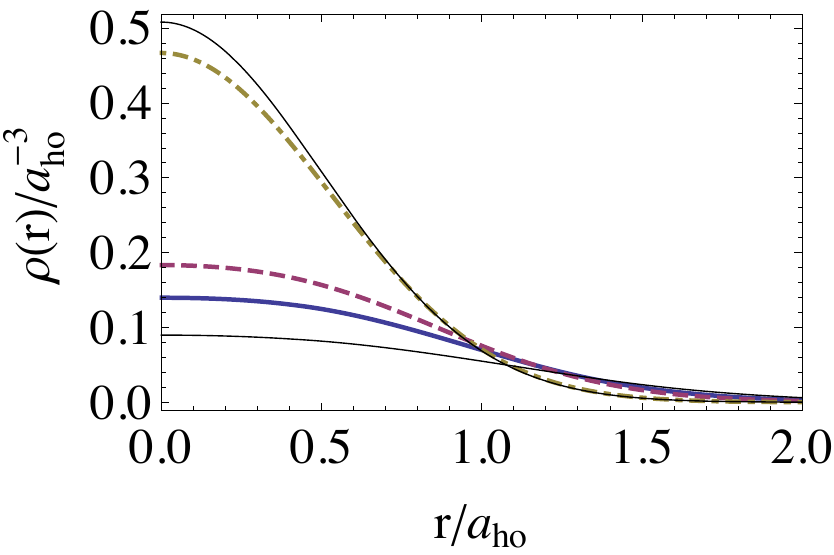}
\caption{Single particle density profiles for interaction strength across the crossover. Thick curves correspond to numerical results: $a=-a_{ho}$ (solid curves), $a=\pm\infty$ (dashed curves) and $a=0.1 a_{ho}$ (dash-dotted curves). Thin curves correspond to analytic limiting solutions: noninteracting four-fermion prediction (less peaked), dimer-dimer solution in the BEC limit (more peaked).}\label{denprofiles} \end{center}
\end{figure}

Figure~\ref{denprofiles} presents the single particle density profiles for different interaction strength across the crossover. The radial density profiles of the equal mass four-fermion system smoothly evolve from the noninteracting solution (less peaked thin black curve) in the BCS limit to the prediction of dimer-dimer model in the BEC limit (more peaked thin black line). In the BEC limit ($a \to 0^+$), the density profile coincides with that of two point like bosonic dimers, i.e.,   $\rho(r)=\exp(-r^2/a_{ho}^{(M)})/ (a_{ho}^{(M)}\sqrt{\pi})^3$. For small but finite positive scattering lengths (dash-dotted curve), the density profiles is broader which can be attributed to both the finite size of the dimers and the effective repulsive dimer-dimer interaction.

Next, we analyze the opposite-spin pair distribution function. For a zero-range pseudopotential, the opposite-spin pair-correlation
function obeys a boundary condition~\cite{lobo2006pce,gior07}
\begin{equation}
\label{paircoorBC} \frac{[r P_{12}(r)]'_{r=0}}{[r
P_{12}(r)]_{r=0}}=-\frac{2}{a}.
\end{equation}
This behavior is a direct consequence of the Bethe-Peierls (B-P)
boundary condition $[r_{12} \Psi(r_{12})]'_{r_{12}=0}/[r_{12}
\Psi(r_{12})]_{r_{12}=0}=-1/a$. The factor of 2 in
Eq.~(\ref{paircoorBC}) appears because the pair
correlation function is proportional to the square of the wave function,
i.e., $P_{12}(r)\propto\Psi(r_{12})^2$. A direct consequence of
Eq.~(\ref{paircoorBC}) is that at unitarity, i.e., when
$|a|=\infty$, $r^2P_{12}(r)$ has zero slope at $r\rightarrow0$.

\begin{figure}[htbp]
\begin{center}
\includegraphics[scale=0.5]{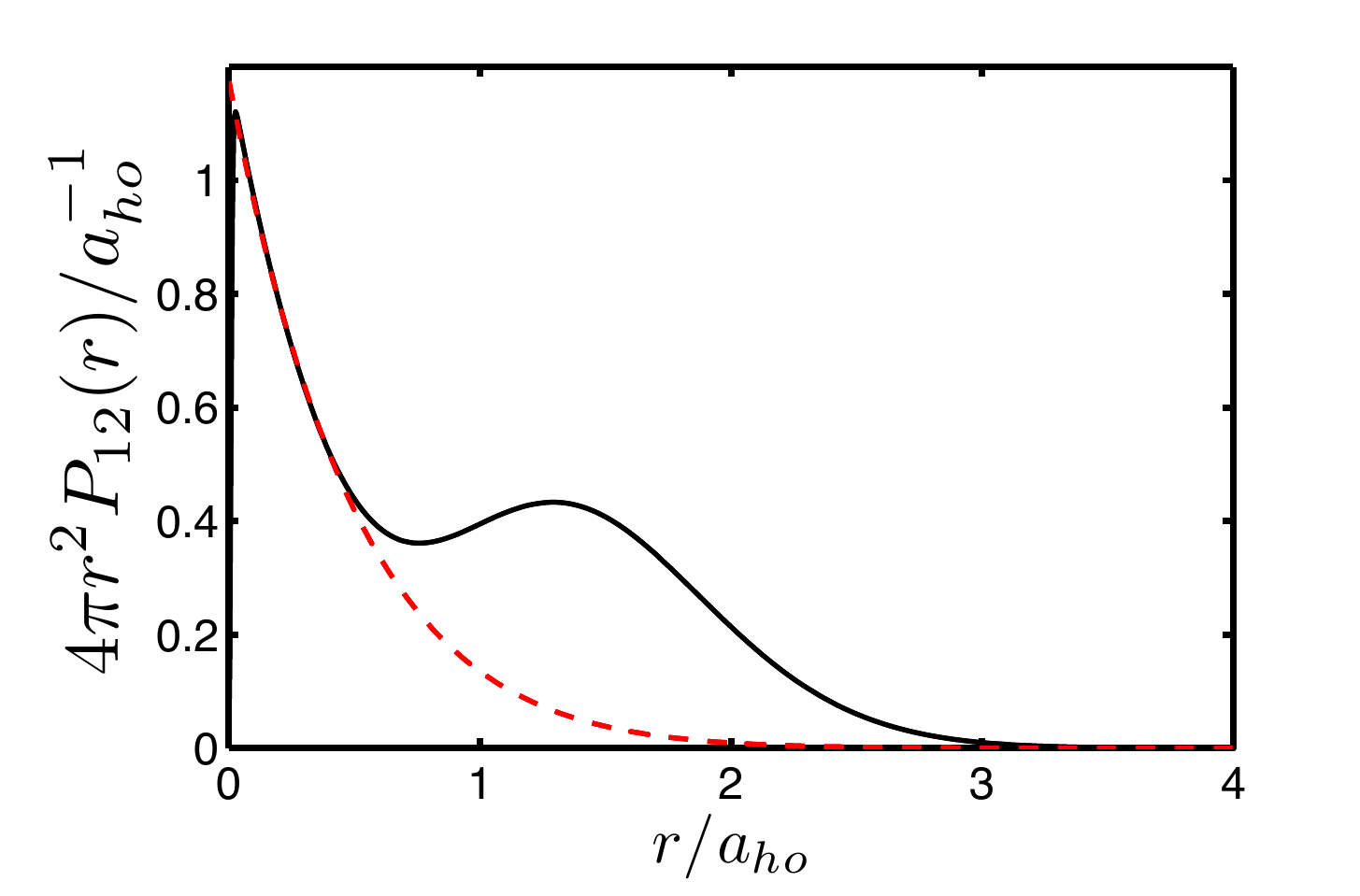}
\caption{  Pair-distribution functions $P_{12}(r)$, multiplied by
$r^2$, for equal-mass--two-component Fermi systems with $N=2$
(dashed curve) and $N=4$ (solid curve). The $N=2$ pair-correlation
function has an arbitrary norm selected to match the $N=4$ pair
correlation in the small $r$ regime.}\label{P1224} \end{center}
\end{figure}

The numerical verification of Eq.~(\ref{paircoorBC}) is challenging. For systems with finite range interactions,
Eq.~(\ref{paircoorBC}) is valid in a narrow regime of $r$ values.
For $a<0$, Eq.~(\ref{paircoorBC}) is valid when $r$ is much larger
than the range of the potential and much smaller than the mean
interparticle distance, i.e., the $r_0\ll r\ll a_{ho}$ regime. For
$a>0$, Eq.~(\ref{paircoorBC}) is valid when $r$ is much larger
than the range of the potential and much smaller than the size of
the dimer (given by $a$) and the mean interparticle distance, i.e.,
the $r_0\ll r\ll \mbox{min}[a, a_{ho}]$ regime. This regime is
almost nonexistent for our numerical calculations, because we consider
only the $a_{ho}/r_0=100$ case.

\begin{figure}[htbp]
\begin{center}
\includegraphics[scale=0.7]{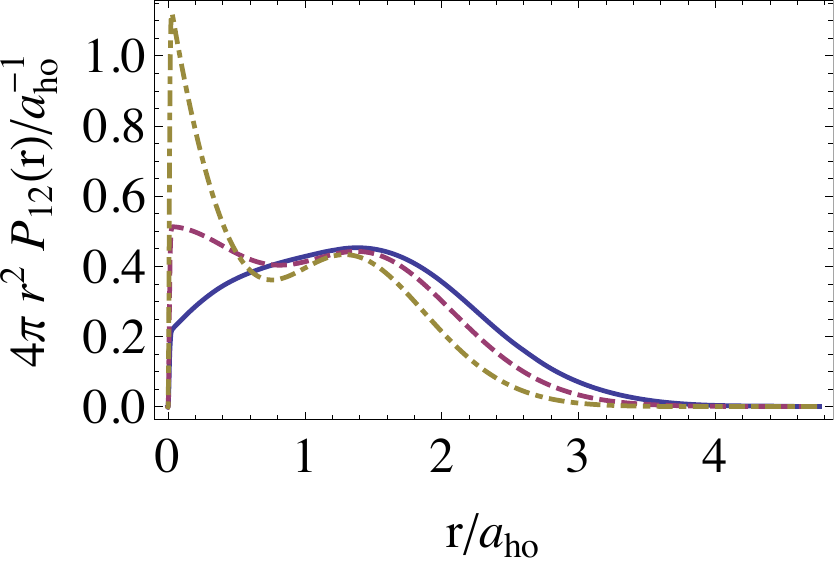}
\caption{Pair-distribution functions $P_{12}(r)$, multiplied by
$r^2$, for equal-mass--two-component Fermi systems with $N=4$
and $J=0$. The solid curve corresponds to $a=-a_{ho}$ (BCS regime),
dashed curve $1/a=0$ (unitarity), and dashed-dotted $a=a_{ho}$ (BEC regime). Adapted from Ref.~\cite{vonstech08a}.
}\label{Corr3and4} \end{center}
\end{figure}

The pair distribution function changes considerably as interactions are tuned across the BCS-BEC crossover. Figure~\ref{Corr3and4} shows the pair distribution function $P_{12}(r)$ along the crossover for equal mass systems [$\kappa=1$] in the BCS (solid lines), unitary (dashed lines) and BEC (dash-dotted lines) regime. As attractive interactions increase, the pair distribution function evolves from a single peak structure to a double peak structure.  The single peak structure is usually associated to solutions where spin-up-spin-down interparticle distances are mainly controlled by a single length scale and all particles are roughly at the same distance. This single peak structure appears in the BCS regime since the attraction is not strong enough to bind the particles and the typical interparticle distance is set mainly by the external trapping potential. The multi peak structures in pair distribution functions are generally associated to solutions were more than one length scale control the interspecies interparticle distance. In this case, a double peak structure already appears at unitarity and becomes more pronounce in the BEC regime. The peak at small $r$
indicates the formation of weakly
bound dimers and its width is associated to the size of the dimers. The broader peak between $1\,a_{ho}$ and $2\,a_{ho}$ is related to the presence of larger atom-atom length scales set approximately by the typical dimer-dimer separation in the harmonic oscillator potential. Understanding the BEC solutions as two dimers in a trap implies that the four-body configurations includes two interspecies distances of the size of the dimer (short) and two interspecies distances of the size of the trap (large). Thus, the probability of finding two  particles from different species at short distances should be equal to the probability of finding them at large distances.
This premise can is verified by comparing the area under first and second peaks in the $P_{12}(r)r^2$ plot.
Finally, note that in moving through the BCS-BEC crossover region, the slope of $r^2P_{12}(r)$ at small $r$ changes from positive to zero to negative, as is predicted by Eq.~(\ref{paircoorBC}).

\begin{figure}[htbp]
\begin{center}
\includegraphics[scale=0.5]{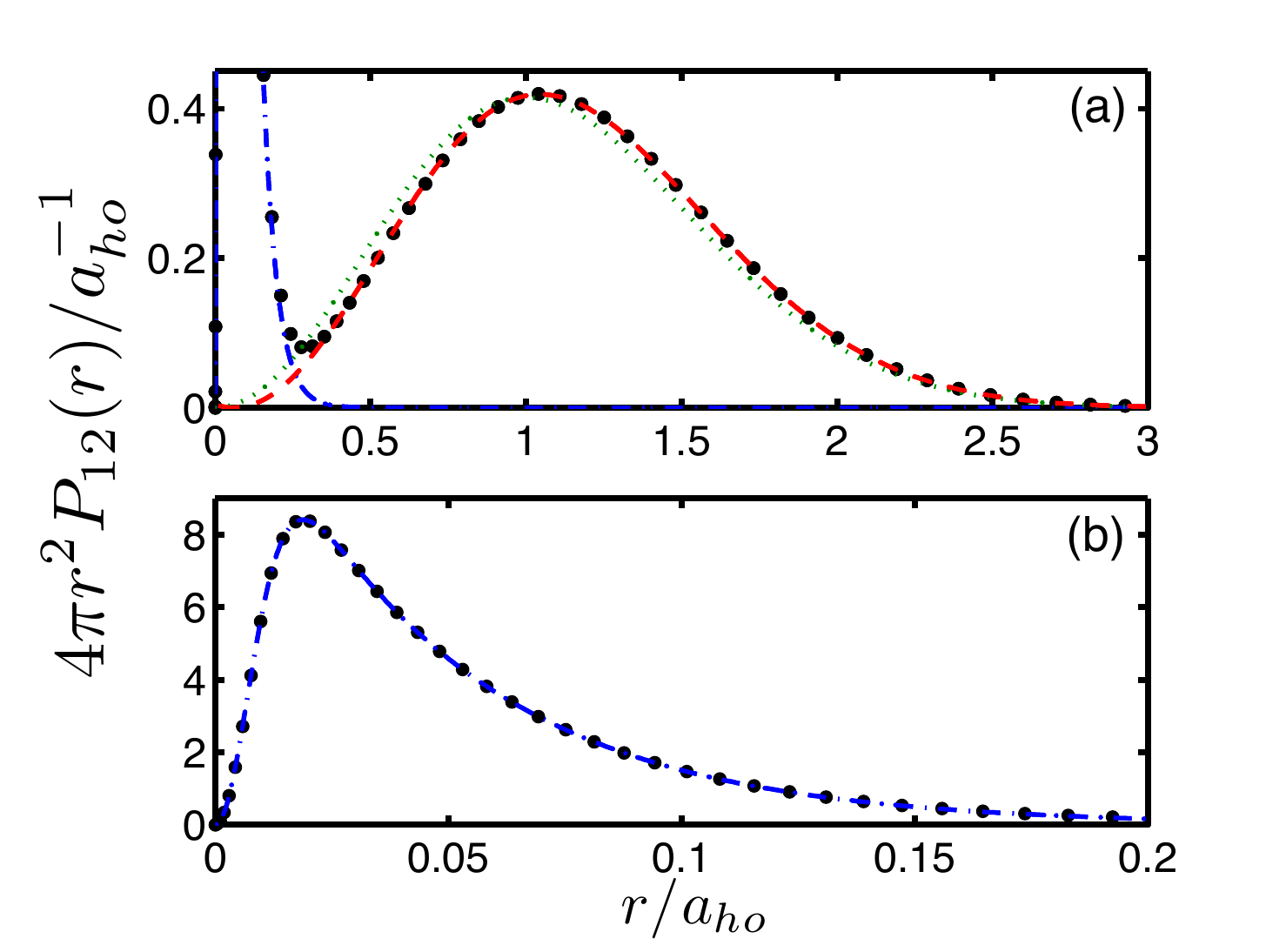}
\caption{  (a) Circles show the pair-distribution function,
$P_{12}(r)$, multiplied by $r^2$, for $a=0.1a_{ho}$ (BEC regime). For comparison, the blue
dash-dotted line shows $P_{12}(r)r^2$ for two atoms of mass $m$ with
the same scattering length but normalized to $1/2$, the red dashed
line shows $P_{12}(r)r^2$ for two trapped bosonic molecules of mass
$2\,m$ interacting through a repulsive effective potential with
$a_{dd}=0.6a$, and the green dotted line shows $P_{12}(r)r^2$ for
two trapped noninteracting bosonic molecules of mass $2\,m$. Panel
(b) shows a blow-up of the small $r$ region. Figure from
Ref.~\cite{vonstech08a}. \label{Corr4BEC}} \end{center}
\end{figure}

The dimer-dimer model can be quantitatively tested in the deep BEC regime ($0<a\ll a_{ho}$). In this regime, the two peaks of the pair distribution function are well separated and can be independently analyzed.
The small-$r$ corresponds to dimer formation and is well described by the pair-distribution function multiplied
by $r^2$ and  normalized to $1/2$, for {\em two} trapped atoms with $a=0.1a_{ho}$ (dash-dotted curve). The large-$r$ peak describes dimer-dimer correlations and is well described by the pair distribution function of two bosonic molecules of mass $2m$ in a harmonic trap (dotted line). The agreement is quite good but it can be improved by including the effective dimer-dimer interaction corrections. The dashed curve, almost indistinguishable from the large $r$ part of the
pair-distribution function for the four-particle system, shows dimer-dimer model prediction including the effective repulsive potential
with dimer-dimer scattering length $a_{dd} \approx
0.6a$~\cite{petr05,vonstechtbp}. Thus, even though the dimer-dimer interaction corrections are small, they are noticeably in the pair distribution function.

\subsubsection{Dynamics across the BCS-BEC crossover region}

The diabatic representation can be used to ramp an initial
configuration through the BCS-BEC crossover region, mimicking experiments
carried out at different laboratories at JILA and Rice University.
The initial configuration is propagated using the time-dependent
Schr\"odinger equation,
\begin{equation}
 \label{SC}
 i\hbar\frac{d\ket{\Psi}}{dt}= \mathcal{H}[\lambda(t)]\ket{\Psi}.
\end{equation}
The time dependence of the Hamiltonian comes entirely from that of $\lambda(t)\equiv 1/a(t)$
term. In this case, we focus on unidirectional ramps. Starting from
the ground state on the BCS side, the parameter $\lambda$ is ramped
through the resonance to the BEC side at different speeds,
$\nu=\frac{d\lambda}{dt}$. The relevant dimensionless speed quantity
is $\xi=a_{ho} \nu/\omega$.
The parameter $\xi$ can be rewritten in terms of $\nu$, the density of the system $\rho$ and the particle mass to relate few-body and many-body predictions~\cite{borca2003tap}. This reformulated version of $\xi$ (see Ref.~\cite{stech07}) agrees with the functional form of the Landau-Zener parameter obtained in Refs.~\cite{goral2004aau,williams2006tfm}. The dependence of the Landau-Zener parameter on $\rho$ has been experimentally verified~\cite{hodby2005peu}.

To propagate the initial configuration, we use the diabatic
representation obtained previously in Sec~\ref{spect}. First, we divide the BCS-BEC
crossover range into sectors. Starting from the BCS side at
$\lambda\approx \lambda_{BCS}$ and finishing at the BEC side at $\lambda\approx
\lambda_{BEC}$, the BCS-BEC crossover is divided into 40--80 sectors.
At the middle of each sector, the time-independent Hamiltonian is diagonalized
using the CG method. For four-body systems, thousands of CG basis functions are usually needed to
describe the spectrum. While in principle this basis set could be used to solve
Eq.~(\ref{SC}), in practice this large basis would make the numerical
propagation very slow. Instead, we use the diabatic representation
obtained at the middle of the $j$-th sector to expand the time-dependent
wave function throughout that sector, i.e.,
\begin{equation}
  \label{psidiaba}
  \ket{\Psi(t)}=\sum_{i}^{N_d} c_i^j(t)\ket{\Psi^{j}_i}.
\end{equation}
Here, $\ket{\Psi^{j}_i}$ is the diabatic basis function $i$ of sector $j$,
and $N_d$ is the number of diabatic states considered. The time
dependence only appears in the complex coefficients $c_i^j(t)$.
Upon selecting only the lowest 20--100 diabatic states at that point, we drastically
reduce the size of the Hamiltonian matrix in Eq.~(\ref{SC}). Since
the inverse scattering length $\lambda$ changes very little in each
sector, the relevant diabatic states are well described by this
reduced basis set throughout the sector.
The time-dependent Schr\"odinger equation, Eq.~(\ref{SC}), is
propagated from one edge of the sector to the other using an
adaptive-step Runge-Kutta method.

To understand the time propagation of this system, consider the way 
the probabilities evolve as the system transits the BCS-BEC crossover region. At each point of
the time propagation, the probability to reside in state $i$ is given
by $p_i(t)=|c_i^j(t)|^2$. Here $j$ denotes the sector that includes
$\lambda(t)$. The probabilities of evolving into a given
family can be found by summing the probabilities of all states belonging to the
same family. For two particles, there are two families, the dimer
family, which only includes the lowest state, and the two-atom
family, which includes the rest of the states. In this case, the appropriate definitions are $p_d(t)=|c_1^j(t)|^2$,  and
$P_{2a}(t)=\sum_{i=2}^{N_d}|c_i^j(t)|^2$.

The $N=4$ system has four relevant families: the
ground state, the excited dimer-dimer states, the dimer--two-atom
states, and the four-atom states. These families are characterized
by the probabilities $p_{g}(t)$, $p_{dd}(t)$, $p_{d2a}(t)$, and
$p_{4a}(t)$, respectively.

\begin{figure}[htbp]
\begin{center}
\includegraphics[scale=0.275]{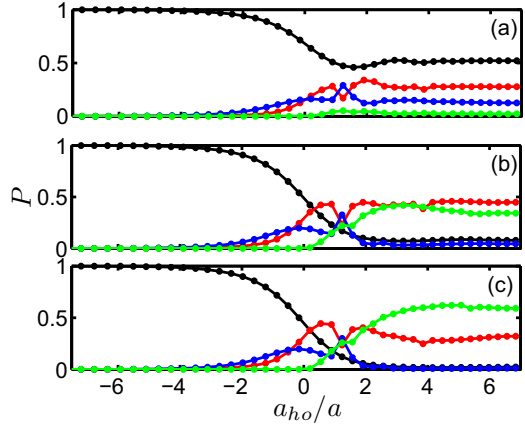}
\caption{Probabilities of different families of the $N=4$ system during a unidirectional
ramp at constant speed $d\lambda/dt$ from the BCS to the BEC side. The initial
configuration is $\ket{\Psi^{BCS}_1}$. The black curve corresponds
to $p_{g}(t)$. The blue curve corresponds to $p_{dd}(t)$. The red
curve corresponds to $p_{d2a}(t)$, and the green curve corresponds
to $p_{4a}(t)$. (a) Probabilities obtained for ramping at a speed of
$\xi\approx13$. (b) $\xi\approx52$. (c)
$\xi\approx128$}. \label{TEtot}
\end{center}
\end{figure}
Figure~\ref{TEtot} presents examples of the numerical time evolution
of an $N=4$ system during a unidirectional ramp at constant speed
from the BCS to the BEC side. As expected, the probability of
staying in the ground state decreases with increasing ramp speed. The probability of
evolving into the final four-atom configuration increases with speed, which
is in agreement with the projection argument. The transfer of
probability occurs mainly in the strongly interacting regime,
$-2\lesssim a_{ho}/a \lesssim 2$. In this region, the diabatic
states are sometimes mixed, producing jumps in the probabilities.
For example, around $a_{ho}/a \approx 1$, the red and blue curves
have a kink due to an avoided crossing between an excited
dimer-dimer state and a dimer--two-atom state.

The probabilities at the end of the time evolution
can also be studied as functions of the speed $\xi$. Before
analyzing these numerical results, however, consider first the simple Landau-Zener model that
provides insights into our numerical calculations.

In its simplest form,  the Landau-Zener model, considers a two-level system whose energy difference depends linearly on the adiabatic parameter $\lambda$, i.e.,
$\epsilon_1-\epsilon_2=\alpha \lambda$, and are coupled by
$\epsilon_{12}$ independent of $\lambda$. The time evolution of this model can be easily solved, yielding the nonadiabatic transition
probability $T_{na}$ to evolve
into a final adiabatic state different from the initial adiabatic
state after the parameter $\lambda$ is varied through an avoided
crossing. To obtain this probability, the time-dependent
Schr\"odinger equation is propagated starting at $t=-\infty$ [with
$\psi(t)=\psi_1$] to $t=+\infty$. The nonadiabatic probability is
then given by $T_{na}=|\braket{\psi(+\infty)|\psi_2}|^2$. Landau and
Zener solved this problem analytically and showed that
\begin{equation}
  \label{TransLZM}
  T_{na}(\nu)=e^{-\delta},
\end{equation}
where $\delta=2\pi \epsilon_{12}^2/(\alpha\nu)$.

Our goal is to use this simple expression to describe each of the important nonadiabatic transitions in the four-fermion description.
However, equation~(\ref{TransLZM}) is  not very useful in its
current form when it comes to analyzing numerical results since it requires a knowledge of nonadiabatic quantities such
as $\alpha$ and $\epsilon_{12}$. $\epsilon_{12}$ can be estimated from the difference between adiabatic energy
curves at the closest approach. But, in the four-body problem, $\alpha$ is very difficult to extract because there is a rich structure of avoided crossings. Nevertheless, Clark~\cite{clarkpmatrix} showed how $\alpha$ can be
obtained from an analysis of the P-matrix coupling between two {\em adiabatic} states. In the Landau-Zener model, the
P-matrix between the two adiabatic states $\ket{\psi_+}$ and
$\ket{\psi_-}$ is
\begin{equation}
  \label{PmatLZ1}
  P_{+-}(\lambda)=\braket{\psi_+|\frac{d
  \psi_-}{d\lambda}}=\frac{\alpha}{4\epsilon_{12}}\frac{1}{1+[\alpha
  \lambda/(2\epsilon_{12})]^2},
\end{equation}
and has a characteristic Lorentzian form whose parameter is directly related to $\alpha$ and $\epsilon_{12}$.

Analysis of the spectrum and the P-matrices permits an identification of the states $\Psi_i$ and $\Psi_j$ involved in the most important nonadiabatic transitions, and a determination of the Landau-Zener transition probability
\begin{equation}
  \label{Trans}
  T_{ij}(\nu)=e^{-\delta_{ij}}=e^{-\frac{\eta_{ij}}{\xi}}.
\end{equation}
The Landau-Zener parameter $\delta_{ij}$ characterizes the
transition and is extracted by analyzing the spectrum, while the P-matrix is obtained from the numerical description of the adiabatic states using the Hellmann-Feynman theorem.

\begin{figure}[htbp]
\begin{center}
\includegraphics[scale=0.4]{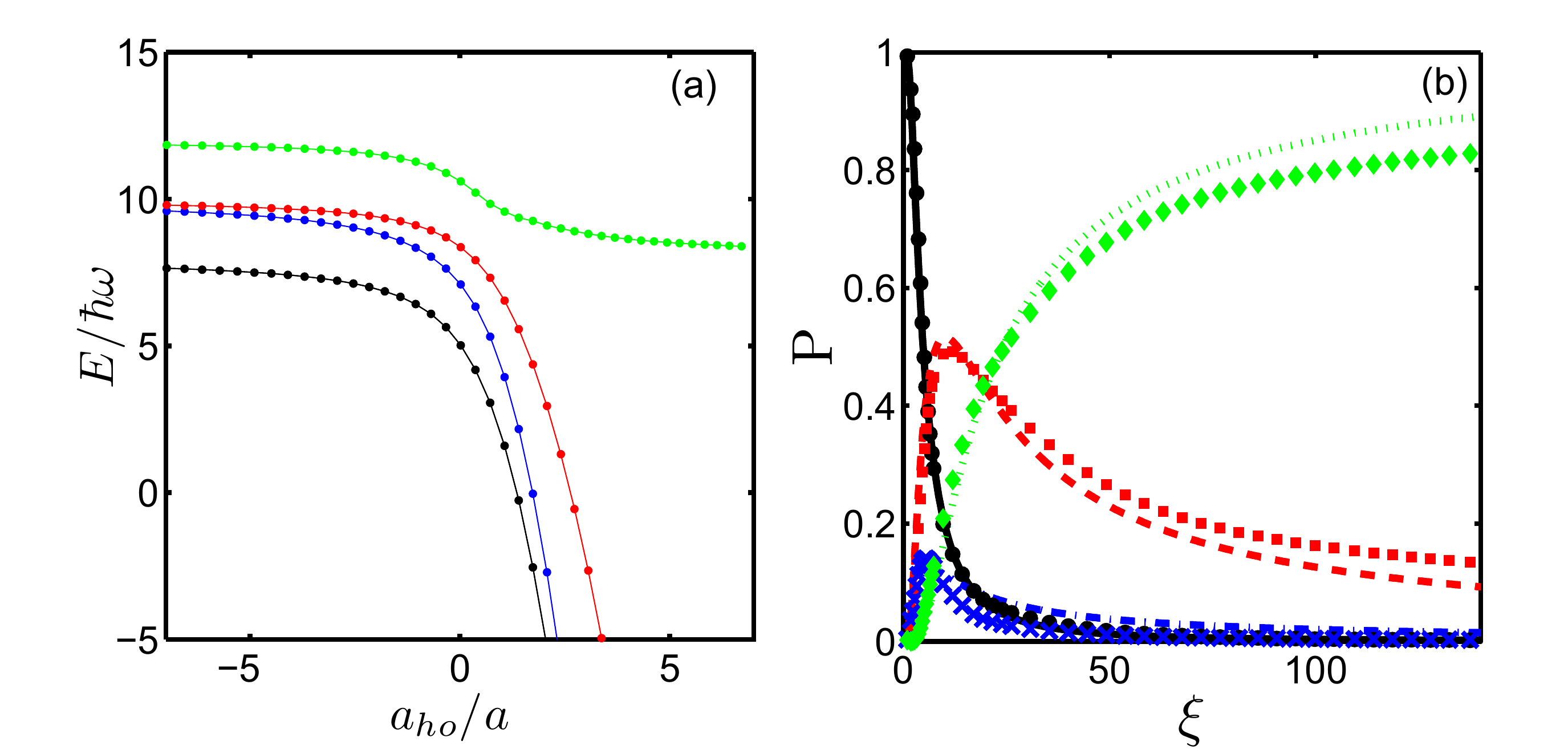}
\caption{ (a) Energy of the most important
states amenable to a Landau-Zener description, through the BCS-BEC crossover regime.The black curve
corresponds to $\ket{\Psi_1}$ which represents the ground state
configuration. The blue curve corresponds to $\ket{\Psi_2}$ which
represents the excited dimer-dimer configuration. The red curve
corresponds to $\ket{\Psi_5}$ which represents
a dimer plus two atoms. The green curve corresponds to
$\ket{\Psi_{13}}$ which represents the lowest four-atom configuration. (b)
Probability of ending up in a given configuration after the ramp, as a function of
the dimensionless speed parameter $\xi$. The symbols correspond to
the numerical evolution, while the curves correspond to the
Landau-Zener approximation. The colors follow the same convention as Figure
(a). Results from Ref.~\cite{stech07}.} \label{Trans4p}
\end{center}
\end{figure}

Figure~\ref{Trans4p}~(b)
displays the results obtained from the numerical time propagation. The black symbols correspond to the
dimer-dimer ground state. The blue symbols correspond to the excited
dimer-dimer family, the red symbols to the dimer--two-atom family,
and the green symbols to the four-atom family. For slow ramps (small
$\xi$), the probability of forming a ``condensate'', i.e., remaining in
the ground state, is large. For intermediate ramps, the greatest
probability is to break one bond and end up with a dimer plus two
particles. For fast ramps (large $\xi$), the probability of staying
in the {\it atomic ground state} on the BEC side is dominant. The probability of the system evolving into an excited
dimer-dimer configuration remains small for all ramping speeds.

To analyze these transitions within the Landau-Zener approximation,
the partially diabatic states have been labeled according to their energies in the BCS
regime. This labeling is arbitrary since many of the states are
almost degenerate. Based on the P-matrix couplings, the possible pathways point towards the states that are most likely important.
Starting from the ground state, note that $\ket{\Psi_1}$ has
important couplings with states $\ket{\Psi_2}$ and $\ket{\Psi_5}$.
Here, $\ket{\Psi_2}$ is the first excited dimer-dimer state, i.e.,
the lowest state of the excited dimer-dimer family. State
$\ket{\Psi_5}$ is the first excited state of the dimer--two-atom
configuration. Since an important probability is transferred to
states $\ket{\Psi_2}$ and $\ket{\Psi_5}$, we analyze the couplings
of these states to follow the flow of probability. State
$\ket{\Psi_2}$ has an important coupling with $\ket{\Psi_5}$, and
state $\ket{\Psi_2}$ has an important coupling with
$\ket{\Psi_{13}}$. Here, $\ket{\Psi_{13}}$ is the lowest state of
the four-atom configuration, i.e., the atomic ground state on the
BEC side. Figure~\ref{Trans4p}~(a) presents the energy curves of
these four states. Conveniently, each of these states represents a
different configuration. For that reason, this is the minimal set of
states that can describe the numerical results. While
more states could be included in the analysis, here only the
simplest possible case is considered.
The P-matrix analysis also reveals the order in which the transition usually occurs.
The
order of the peaks reveals the following sequence: the first
transition is $1\rightarrow 2$, then $2\rightarrow 5$, then
$1\rightarrow 5$, and finally $5\rightarrow 13$. The Landau-Zener
prediction for this sequence is
\begin{gather}
\label{probs}
p_1=(1-T_{1,2})(1-T_{1,5}),\nonumber \\
p_2=T_{1,2}(1-T_{2,5}),\nonumber \\
p_5=((1-T_{1,2})T_{1,5}+T_{2,5}T_{1,2})(1-T_{5,13}), \\
p_{13}=((1-T_{1,2})T_{1,5}+T_{2,5}T_{1,2})T_{5,13}.\nonumber
\end{gather}
Again, the sum of all these probabilities is, by construction, unity.
The Landau-Zener parameters obtained from the P-matrix analysis are
$\eta_{12} \approx 5.4$, $\eta_{15} \approx 6.6$, $\eta_{25} \approx
2.1$ and $\eta_{5,13} \approx 13.8$.
The sequence of Landau-Zener transitions in the
model shows good agreement with the numerical results even
though many possible transitions have been neglected in the approximate model.

\subsubsection{Universal properties}

The universal properties of two-component Fermi gases interacting through s-wave collisions has been intensively studied in recent years.
Some particularly important research has been carried out concerning the universal properties of four-fermion systems. Here we select two studies that benchmark the universal behavior of four-fermion systems.

The specific point in the strongly-interacting region where the
s-wave interaction strength reaches its maximal value, is usually
called unitarity. Unitarity is alternatively characterized by a divergent s-wave
scattering length, $|a|=\infty$.  If inelastic two-body scattering channels are energetically open, the scattering length is complex and it does not diverge all the way to infinity, but the following discussion assumes that such inelastic processes can be neglected. In this situation, if the range of
the interaction is much smaller than the typical interparticle
distance and if the scattering length is divergent, then no
relevant length scale exists that can characterize the interaction. This
situation is similar to the noninteracting limit, where the absence
of interactions implies, of course, the absence of a length scale
that describes the interaction. The absence of a length scale that
describes the interaction allows us to extract the functional form
of various quantities via dimensional analysis. Furthermore, it
allows us to relate quantities at unitarity to those in the
noninteracting limit.

Dimensional analysis becomes particularly
simple in the hyperspherical framework for a noninteracting or unitary system in free space, where
the hyperradius is the only coordinate with dimensions of length. Since
the potential curves have units of energy and the only length scale
is given by $R$, it follows that $U(R)\propto 1/R^2$. This is
equivalent to saying that the potential curves at unitarity are proportional to the noninteracting potential curves, i.e.,$U(R)\propto U_{NI}(R)$, since the
noninteracting potential curves have the form $1/R^2$. The resulting predictions have been derived in Refs.~\cite{tan05,wern06}.  

\begin{figure}[htbp]
\begin{center}
\includegraphics[scale=0.5,angle=0]{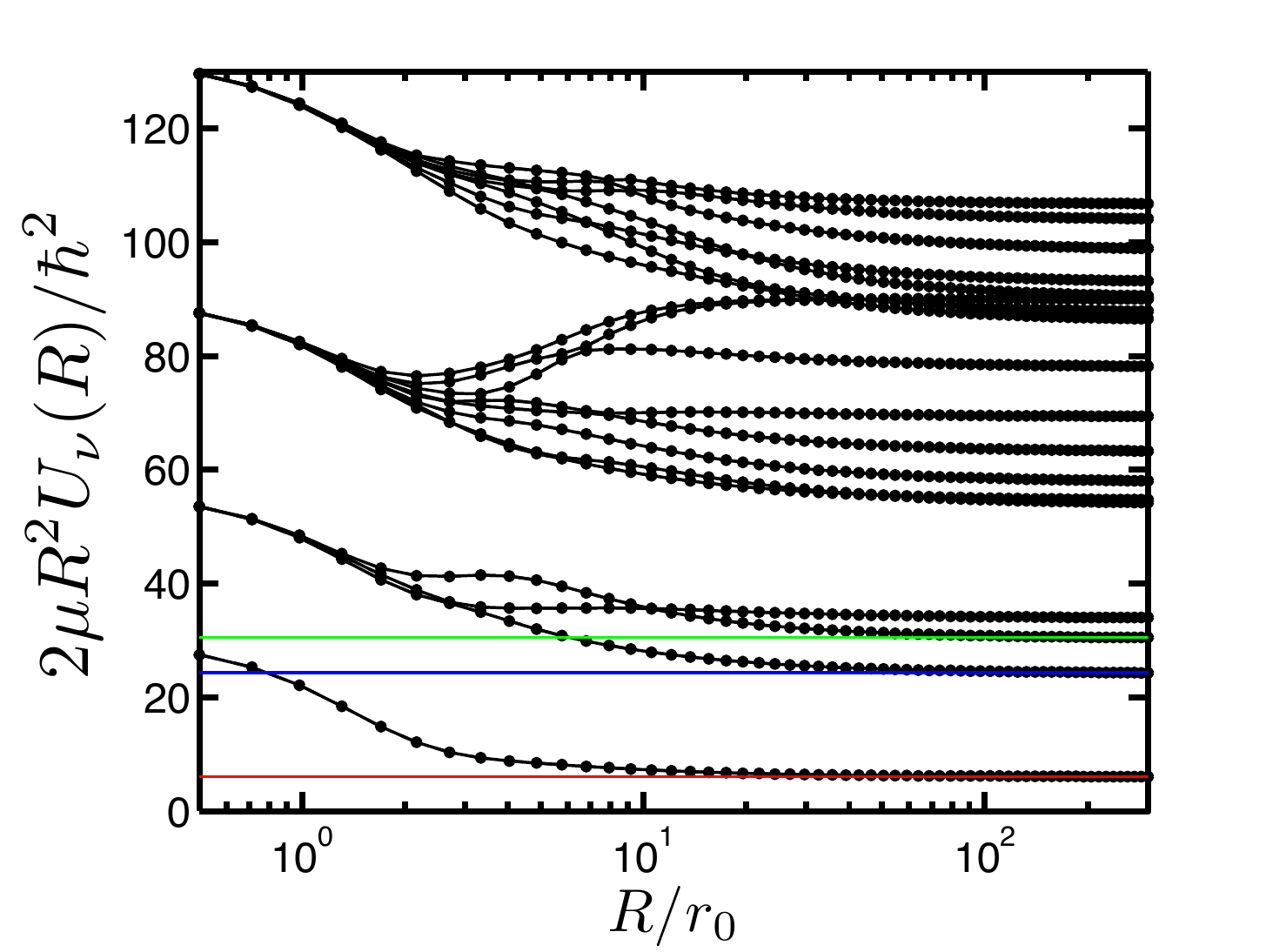}
\caption{Hyperspherical potential curves at unitarity for the 4-fermion system
multiplied by $2\mu R^2/\hbar^2$. The solid lines represent the
predictions extracted from the spectrum obtained with the CG method.
The symbols correspond to direct evaluation of the potential curves
with the CGHS method. Solid lines correspond to predictions of the large $R$ behavior of the potential curves extracted from the analysis of the excitation spectrum of the four-fermion trapped system. } \label{UnitPotCurves}
\end{center}
\end{figure}

The four-fermion potential curves calculated at unitarity using CGHS methods allow us to
test the premise of universality at the four-body level. Figure~\ref{UnitPotCurves}
presents predictions for the lowest 20 adiabatic potential curves for a Gaussian interaction model potential.
At large $R$ the potential curves become proportional to $1/R^2$ as predicted by the universal theory.
This behavior of the potential curves can be used to understand universal predictions for trapped systems
such as the virial theorem and the energy spacing of the trapped spectrum.

The notion of universality extends beyond the unitarity regime and can be applied to any finite scattering length.
Recently, Tan was able to derive a series of relations between different observables in two-component Fermi systems~\cite{tan2008generalized,tan2008energetics,tan2008large}.
These relations were obtained under the premise of universality and are a consequence of the wavefunction behavior when
two particles come close together (which is well described by the Bethe-Peierls boundary condition).
The universal Tan relations connect the energy, the expectation value of the trapping potential, the pair and momentum distribution functions
 of two-component Fermi systems through a quantity termed the integrated contact intensity $I(a)$.

\begin{figure}[htbp]
\begin{center}
\includegraphics[scale=0.5,angle=0]{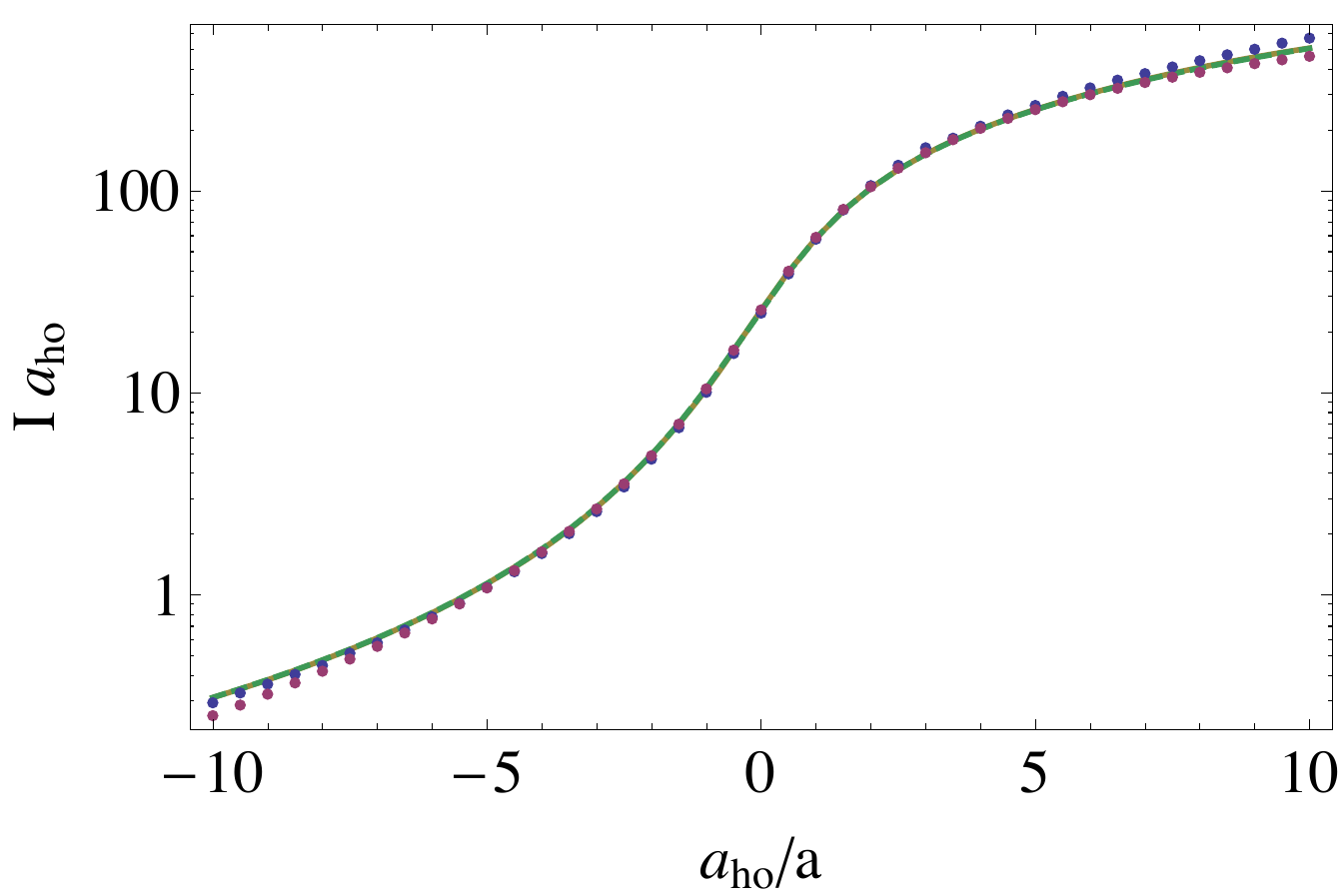}
\caption{Integrated contact intensity obtained in four different ways: through the analysis of the energy dependence on $a$ (solid curve), virial theorem (dashed curve), momentum distribution (blue symbols), and pair
correlation function (red symbols). The solid and dashed curves are
essentially indistinguishable on this scale.  Figure adapted from Ref.~\cite{blume2009universal}.
} \label{Intensity}
\end{center}
\end{figure}

Blume and Daily investigated the Tan relations for a trapped four-fermion system\cite{blume2009universal}. Using the CG method, they extracted the spectrum, the pair correlation function, the momentum distribution and the external potential expectation value for determining $I(a)$ in four different ways. Figure~\ref{Intensity} presents a comparison of the different predictions for the contact intensity. The excellent agreement between the different predictions numerically demonstrates the validity of the Tan relations. Furthermore, it quantifies the contact intensity for the four-fermion system, which can serve as a benchmark for further studies.

\section{Summary}

This review has concentrated on recent developments in the four-body problem, emphasizing those insights that have either used hyperspherical coordinate techniques directly, or else which have benefited indirectly from those insights.  The utility of formulating the few-body or even the many-body problem in hyperspherical coordinates was glimpsed early on by some of the pioneers in the field such as Delves\cite{delves1959tag, delves1960tag}, Smirnov and Shitikova\cite{Shitikova77}, Macek\cite{MacekJPB1968}, Lin\cite{CDLReview}, Fano\cite{fano1976dee}, Kuppermann\cite{kuppermann1997rsr}, and Aquilanti\cite{aquilanti1997qmh}.  This class of methods has been especially valuable for studying ultracold collisions in recent years, and ultracold applications have been the focus of this overview.

Bosonic four-body systems could not be discussed in much detail in this article, owing to length constraints, but they have provided an important proving ground for many of the methods discussed in the present review.  Early ideas on 3-body recombination for three bosons or for three nonidentical fermions that arose first from an adiabatic hyperspherical perspective\cite{nielsen1999ler, EsryGreeneBurke1999} have since been confirmed and developed further, with many useful new insights from independent theoretical perspectives. The few-body system receiving the most attention over the years has been the three-body problem, both with and without long range Coulombic interactions.  The exciting headway represented by that body of literature has had some extensions to handle nontrivial reactive processes for systems with four interacting particles,\cite{Greene2010} and in a handful of studies, for systems with many more particles.\cite{bohn_esry_greene_hsbec,blume00, kim2000elt, kushibe2004aha, sorensen2002ctb, rittenhouse2006hdd, rittenhouse2008dfg}

Numerous questions still remain to be addressed in this field, aiming in the long run towards not only answering questions of universality in the ultracold, but also towards developing systematic ways of handling four-body chemical reaction dynamics at the fully quantum level.  A large parameter space also remains to be explored just in the ultracold limit of the four-body problem, such as varying mass ratios for fermionic and for bosonic systems as well as mixed Bose-Fermi systems.

\section{Acknowledgments}

Many of the results described in this review were developed in close collaboration with Doerte Blume, and we deeply appreciate her contributions and insights.  We also thank Brett Esry for numerous informative discussions and for access to some of his results prior to publication.  This project was supported by NSF.

\appendix

\section{Hyperangular coordinates}
\label{App:Hyper_angs}

As with Jacobi coordinates, there is no unique way to construct the
hyperangles of a system. In this section we construct the hyperangular
coordinates used in the four-fermion problem. The choice of hyperangular
parameterization has physical meaning. Different parameterizations can be used
to describe different correlations within the system. Also, in the case of
body-fixed coordinates, hyperspherical coordinates can be used to remove the
Euler angles of solid rotation, reducing the dimensionality of the system.

\subsection{Delves coordinates}

Unfortunately, or possibly fortuitously depending on your view point, there is
no unique way to define the hyperangles in a given system. Here we use a
simple, standardized method of defining them used by many others
\cite{MacekJPB1968,cavagnero1986eca,bohn_esry_greene_hsbec,SmirnovShitikova,rittenhouse2006hdd,rittenhouse2008dfg}
in the form of the so called Delves coordinates
\cite{delves1959tag,delves1960tag}. We will begin by examining a well known
example of hyperspherical coordinates, that of normal spherical polar
coordinates. Clearly these coordinates can be used to describe the relative
motion of 2 particles in 3 dimensions or the position of a single particle in
a trap-centered coordinate system, but it can also be used in less obvious
ways. For instance, spherical polar angles may be used to describe the
relative motion of 4 particles in 1 dimension.

\begin{figure}[t]
\begin{center}
\includegraphics[width=1.5in]{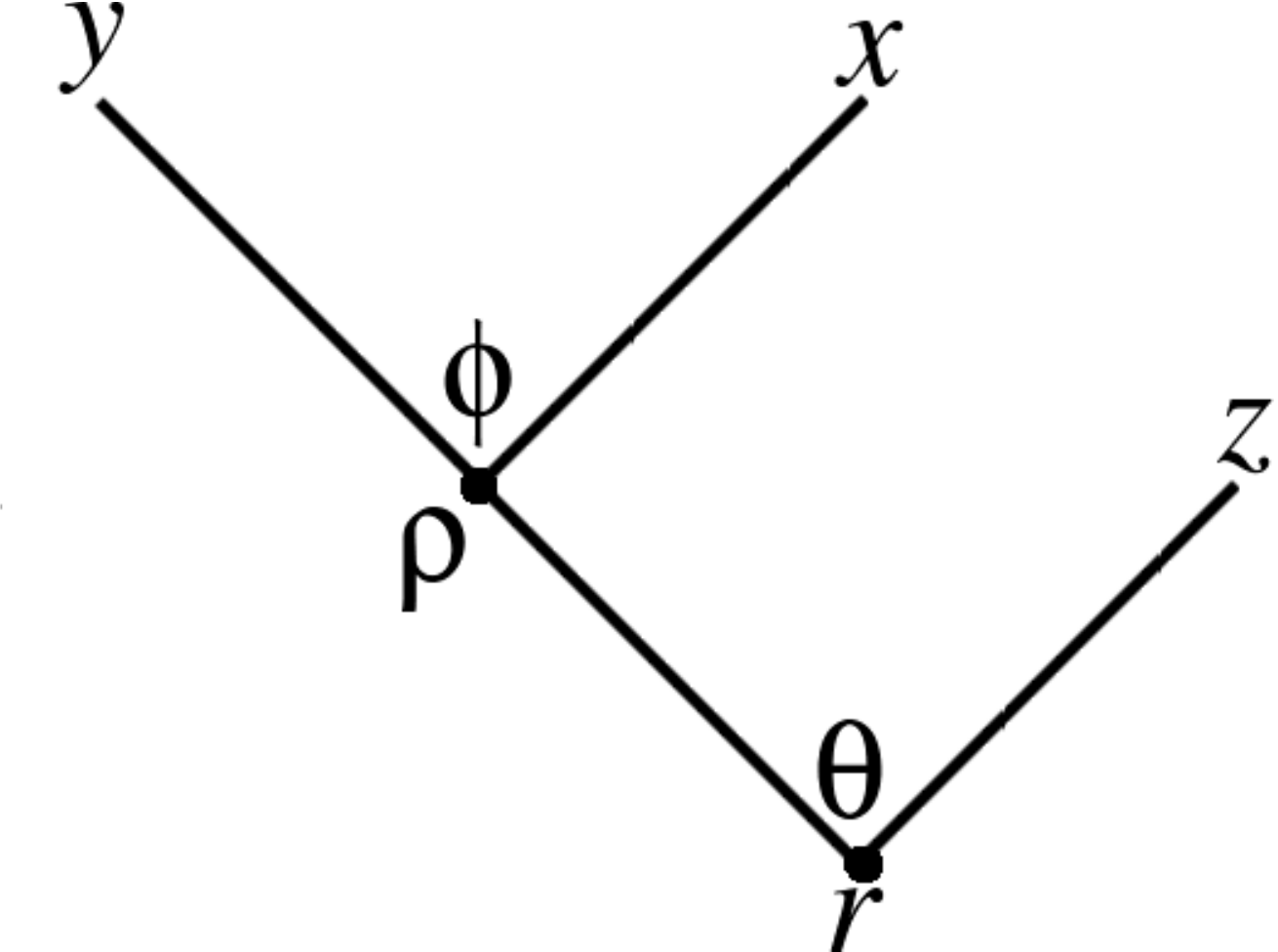}
\end{center}
\caption{The tree that gives the standard spherical coordinates for a 3
dimensional system is shown.}%
\label{delvesfigspherical}%
\end{figure}

The components of a three dimensional vector, $\bm{r}$, can be written in
terms of a radius and two angles as%
\begin{align}
x  &  =r\cos\phi\sin\theta,\label{Eq:spherical_x}\\
y  &  =r\sin\phi\sin\theta,\label{Eq:spherical_y}\\
z  &  =r\cos\theta. \label{Eq:spherical_z}%
\end{align}
This parameterization can be represented in a simple tree structure shown in
Fig. \ref{delvesfigspherical}. The end points of the tree represent each component of the vector
$\bm{r}$, and each node in the tree is represents an angle. Also associated
with each node is a sub radius. For the lowest node the \textquotedblleft
subradius" is merely the total length of the vector, $r$. For the upper node
the subradius is merely the cylindrical radius $\rho=\sqrt{x^{2}+y^{2}}$.
Using the tree structure from Fig. \ref{delvesfigspherical}, a set of rules
can be developed for extracting the parameterization of Eqs.
(\ref{Eq:spherical_x}), (\ref{Eq:spherical_y}), and (\ref{Eq:spherical_z}). Starting
at the bottom node with total radius, $r$, move up through the tree to the
desired coordinate. For each move through the tree, if you move to the left
(right) from a node multiply by the sine (cosine) of the angle associated with
that node. Continue until you reach the Cartesian component.

\begin{figure}[t]
\begin{center}
\includegraphics[width=2.5in]{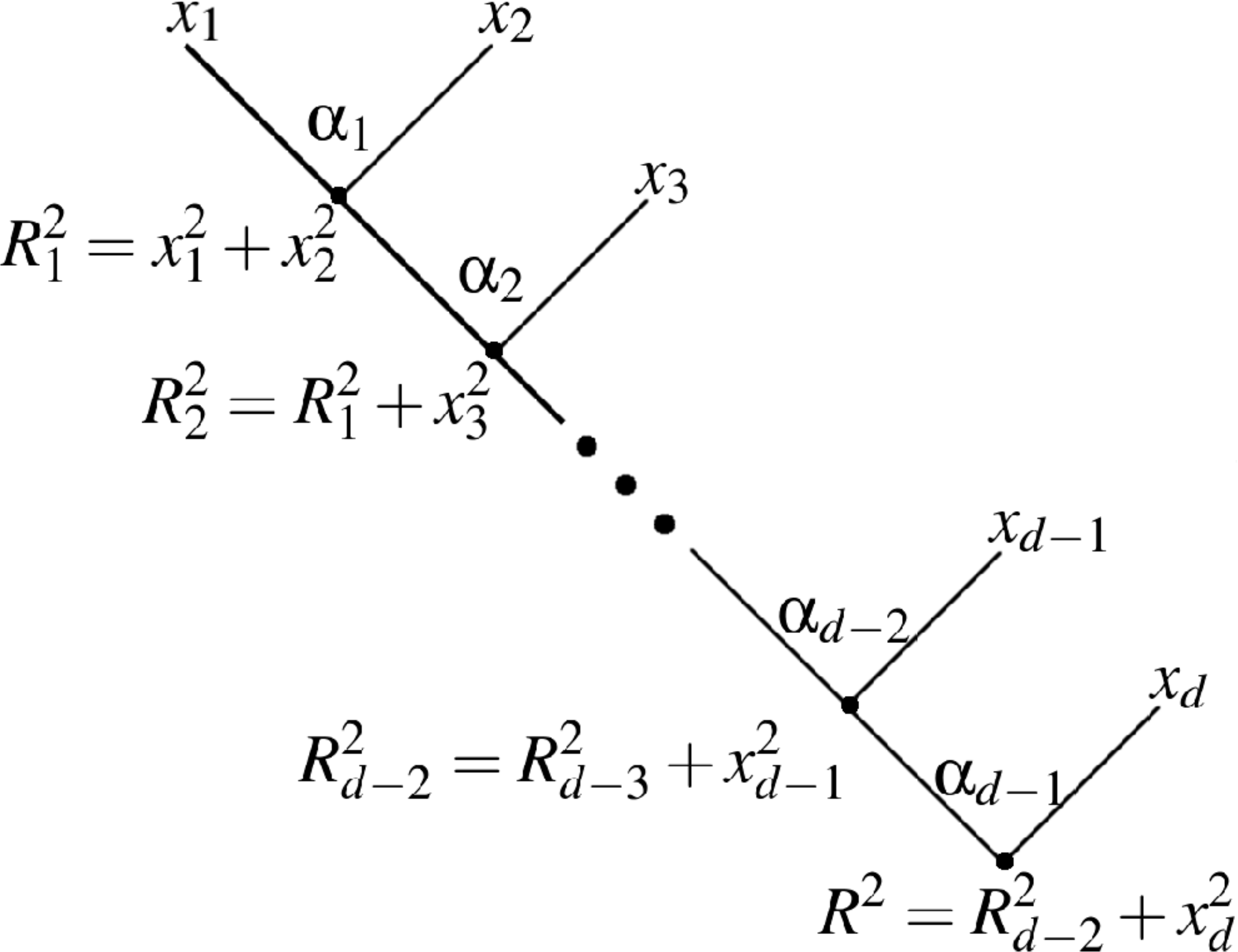}
\end{center}
\caption{\label{delvesfig1}The canonical tree that gives a hyperangular parameterization for a
$d$ dimensional system is shown.}%
\end{figure}

This procedure can be generalized readily from three to $d$ dimensions. Start
by building a tree with $d$ free ends and $d-1$ nodes, associate an angle with
each node and follow the above rules. Using the tree structure, starting at
the bottom node with total hyperradius, $R$, move up through the tree to the
desired coordinate. If you move to the left (right) from a node, multiply by
the sine (cosine) of the angle associated with that node. Continue until you
have reach the desired Cartesian component. A specific tree for $d$ dimensions
is shown in Fig. \ref{delvesfig1}. Following the rules this tree gives the
hyperangular representation
\begin{align}
x_{n}  &  =R\cos\alpha_{n-1}\prod\limits_{j=n}^{d-1}%
\sin\alpha_{j},\label{Eq:hypang_canonical}\\
0  &  \leq\alpha_{j}\leq\pi,\text{ }j=2,...,d-1\nonumber\\
0  &  \leq\alpha_{1}\leq2\pi\nonumber
\end{align}
where $\cos\alpha_{0}\equiv1$ and $\prod\limits_{j=d}^{d-1}\sin\alpha
_{j}\equiv1$. This can also be written as%
\begin{align}
\tan\alpha_{n}  &  =\dfrac{\sqrt{\sum_{j=1}^{n}x_{j}^{2}}}{x_{n+1}%
},\label{hypera}\\
n  &  =1,2,3,...,d-1.\nonumber
\end{align}
This hyperspherical tree has been dubbed the canonical tree
\cite{Avery,SmirnovShitikova} as it is simple to construct and very easy to
add more dimensions to.

To avoid double counting, the range that the hyperangles take on is restricted
depending on how many free branches are attached to the node corresponding to
a given angle. If the node has two free branches, then the angle takes on the
full range $0$ to $2\pi$. If the node has one free branch attached, the angle
goes from $0$ to $\pi$. If the node has no free branches attached to it, the
associated angle goes from $0$ to $\pi/2$. Following these rules for the
canonical tree gives the ranges of the angles $\alpha_{i}$,%
\begin{align*}
0  &  \leq\alpha_{1}\leq2\pi,\\
0  &  \leq\alpha_{i}\leq\pi,i=2,...,d-1.
\end{align*}

\begin{figure}[t]
\begin{center}
\includegraphics[width=2in]{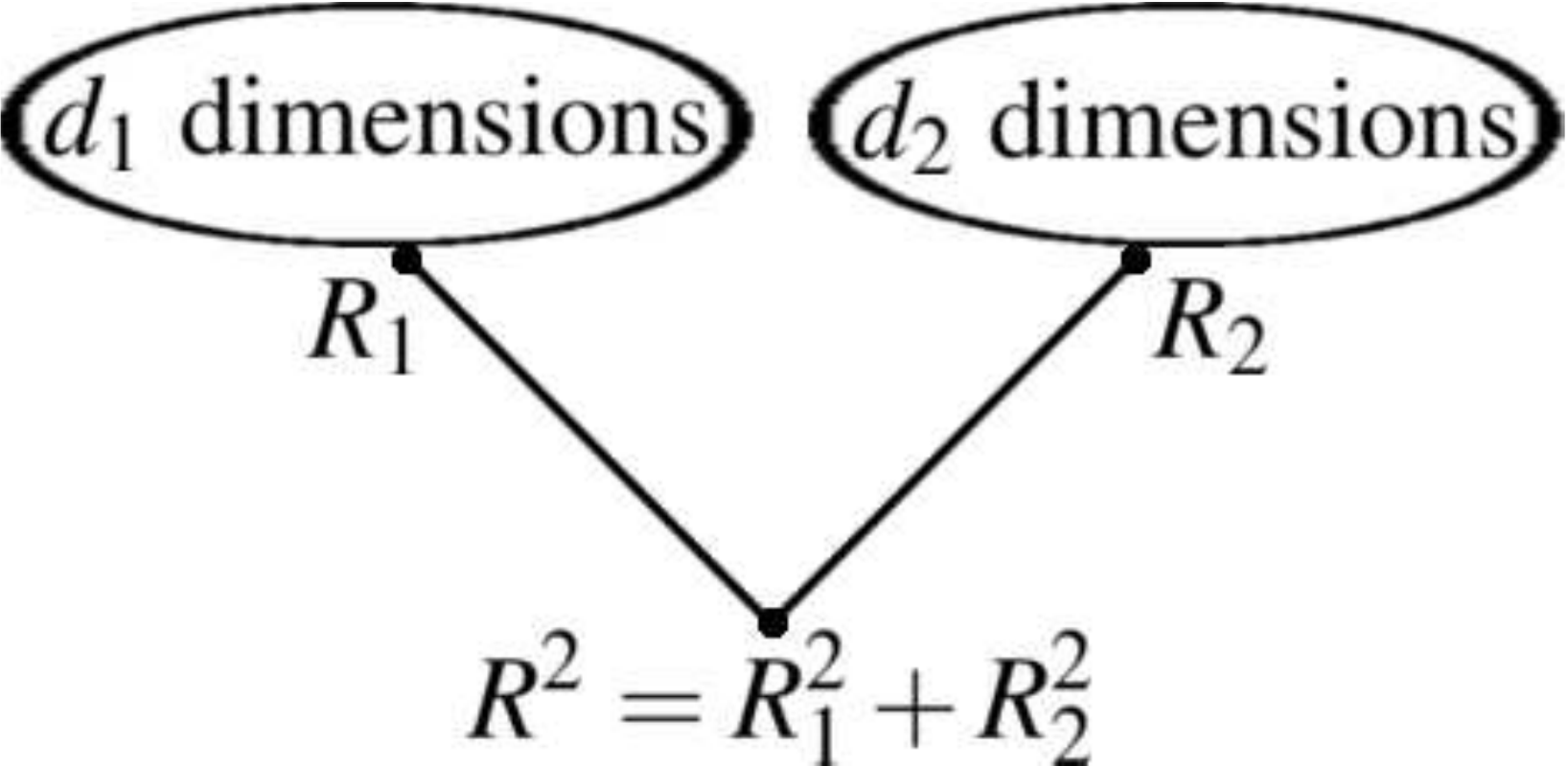}
\end{center}
\caption{The tree structure used to correlate two subspaces to a single
hyperradius.}%
\label{delvesfig2}%
\end{figure}

Another slightly more abstract way of considering this construction is to
start by breaking the $d$ dimensional space into two subspaces of dimension
$d_{1}$ and $d_{2}$, and assuming that these two subspaces are already
described by two sets of sub-hyperspherical coordinates, $\left(  R_{1}%
,\Omega_{1}\right)  $ and $\left(  R_{2},\Omega_{2}\right)  $. With these
assumptions all that remains is to correlate the sub-hyperradii. This is done
by following the type of procedure described above using the tree structure
shown in Fig. \ref{delvesfig2},%
\begin{align}
R_{1}  &  =R\sin\alpha,\label{subradiuscoor}\\
R_{2}  &  =R\cos\alpha,\nonumber
\end{align}
where $\alpha$ is now the final hyperangle in the system. Using
this procedure recursively, one can define the hyperangles in the subspaces
until the only remaining subspaces are the individual Cartesian components of
the total $d$ dimensional space. The concept of dividing the total space up
into subspaces will prove very or the purpose of constructing basis functions.

\begin{figure}[t]
\begin{center}
\includegraphics[width=2.5in]{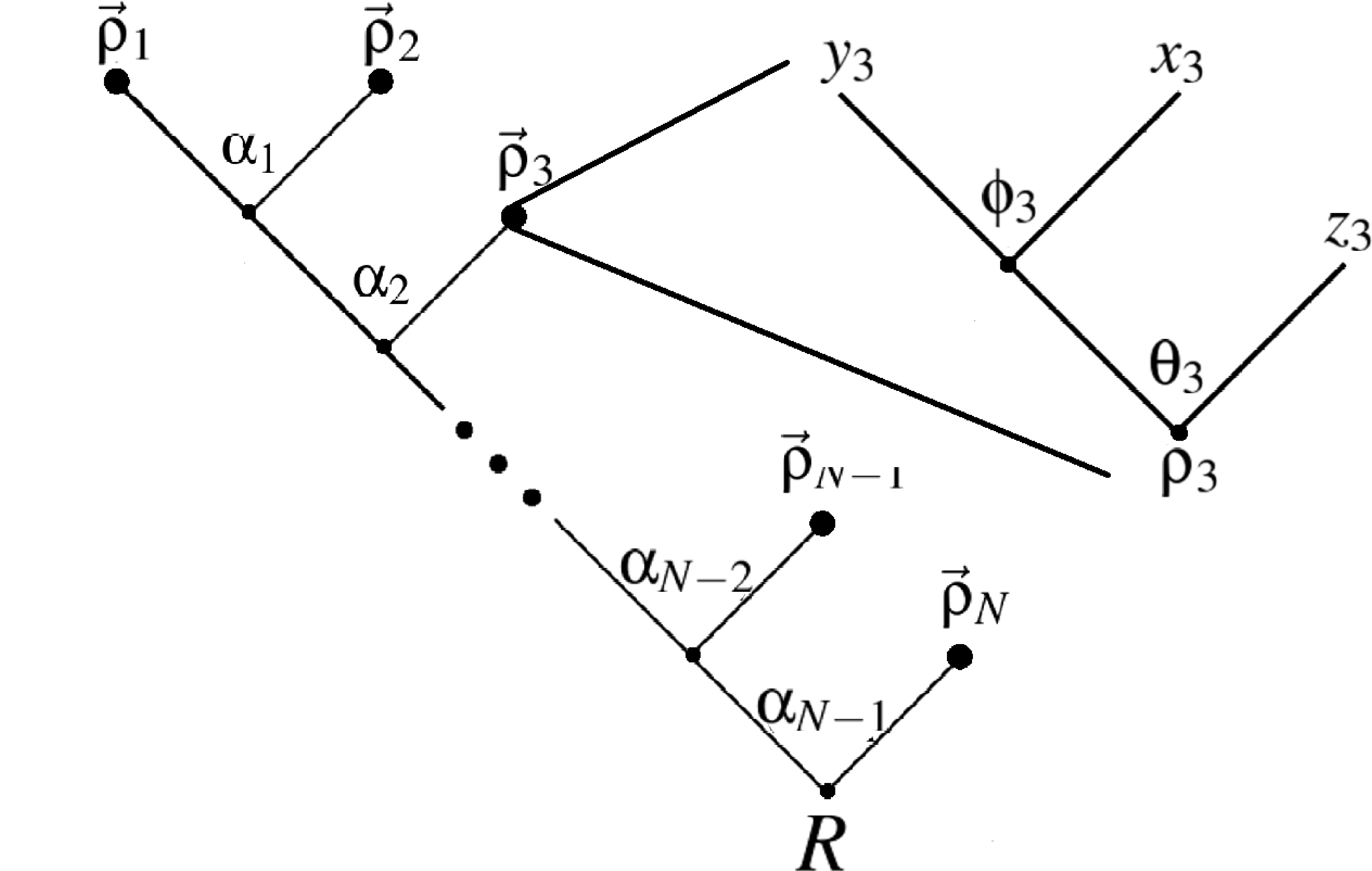}
\end{center}
\caption{\label{delves_3}The tree structure used to parameterize the hyperangles for $N$ three
dimensional vectors is shown. Note that the dot at the end of each branch in
the tree on the left stands for a spherical polar tree.}%
\end{figure}

As a final example of hyperangular parameterizations, we introduce a
parameterization for $N$ 3-dimensional vectors $\left\{  \bm{\rho}%
_{i}\right\}  _{i=1}^{N}$. One could break each vector up into its individual
components and use the canonical parameterization from Eq.
(\ref{Eq:hypang_canonical}), but this removes much of the spatial physical
intuition that one could bring to bear, such as the individual spatial angular
momentum corresponding to each vector. Instead one can use a variation on the
canonical tree shown in Fig. \ref{delves_3}. On first glance, this tree might
seem the same as the canonical tree shown in Fig. \ref{delvesfig1}. In this
case, though, the large dot at the end of each branch represents the spherical
polar sub-tree of the form shown in Fig. \ref{delvesfigspherical} for each
vector $\bm{\rho}_{i}$. Using this tree structure and following the rules outlined above,
$2N$ of the $3N-1$ hyperangles are given by the normal spherical polar angles
for each vector, $\left(  \theta_{1},\phi_{1},\theta_{2},\phi_{2}%
,...,\theta_{N},\phi_{N}\right)  $. The remaining $N-1$ hyperangles are given
by%
\begin{align}
\tan\alpha_{i}  &  =\dfrac{\sqrt{\sum_{j=1}^{i}\rho_{j}^{2}}%
}{\rho_{i+1}},\label{Eq:N_bod_hyp_ang}\\
0  &  \leq\alpha_{i}\leq\dfrac{\pi}{2},\nonumber\\
i  &  =1,2,3,...,N-1.\nonumber
\end{align}
where $\rho_{i}$ is the length of the $i$th vector. It will be shown in the
next section that this parameterization is useful when spatial angular
momentum plays a role in the problem of interest. For completeness, following
the rules laid out in Appendix A, the hyperangular volume element that results
from this parameterization is given by
\[
d\Omega=\left(  \prod_{i=1}^{N}d\omega_{i}\right)  \left(  \prod_{j=1}%
^{N-1}\cos^{2}\alpha_{j}\sin^{3j-1}\alpha_{j}\right)
\]
where $\omega_{i}$ is the normal spherical polar differential volume for
$\bm{\rho}_{i}$.

\begin{figure}[htbp]
\begin{center}
\includegraphics[width=1.5in]{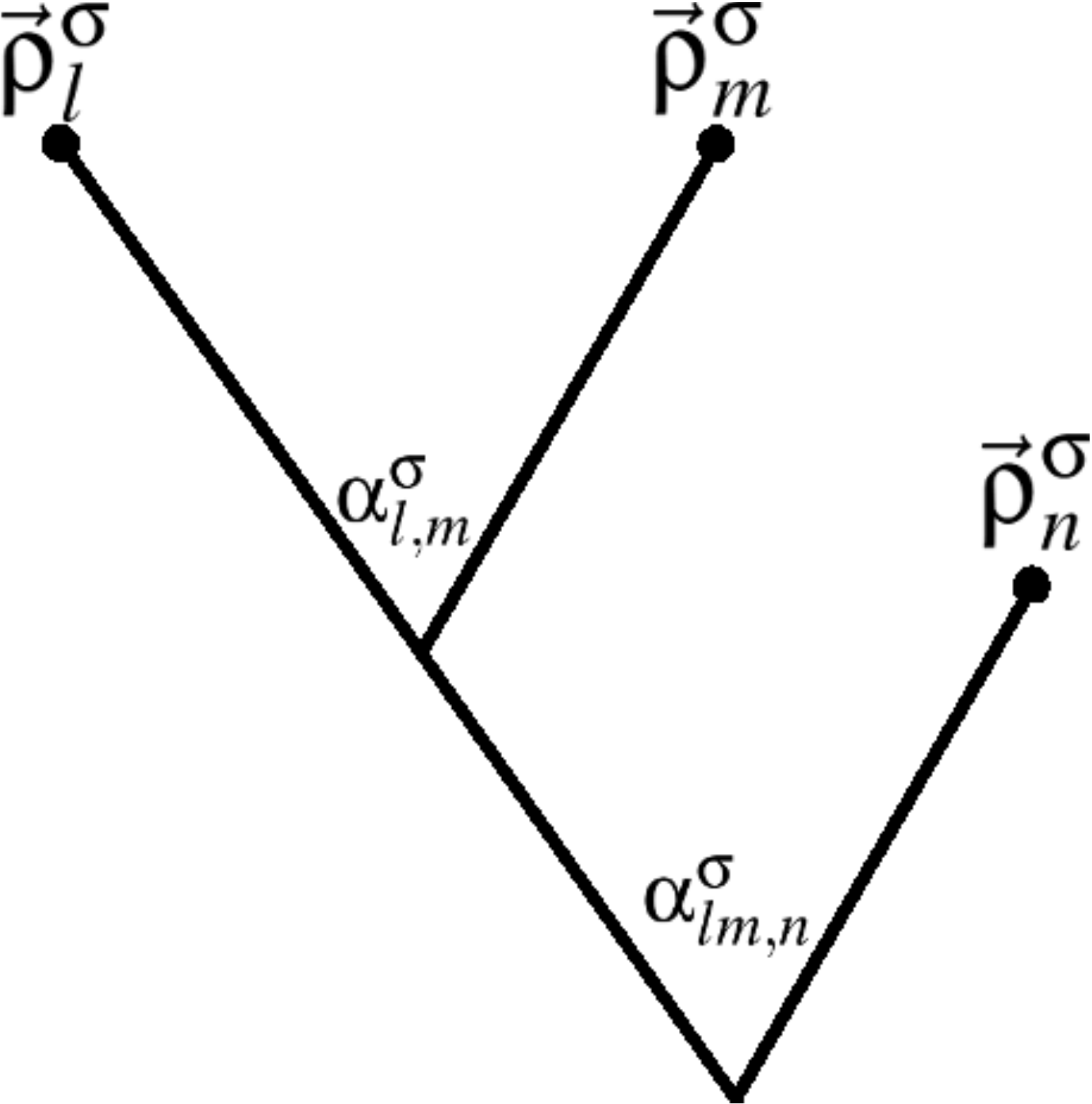}
\end{center}
\caption{The hyperspherical tree used to parameterize the hyperangular
coordinates in the four-fermion problem is shown. See Chapter 2 or Ref.
\cite{SmirnovShitikova} for details.}%
\label{delves_4bod}%
\end{figure}

The first hyperangular parameterization used here is in the form of Eq.
(\ref{Eq:N_bod_hyp_ang}) for three 3D vectors. The hyperradius is defined in the
same way as in Eq. (\ref{Eq:hyperr_def}),%
\begin{equation}
R^{2}=\sum_{i=1}^{d}x_{i}^{2}, \label{hyperrad_def_d}%
\end{equation}
where each $x_{i}$ is a Cartesian component of one of the Jacobi vectors. The
hyperspherical trees that will be used for the four fermion problem will of
the type in Fig. \ref{delves_3} in which $N$ vectors are described by the
spherical polar angles of each vector and a set of hyperangles correlating the
lengths of each vector. Specifically, the hyperangles are defined by the tree
shown in Fig. \ref{delves_4bod} combined with the spherical polar angles of
each Jacobi vector. Following the rules described above gives the hyperangles,%
\begin{align}
\alpha_{l,m}^{\sigma}  &  =\tan^{-1}\dfrac{\left\vert \bm{\rho}_{l}^{\sigma
}\right\vert }{\left\vert \bm{\rho}_{m}^{\sigma}\right\vert }%
,\label{symm_hyp_angs}\\
\alpha_{lm,n}^{\sigma}  &  =\tan^{-1}\dfrac{\sqrt{\left\vert \bm{\rho}%
_{l}^{\sigma}\right\vert ^{2}+\left\vert \bm{\rho}_{m}^{\sigma}\right\vert
^{2}}}{\left\vert \bm{\rho}_{n}^{\sigma}\right\vert }.\nonumber
\end{align}
Here the superscript $\sigma=s,i1,i2$ indicates which Jacobi system is being
used, while $l,m,n$ indicate the three Jacobi vectors from that system. In
principle there are six of these hyperangular coordinate systems that can be
constructed from each set of Jacobi vectors giving a total of 18 different
hyperangular systems. Fortunately we will not need all of these to tackle the
four fermion problem. If one considers two particles, say 1 and 2, the tree
defined by Fig. \ref{delves_4bod} with $\sigma=i1,$ $l=2,m=3$ and $n=1$ can be
decomposed into two subtrees. The right branch describes the hyperangular
behavior of the dimer alone, while the left branch describes the behavior of
the remaining three body system composed of a dimer and two free particles.
This type of decomposition will be important for evaluating kinetic energy
matrix elements and defining basis functions.


Using the recursive definition of the hyperangular momentum operator in Eq.
(\ref{Eq:Hyperangmoment_delves}) the total hyperangular momentum operator can be
written in terms of the hyperangular coordinates as%
\begin{align}
\Lambda^{2}=  &  \Delta_{1}\left(  \alpha_{lm,n}^{\sigma}\right)  -\dfrac
{1}{\sin^{2}\alpha_{lm,n}^{\sigma}\sin\alpha_{l,m}^{\sigma}\cos\alpha
_{l,m}^{\sigma}}\left[  \dfrac{\partial}{\partial\alpha_{l,m}^{\sigma}%
}\right]  ^{2}\sin\alpha_{l,m}^{\sigma}\cos\alpha_{l,m}^{\sigma}%
\label{hyper_ang_moment1}\\
&  +\dfrac{\hat{l}_{l}^{2}}{\sin^{2}\alpha_{lm,n}^{\sigma}\sin^{2}\alpha
_{m,n}^{\sigma}}+\dfrac{\hat{l}_{m}^{2}}{\sin^{2}\alpha_{lm,n}^{\sigma}%
\cos^{2}\alpha_{m,n}^{\sigma}}+\dfrac{\hat{l}_{n}^{2}}{\cos^{2}\alpha
_{lm,n}^{\sigma}},\nonumber\\
\Delta_{1}\left(  \alpha_{lm,n}^{\sigma}\right)  =  &  \dfrac{-1}{\sin
^{2}\alpha_{lm,n}^{\sigma}\cos\alpha_{lm,n}^{\sigma}}\dfrac{1}{\sin
\alpha_{lm,n}^{\sigma}}\dfrac{\partial}{\partial\alpha_{lm,n}^{\sigma}}%
\sin\alpha_{lm,n}^{\sigma}\dfrac{\partial}{\partial\alpha_{lm,n}^{\sigma}}%
\sin^{2}\alpha_{lm,n}^{\sigma}\cos\alpha_{lm,n}^{\sigma},\nonumber
\end{align}
where $\hat{l}_{l},\hat{l}_{m}$ and $\hat{l}_{n}$ are the normal spatial
angular momentum operators for each Jacobi vector. This can also be written
directly from Eq. (\ref{Eq:Hyperangmoment_delves}) as
\begin{align}
\Lambda^{2}  &  =\Delta_{1}\left(  \alpha_{lm,n}^{\sigma}\right)
+\dfrac{\Lambda_{l,m}^{2}}{\sin^{2}\alpha_{lm,n}}+\dfrac{\hat{l}_{n}^{2}}%
{\cos^{2}\alpha_{lm,n}^{\sigma}},\label{hyper_ang_moment2}\\
\Lambda_{l,m}^{2}  &  =-\dfrac{1}{\sin\alpha_{l,m}^{\sigma}\cos\alpha
_{l,m}^{\sigma}}\left[  \dfrac{\partial}{\partial\alpha_{l,m}^{\sigma}%
}\right]  ^{2}\sin\alpha_{l,m}^{\sigma}\cos\alpha_{l,m}^{\sigma}+\dfrac
{\hat{l}_{l}^{2}}{\sin^{2}\alpha_{m,n}^{\sigma}}+\dfrac{\hat{l}_{m}^{2}}%
{\cos^{2}\alpha_{m,n}^{\sigma}},\nonumber
\end{align}
where all of the hyperangular behavior above the second node in
Fig. \ref{delves_4bod} is described by a sub-hyperangular momentum,
$\Lambda_{lm,n}^{2}$.

Constructing the hyperspherical harmonics for the four-body system is
accomplished following the procedure in Section~\ref{Sec:Hyperharms} giving%

\begin{align}
Y_{\left[  \lambda\lambda_{l,m}l_{l},l_{m},l_{n}\right]  }^{\left(  4b\right)
}\left(  \Omega\right)  =  &  N_{l_{l}l_{m}\lambda_{l,m}}^{33}N_{\lambda
_{l,m}l_{n},\lambda}^{63}\sin^{\lambda_{l,m}}\left(  \alpha_{lm,n}\right)
\cos^{l_{n}}\left(  \alpha_{lm,n}\right)  P_{\left(  \lambda-\lambda
_{l,m}-l_{n}\right)  /2}^{\lambda_{l,m}+5/2,l_{n}+1}\left(  \cos2\alpha
_{lm,n}\right) \label{4bod_harm}\\
&  \times N_{l_{l},l_{m}}^{\lambda_{l,m}}\sin^{l_{l}}\left(  \alpha
_{l,m}\right)  \cos^{l_{m}}\left(  \alpha_{l,m}\right)  P_{\left(
\lambda_{l,m}-l_{l}-l_{m}\right)  /2}^{l_{l}+1,l_{m}+1}\left(  \cos
2\alpha_{l,m}\right) \nonumber\\
&  \times y_{l_{l}m_{l}}\left(  \omega_{l}\right)  y_{l_{m}m_{m}}\left(
\omega_{m}\right)  y_{l_{n}m_{n}}\left(  \omega_{n}\right)  ,\nonumber
\end{align}
where $P_{\gamma}^{\alpha,\beta}\left(  x\right)  $ is a Jacobi polynomial of
order $\gamma$, $y_{lm}\left(  \omega\right)  $ is a normal spherical harmonic
with spherical polar solid angle $\omega,$ and $N_{abc}^{de}$ is a
normalization constant \cite{SmirnovShitikova,Avery}:%
\[
N_{abc}^{de}=\left[  \dfrac{\left(  2c+d+e-2\right)  \Gamma\left(
\tfrac{a+b+c+d+e-2}{2}\right)  \left(  \tfrac{c-a-b}{2}\right)  !}%
{\Gamma\left(  \tfrac{c+a-b+d}{2}\right)  \Gamma\left(  \tfrac{c+b-a+e}%
{2}\right)  }\right]  ^{1/2}.
\]
In Eq. (\ref{4bod_harm}) the degeneracy quantum number $\mu$ has been replaced
with an explicit tabulation of the hyperangular momentum quantum numbers, i.e.
$\lambda\mu\rightarrow\left[  \lambda\lambda_{l,m}l_{l},l_{m},l_{n}\right]  $.
The total four-body hyperspherical harmonics satisfy the eigenvalue equation
$\Lambda^{2}Y_{\left[  \lambda\lambda_{l,m}l_{l},l_{m},l_{n}\right]
}^{\left(  4b\right)  }\left(  \Omega\right)  =\lambda\left(  \lambda
+7\right)  Y_{\left[  \lambda\lambda_{l,m}l_{l},l_{m},l_{n}\right]  }^{\left(
4b\right)  }\left(  \Omega\right)  $. The sub-harmonics that are
eigenfunctions of $\Lambda_{l,m}^{2}$ can be found as well:%
\begin{align}
Y_{\left[  \lambda_{l,m}l_{l},l_{m}\right]  }^{\left(  3b\right)  }\left(
\Omega_{l,m}\right)  =  &  N_{l_{l},l_{m}\lambda_{l,m}}^{33}\sin^{l_{l}%
}\left(  \alpha_{l,m}\right)  \cos^{l_{m}}\left(  \alpha_{l,m}\right)
P_{\left(  \lambda_{l,m}-l_{l}-l_{m}\right)  /2}^{l_{l}+1,l_{m}+1}\left(
\cos2\alpha_{l,m}\right) \label{3bod_harm}\\
&  \times y_{l_{l}m_{l}}\left(  \omega_{l}\right)  y_{l_{m}m_{m}}\left(
\omega_{m}\right)  .\nonumber
\end{align}
Here the superscript, (3b), indicates that this eigenfunction behaves as a
3-body hyperspherical harmonic. For instance if a hyperspherical tree is used
with Jacobi vectors defined in the $i1$ interaction coordinate system and
$l=1$, $m=3$, and $n=2$, this three-body harmonic describes the free-space
behavior of a dimer with two free particles. The three body harmonics obey the
eigenvalue equation, $\Lambda_{l,m}^{2}Y_{\left[  \lambda_{l,m}l_{l}%
,l_{m}\right]  }^{\left(  3b\right)  }\left(  \Omega_{l,m}\right)
=\lambda_{l,m}\left(  \lambda_{l,m}+4\right)  Y_{\left[  \lambda_{l,m}%
l_{l},l_{m}\right]  }^{\left(  3b\right)  }\left(  \Omega_{l,m}\right)  $. The
restrictions on the values of $\lambda$ and $\lambda_{l,m}$ are%
\begin{align}
\lambda_{l,m}  &  =l_{l}+l_{m}+2j,\label{lamm_nums}\\
\lambda &  =\lambda_{l,m}+l_{n}+2k\nonumber\\
&  =l_{l}+l_{m}+l_{n}+2j+2k,\nonumber
\end{align}
where $j,k=0,1,2,...$. The quantum numbers $l_{l},l_{m}$ and $l_{n}$ are the
spatial angular momentum quantum numbers associated with Jacobi vectors
$\bm{\rho}_l$, $\bm{\rho}_m$ and $\bm{\rho_n}$ respectively,
and each has a z-projection quantum number associated with it which we have
suppressed in Eqs. (\ref{4bod_harm}) and (\ref{3bod_harm}).


\subsection{Democratic coordinates}

Using Delves coordinates greatly simplifies evaluating hyperangular momentum
matrix elements, but it still leaves the 8 dimensional space of hyperangles. I
will only be considering systems with total angular momentum $L=0$. Therefore
it is convenient to move into a body-fixed coordinate system, as the final
wavefunction for the four-body problem will not depend on the Euler angles
that produce a solid rotation of the system. Removing the Euler angle
dependence is accomplished by transforming into the so-called democratic, or
body-fixed coordinates. Four-body democratic coordinates are developed in
several references (see Refs.
\cite{aquilanti1997qmh,kuppermann1997rsr,littlejohn1999qdk}). In this work we
use the parameterization of Aquilanti and Cavalli. For a detailed derivation
of the coordinate system see their work in Ref. \cite{aquilanti1997qmh}.

At the heart of democratic coordinates is a rotation from a space fixed frame
to a body fixed frame:%
\begin{equation}
\bm{\varrho}=\bm{D}^T\left(  \alpha,\beta,\gamma\right)  \bm{\varrho}_{bf}
\label{BF_rot}%
\end{equation}
where $\bm{\varrho}$ is the matrix of Jacobi vectors defined in Eq.
(\ref{coord_mats}), $\bm{\varrho}_{bf}$ is the set of body fixed Jacobi
coordinates, and $D\left(  \alpha,\beta,\gamma\right)  $ is an Euler rotation
matrix defined in the standard way as%
\begin{equation}
\bm{D}=\left[
\begin{array}
[c]{ccc}%
\cos\alpha & -\sin\alpha & 0\\
\sin\alpha & \cos\alpha & 0\\
0 & 0 & 1
\end{array}
\right]  \left[
\begin{array}
[c]{ccc}%
\cos\beta & 0 & \sin\beta\\
0 & 1 & 0\\
-\sin\beta & 0 & \cos\beta
\end{array}
\right]  \left[
\begin{array}
[c]{ccc}%
\cos\gamma & -\sin\gamma & 0\\
\sin\gamma & \cos\gamma & 0\\
0 & 0 & 1
\end{array}
\right]  . \label{Euler_rot}%
\end{equation}
The \textquotedblleft$\symbol{126}$" in Eq. (\ref{BF_rot}) indicates a transpose
has been taken.

The body-fixed coordinates are defined in a system whose axes are defined by
the principle moments of inertia, $I_{1},I_{2}$ and $I_{3}$. In this
coordinate system the body-fixed Jacobi coordinates are given by%
\begin{equation}
\bm{\varrho}_{bf}=\bm{\Pi}\bm{D}\left(  \phi_{1},\phi_{2},\phi_{3}\right)  ,
\label{Dem_BF}%
\end{equation}
where $\bm{D}$ is defined in the same way as in Eq. (\ref{Euler_rot}) with
$\phi_{1},$ $\phi_{2}$ and $\phi_{3}$ replacing $\alpha,$ $\beta,$ and
$\gamma$. $\Pi$ is a $3\times3$ diagonal matrix whose diagonals are given by
$\xi_{1},$ $\xi_{2}$ and $\xi_{3}$ which are parameterized by the hyperradius
and two hyperangles $\Theta_{1}$ and $\Theta_{2}$:%
\begin{align}
\xi_{1}  &  =\dfrac{R}{\sqrt{3}}\cos\Theta_{1},\nonumber\\
\xi_{2}  &  =\dfrac{R}{\sqrt{3}}\sqrt{3\sin^{2}\Theta_{1}\sin^{2}\Theta
_{2}+\cos^{2}\Theta_{1}},\label{Pi_diag}\\
\xi_{3}  &  =\dfrac{R}{\sqrt{3}}\sqrt{3\sin^{2}\Theta_{1}\cos^{2}\Theta
_{2}+\cos^{2}\Theta_{1}}.\nonumber
\end{align}
To avoid double counting and to allow for different chiralities, $\Theta_{1}$
and $\Theta_{2}$ are restricted to $0\leq\Theta_{1}\leq\pi$ and $0\leq
\Theta_{2}\leq\pi/4.$ With this parameterization the moments of inertia are
given by%
\begin{align}
\dfrac{I_{1}}{\mu}  &  =\xi_{2}^{2}+\xi_{3}^{2}=\dfrac{R^{2}}{3}\left(
2+\sin^{2}\Theta_{1}\right)  ,\nonumber\\
\dfrac{I_{2}}{\mu}  &  =\xi_{1}^{2}+\xi_{3}^{2}=\dfrac{R^{2}}{3}\left(
3\sin^{2}\Theta_{1}\cos^{2}\Theta_{2}+2\cos^{2}\Theta_{1}\right)
,\label{moments_inert}\\
\dfrac{I_{3}}{\mu}  &  =\xi_{1}^{2}+\xi_{2}^{2}=\dfrac{R^{2}}{3}\left(
3\sin^{2}\Theta_{1}\sin^{2}\Theta_{2}+2\cos^{2}\Theta_{1}\right)  .\nonumber
\end{align}
The hyperradius in terms of the principle moments of inertia can be then
written as $R^{2}=\xi_{1}^{2}+\xi_{2}^{2}+\xi_{3}^{2}=\left(  I_{1}%
+I_{2}+I_{3}\right)  /2\mu.$

With this parameterization, all 8 hyperangles have been defined. The first
three are the Euler angles $\left\{  \alpha,\beta,\gamma\right\}  $, which are
external degrees of freedom describing solid rotations of the four body
system. The two angles, $\Theta_{1}$ and $\Theta_{2}$, defined in Eq.
(\ref{Pi_diag}), describe the overall $x,$ $y$ and $z$ extent of the four-body
system in the body fixed frame. From Eq. (\ref{moments_inert}), if $\Theta
_{1}=0,\pi,$ then the principle moments of inertia are all equal, i.e.
$I_{1}=I_{2}=I_{3}$, meaning that the four particles are arranged at the
vertices of a regular tetrahedron. When $\Theta_{1}=\pi/2$, Eq. (\ref{Pi_diag})
shows that the particles are in a planar configuration. The remaining angles,
$\left\{  \phi_{1},\phi_{2},\phi_{3}\right\}  $, are kinematic rotations
within the system, and coalescence points and operations like particle
exchange are described in these angles. Broadly speaking, the democratic
angles $\Theta_{1}$ and $\Theta_{2}$ can be thought of as correlating the
overall $x$, $y$, and $z$ spatial extent of the four-body system in the
body-fixed frame, while the kinematic angles $\phi_{1}$, $\phi_{2}$, and
$\phi_{3}$ parameterize the internal configuration of the particles.

Since transformations from one Jacobi set to another are merely rotation
matrices (sometimes combined with an inversion), the democratic
parameterization can always be written in the same form for any given Jacobi
coordinate system. For example, the symmetry coordinates [Eq.
(\ref{symm_coords})] can be transformed into the second set interaction
coordinates [Eq. (\ref{interact_coords2})] using the kinematic rotation defined
by Eq. (\ref{Kin_rot_s_i2}). If the democratic parameterization defined in Eqs.
(\ref{BF_rot}) and (\ref{Dem_BF}) is used, this transformation reads%
\begin{align*}
\bm{\varrho}^{i2}  &  =\bm{D}\left(  \alpha,\beta,\gamma\right)  \bm{\Pi
}\bm{D}^T\left(  \phi_{1},\phi_{2},\phi_{3}\right)  \bm{U}_{s\rightarrow
i2},\\
&  =\bm{D}\left(  \alpha,\beta,\gamma\right)  \bm{\Pi}\bm{D}^T\left(
\phi_{1}^{\prime},\phi_{2}^{\prime},\phi_{3}^{\prime}\right)
\end{align*}
where we have used the fact that the product of two rotations in 3D is itself a
rotation. From this it is clear that within a given type of Jacobi coordinate
(H-type or K-type), all coalescence points are equally well described. This is
an important feature as it does not appear in Delves type coordinates, which
are strongly dependent on which Jacobi tree is used to define them. For the
purposes of this paper, to make symmetrization of the wave function easier, we
will always define the body fixed coordinates in terms of the symmetry Jacobi
system $\bm{\varrho}^{s}$.

Putting this all together, the body-fixed Jacobi vectors in terms of the
internal hyperangles can be defined:%
\begin{align}
\rho_{1x}^{s}  &  =\dfrac{R}{\sqrt{3}}\cos\Theta_{1}\left(  \cos\phi_{1}%
\cos\phi_{2}\cos\phi_{3}-\sin\phi_{1}\sin\phi_{3}\right)  ,\nonumber\\
\rho_{1y}^{s}  &  =\dfrac{R}{\sqrt{3}}\sqrt{\sin^{2}\Theta_{1}\sin^{2}%
\Theta_{2}+\cos^{2}\Theta_{1}}\left(  \sin\phi_{1}\cos\phi_{2}\cos\phi
_{3}+\cos\phi_{1}\sin\phi_{3}\right)  ,\nonumber\\
\rho_{1z}^{s}  &  =\dfrac{-R}{\sqrt{3}}\sqrt{\sin^{2}\Theta_{1}\cos^{2}%
\Theta_{2}+\cos^{2}\Theta_{1}}\sin\phi_{2}\cos\phi_{3},
\label{Symm_dem_coords}\\
\rho_{2x}^{s}  &  =\dfrac{-R}{\sqrt{3}}\cos\Theta_{1}\left(  \cos\phi_{1}%
\cos\phi_{2}\sin\phi_{3}+\sin\phi_{1}\cos\phi_{3}\right)  ,\nonumber\\
\rho_{2y}^{s}  &  =\dfrac{-R}{\sqrt{3}}\sqrt{\sin^{2}\Theta_{1}\sin^{2}%
\Theta_{2}+\cos^{2}\Theta_{1}}\left(  \sin\phi_{1}\cos\phi_{2}\sin\phi
_{3}-\cos\phi_{1}\cos\phi_{3}\right)  ,\nonumber\\
\rho_{2z}^{s}  &  =\dfrac{R}{\sqrt{3}}\sqrt{\sin^{2}\Theta_{1}\cos^{2}%
\Theta_{2}+\cos^{2}\Theta_{1}}\sin\phi_{2}\sin\phi_{3},\nonumber\\
\rho_{3x}^{s}  &  =\dfrac{R}{\sqrt{3}}\cos\Theta_{1}\cos\phi_{1}\sin\phi
_{2},\nonumber\\
\rho_{3y}^{s}  &  =\dfrac{R}{\sqrt{3}}\sqrt{\sin^{2}\Theta_{1}\sin^{2}%
\Theta_{2}+\cos^{2}\Theta_{1}}\sin\phi_{1}\sin\phi_{2},\nonumber\\
\rho_{3z}^{s}  &  =\dfrac{R}{\sqrt{3}}\sqrt{\sin^{2}\Theta_{1}\cos^{2}%
\Theta_{2}+\cos^{2}\Theta_{1}}\cos\phi_{2},\nonumber\\
0  &  \leq\Theta_{1}\leq\pi;0\leq\Theta_{2}\leq\dfrac{\pi}{4},\nonumber\\
0  &  \leq\phi_{1},\phi_{2},\phi_{3}\leq\pi.\nonumber
\end{align}
The restriction on the range of the internal hyperangles is to avoid double
counting configurations and allows for configurations of different chirality .

By moving into democratic coordinates, the dimensionality of the four-body
problem can be decreased from 9 to 6, but, as with many simplifications, there
is a cost. This cost comes in the form of the differential volume element
$d\Omega$ \cite{aquilanti1997qmh}:
\begin{align}
d\Omega &  =\left(  d\alpha\sin\beta d\beta d\gamma\right)  \dfrac{\sqrt
{3}\cos^{3}\Theta_{2}\sin^{3}\Theta_{2}\cos2\Theta_{2}\sin^{9}\Theta_{1}%
}{\left[  \left(  \cos^{2}\Theta_{2}+3\sin^{2}\Theta_{1}\cos^{2}\Theta
_{2}\right)  \left(  \cos^{2}\Theta_{2}+3\sin^{2}\Theta_{1}\sin^{2}\Theta
_{2}\right)  \right]  ^{1/2}}\label{volume_el}\\
&  \times d\Theta_{1}d\Theta_{2}d\phi_{1}\sin\phi_{2}d\phi_{2}d\phi
_{3}.\nonumber
\end{align}
The first factor is purely from the Euler angle rotation and will always yield
a factor of $8\pi^{2}$ for functions that are independent of $\alpha,$ $\beta$
and $\gamma$.

Another price that is paid using democratic coordinates comes in the form of
the hyperangular momentum operator, $\Lambda^{2}$. In terms of the democratic
hyperangles, $\Lambda^{2}$ is quite complex and can be found in Ref.
\cite{aquilanti1997qmh}:%
\begin{align*}
\Lambda^{2}=  &  -\Delta\left(  \Theta_{1},\Theta_{2}\right)  +2\mu
R^{2}\left\{  \dfrac{I_{1}}{2\left(  I_{2}-I_{3}\right)  ^{2}}\left(
L_{1}^{2}+J_{1}^{2}\right)  \right. \\
&  +\dfrac{I_{2}}{2\left(  I_{1}-I_{3}\right)  ^{2}}\left(  L_{2}^{2}%
+J_{2}^{2}\right)  +\dfrac{I_{3}}{2\left(  I_{1}-I_{2}\right)  ^{2}}\left(
L_{3}^{2}+J_{3}^{2}\right) \\
&  +\dfrac{2\left[  I_{1}^{2}-\left(  I_{2}-I_{3}\right)  ^{2}\right]  ^{1/2}%
}{\left(  I_{2}-I_{3}\right)  ^{2}}L_{1}J_{1}\\
&  +\dfrac{2\left[  I_{2}^{2}-\left(  I_{1}-I_{3}\right)  ^{2}\right]  ^{1/2}%
}{\left(  I_{1}-I_{3}\right)  ^{2}}L_{2}J_{2}\\
&  +\left.  \dfrac{2\left[  I_{3}^{2}-\left(  I_{1}-I_{2}\right)  ^{2}\right]
^{1/2}}{\left(  I_{1}-I_{2}\right)  ^{2}}L_{3}J_{3}\right\}  ,
\end{align*}
where $\bm{J}$ is the total angular momentum operator, and
\begin{align*}
\Delta\left(  \Theta_{1},\Theta_{2}\right)  =  &  \dfrac{1}{\sin^{7}\Theta
_{1}}\dfrac{\partial}{\partial\Theta_{1}}\sin^{7}\Theta_{1}\dfrac{\partial
}{\partial\Theta_{1}}+\dfrac{2}{\sin^{2}\Theta_{1}}\left[  \dfrac{\partial
^{2}}{\partial\Theta_{2}^{2}}+\cot\Theta_{1}\left(  \dfrac{4}{\sin^{2}%
2\Theta_{2}}-1\right)  \dfrac{\partial}{\partial\Theta_{1}}\right] \\
&  +\dfrac{4}{\sin^{2}\Theta_{1}}\left\{  \dfrac{1}{4\sin4\Theta_{2}}%
\dfrac{\partial}{\partial\Theta_{2}}\sin4\Theta_{2}\dfrac{\partial}%
{\partial\Theta_{2}}+\dfrac{2}{3}\cot^{2}\Theta_{1}\left[  \dfrac{1+3\cos
^{2}2\Theta_{2}}{\sin^{2}2\Theta_{2}}\right.  \right. \\
&  \times\left.  \left.  \left(  \dfrac{1}{4}\dfrac{\partial^{2}}%
{\partial\Theta_{2}^{2}}+\dfrac{\cot2\Theta_{2}}{2}\dfrac{\partial}%
{\partial\Theta_{2}}\right)  -\dfrac{1}{\sin4\theta_{2}}\dfrac{\partial
}{\partial\Theta_{2}}\right]  +\cot\Theta_{1}\cot2\Theta_{2}\dfrac{\partial
}{\partial\Theta_{2}}\dfrac{\partial}{\partial\Theta_{2}}\right\}  .
\end{align*}
Terms in $J_{i}^{2}$ are centrifugal contributions, terms in $L_{i}J_{i}$ are
Coriolis contributions and terms in $L_{i}^{2}$ are contributions from
internal kinematic angular momentum with%
\[
\bm{L}=i\hbar\left[
\begin{array}
[c]{ccc}%
\sin\phi_{1}\cot\phi_{2} & \cos\phi_{1} & -\dfrac{\sin\phi_{1}}{\cos\phi_{2}%
}\\
\cos\phi_{1}\cot\phi_{2} & -\sin\phi_{1} & -\dfrac{\sin\phi_{1}}{\sin\phi_{2}%
}\\
1 & 0 & 0
\end{array}
\right]  \left[
\begin{array}
[c]{c}%
\dfrac{\partial}{\partial\phi_{1}}\\
\dfrac{\partial}{\partial\phi_{2}}\\
\dfrac{\partial}{\partial\phi_{3}}%
\end{array}
\right]  .
\]
Fortunately, the methods for evaluating matrix elements in what follows will
not directly require this form of the hyperangular momentum, but it is
included here for completeness.

The final element needed from the democratic coordinates is the inter-particle
spacing. The ability to define these will be necessary to describe pairwise
interactions and correlations. Using Eqs. (\ref{Jac_coord_Htype}),
(\ref{tranform_Eq}) and (\ref{Dem_BF}),
\begin{align}
\left\vert \bm{r}_{12}\right\vert ^{2}  &  =\sqrt{\dfrac{\mu}{\mu_{12}}%
}\left[  \left(  \bm{\varrho}_{bf}^{s}\bm{U}_{s\rightarrow i1}\right)
^{\dag}\left(  \bm{\varrho}_{bf}^{s}\bm{U}_{s\rightarrow i1}\right)
\right]  _{11},\label{interpart12}\\
\left\vert \bm{r}_{13}\right\vert ^{2}  &  =\sqrt{\dfrac{\mu}{\mu_{13}}%
}\left[  \bm{\varrho}_{bf}^{s\dag}\bm{\varrho}_{bf}^{s}\right]
_{11},\label{interpart13}\\
\left\vert \bm{r}_{14}\right\vert ^{2}  &  =\sqrt{\dfrac{\mu}{\mu_{14}}%
}\left[  \left(  \bm{\varrho}_{bf}^{s}\bm{U}_{s\rightarrow i2}\right)
^{\dag}\left(  \bm{\varrho}_{bf}^{s}\bm{U}_{s\rightarrow i2}\right)
\right]  _{11},\label{interpart14}\\
\left\vert \bm{r}_{23}\right\vert ^{2}  &  =\sqrt{\dfrac{\mu}{\mu_{23}}%
}\left[  \left(  \bm{\varrho}_{bf}^{s}\bm{U}_{s\rightarrow i2}\right)
^{\dag}\left(  \bm{\varrho}_{bf}^{s}\bm{U}_{s\rightarrow i2}\right)
\right]  _{22},\label{interpart23}\\
\left\vert \bm{r}_{24}\right\vert ^{2}  &  =\sqrt{\dfrac{\mu}{\mu_{24}}%
}\left[  \bm{\varrho}_{bf}^{s\dag}\bm{\varrho}_{bf}^{s}\right]
_{22},\label{interpart24}\\
\left\vert \bm{r}_{34}\right\vert ^{2}  &  =\sqrt{\dfrac{\mu}{\mu_{23}}%
}\left[  \left(  \bm{\varrho}_{bf}^{s}\bm{U}_{s\rightarrow i1}\right)
^{\dag}\left(  \bm{\varrho}_{bf}^{s}\bm{U}_{s\rightarrow i1}\right)
\right]  _{22}, \label{interpart34}%
\end{align}
where $\left[  \text{ }\right]  _{ij}$ indicates the $ij$th element of a
matrix. In this equation, only the body fixed Jacobi coordinates from Eq.
(\ref{Dem_BF}) are used. This is because the unitary Euler rotation used to
rotate into the body fixed frame is the same for all Jacobi coordinates
canceling out the $\left\{  \alpha,\beta,\gamma\right\}  $ dependence in the
inter-particle spacings.

Figure \ref{coalplots} shows the surfaces in $\left\{  \phi_{1},\phi_{2}%
,\phi_{3}\right\}  $ for constant $r_{ij}$ in a planar configuration for
$\Theta_{2}=\pi/4,\pi/6,$ and $\pi/12$ for equal mass particles. The $\phi
_{1}$ coordinate axis has been transformed to $\phi_{1}-\pi\Theta\left(
\phi_{1}-\pi/2\right)  $, where $\Theta\left(  x\right)  $ is the unit step
function, to emphasize the symmetry of the surfaces. The red surfaces
correspond to the interacting particles in the four-fermion system
($r_{12},r_{14},r_{23}$ and $r_{34}$) while the blue surfaces correspond to
the identical fermions ($r_{13}$ and $r_{24}$). The identical particle
surfaces surround a coalescence point that must be a Pauli exclusion node in
the final four-body wave function. The simple nature of these coalescence
points makes clear the reason for choosing to base the democratic coordinates
on the symmetry Jacobi vectors. The red surfaces will play an important role
in the pairwise interaction as these surfaces outline the valleys of the
potential. As the system becomes more linear ($\Theta_{2}$ becomes smaller) it
can be seen that the surfaces become broader in the $\phi_{1}$ direction. In
fact, when $\Theta_{2}=0$ (in perfectly linear configurations) these surfaces
become independent of $\phi_{1}$.
\section{Hyperspherical harmonics}
\label{Sec:Hyperharms}

Strictly speaking an overview of hyperspherical coordinates is not really
needed for this thesis, as there are many excellent existing works on the
subject (see for instance
\cite{delves1959tag,delves1960tag,Avery,SmirnovShitikova,cavagnero1986eca,bohn_esry_greene_hsbec}%
). This section is included here for completeness in order to provide a more complete foundation for the
 variational basis functions used in this review.  

To begin, consider a $d$
dimensional Cartesian space whose coordinate axes are given by $\left\{
x_{i}\right\}  _{i=1}^{d}$. For the majority of this review these coordinates
are considered to be the components of a set of Jacobi vectors or the
components of a set of trap centered vectors, but for now we proceed with
a more abstract approach. 
The basic concept of the hyperradius is introduced here as%
\begin{equation}
R^{2}=\sum_{i=1}^{d}x_{i}^{2}. \label{Eq:hyperr_def}%
\end{equation}
While this definition is used here, often a mass scaling will be
inserted. For instance in a trap centered system of equal mass atoms an extra
factor of $1/N$, where $N$ is the number of atoms, will be used to simplify
the interpretation of the hyperradius. For the purposes of this section,
though, this definition will be adequate. With Eq. (\ref{Eq:hyperr_def}), the
$d$ dimensional Laplacian can be rewritten in terms of the hyperradius
\cite{Avery,SmirnovShitikova}:%
\begin{equation}
\nabla^{2}=\sum_{i=1}^{d}\dfrac{\partial^{2}}{\partial x_{i}^{2}}=\dfrac
{1}{R^{d-1}}\dfrac{\partial}{\partial R}R^{d-1}\dfrac{\partial}{\partial
R}-\dfrac{{\Lambda}^{2}}{R^{2}}. \label{Eq:Laplacian}%
\end{equation}
In this equation ${\Lambda}$ is called the hyperangular momentum, or
grand angular momentum operator, the square of which is given by%
\begin{align}
{\Lambda}^{2}  &  =\sum_{i>j}-\left\vert \Lambda_{ij}\right\vert
^{2},\label{Eq:hyperangmoment_def}\\
\Lambda_{ij}  &  =x_{i}\dfrac{\partial}{\partial x_{j}}-x_{j}\dfrac{\partial
}{\partial x_{i}}.\nonumber
\end{align}
Already the $d$ dimensional Laplacian has a rather pleasing form reminiscent
of its 3D counterpart. In fact if $d$ is taken to $3$, Eq.
(\pageref{Eq:Laplacian}) reduces exactly to the three dimensional Laplacian in
spherical coordinates, and ${\Lambda}$ becomes merely the normal
spatial angular momentum operator. To proceed from here a way of defining the
remaining $d-1$ degrees of freedom in terms of angles is needed.

The hyperangular momentum operator in terms of hyperangular coordinates can be
found by using the fact that each subset of Cartesian components is itself a
Cartesian vector space. With that in mind, consider the hyperspherical tree
given by Fig. \ref{delvesfig2}. By writing the Laplacian for each subspace in
terms of the sub-hyperradii $R_{1}$ and $R_{2}$ and the sub-hyperangular
momentum operators ${\Lambda}_{1}$ and ${\Lambda}_{2}$ the total
hyperangular momentum operator can be extracted \cite{Avery}. It is%
\begin{align}
{\Lambda}^{2}  &  =\dfrac{-1}{\sin^{\left(  d_{1}-1\right)  /2}%
\alpha\cos^{\left(  d_{2}-1\right)  /2}\alpha}\dfrac{\partial^{2}}%
{\partial\alpha^{2}}\sin^{\left(  d_{1}-1\right)  /2}\alpha\cos^{\left(
d_{2}-1\right)  /2}\alpha\label{Eq:Hyperangmoment_delves}\\
&  +\dfrac{{\Lambda}_{1}^{2}+\left(  d_{1}-1\right)  \left(
d_{1}-3\right)  /4}{\sin^{2}\alpha}+\dfrac{{\Lambda}_{2}^{2}+\left(
d_{2}-1\right)  \left(  d_{2}-3\right)  /4}{\cos^{2}\alpha}-\dfrac{\left(
d-1\right)  \left(  d-3\right)  +1}{4}.\nonumber
\end{align}
where $\alpha$ is defined as in Eq. (\ref{subradiuscoor}) and ${\Lambda
}_{i}$ is the sub-hyperangular momentum of the subspace of dimension $d_{i}$.
If one of the subspaces corresponds to a single Cartesian component then the
sub-hyperangular momentum for that space is zero, i.e. if $d_{i}=1$ then
${\Lambda}_{i}^{2}=0$. To find ${\Lambda}_{1}^{2}$
(${\Lambda}_{2}^{2}$), ones needs only to apply Eq.
(\ref{Eq:Hyperangmoment_delves}) recursively to each subspace. In this way,
there is a sub-hyperangular momentum operator associated with each node in any
given hyperspherical tree.

It is useful to be able to diagonalize the hyperangular
momentum operator. The eigenfunctions of ${\Lambda}^{2}$ are detailed
in several references (See Refs.
\cite{MacekJPB1968,cavagnero1986eca,SmirnovShitikova} for example), and the
method of constructing them is given in Appendix A. These
functions, $Y_{\lambda\mu}\left(  \Omega\right)  $, are called hyperspherical
harmonics. Their eigenvalue equation is%
\begin{equation}
{\Lambda}^{2}Y_{\lambda\mu}\left(  \Omega\right)  =\lambda\left(
\lambda+d-2\right)  Y_{\lambda\mu}\left(  \Omega\right)  ,
\label{Eq:hyperharm_eigeq}%
\end{equation}
where $\lambda=0,1,2,...$ is the hyperangular momentum quantum number. The
index $\mu$ enumerates the degeneracy for each $\lambda$ and can be thought of
as the collection of sub-hyperangular momentum quantum numbers that result
from a given tree. Hyperspherical harmonics are also constructed as to
diagonalize the sub-hyperangular momenta of each node in a given
hyperspherical tree, e.g.%
\begin{equation}
{\Lambda}_{1}^{2}Y_{\lambda\mu}\left(  \Omega\right)  =\lambda
_{1}\left(  \lambda_{1}+d_{1}-2\right)  Y_{\lambda\mu}\left(  \Omega\right)  ,
\label{Eq:subhyperharm_eigeq}%
\end{equation}
where $\lambda_{1}=0,1,2,...$ is the sub-hyperangular momentum quantum number
associated with $\Lambda_{1}^{2}$. The total hyperangular momentum quantum
number $\lambda$ is limited by the relation%
\begin{equation}
\lambda=\left\vert \lambda_{1}\right\vert +\left\vert \lambda_{2}\right\vert
+2n, \label{Eq:lamda_restrict}%
\end{equation}
where $n$ is a non-negative integer. The absolute values in this case are
there to allow for when either $d_{1}$ or $d_{2}$ are $2.$ In this special
case the hyperangular momentum quantum number $\lambda_{i}$ associated with
the two dimensional subspace can be negative, as with the magnetic quantum
number, $m$, in spherical polar coordinates. Equation (\ref{Eq:lamda_restrict})
only applies if both $d_{1}$ and $d_{2}$ are greater than $1$. If, for
instance $d_{2}=1$, then the restriction takes on the form%
\[
\lambda=\left\vert \lambda_{1}\right\vert +n.
\]

The behavior illustrated in Eq. (\ref{Eq:subhyperharm_eigeq}) clearly
demonstrates why the parameterization shown in Fig. \ref{delves_3} is useful.
Each three dimensional spherical polar subtree will have a spatial angular
momentum and z-projection associated with it, e.g.%
\[
l_{i}^{2}Y_{\lambda\mu}\left(  \Omega\right)  =l_{i}\left(  l_{i}+1\right)
Y_{\lambda\mu}\left(  \Omega\right)  ,
\]
where $l_{i}^{2}$ is the square of the angular momentum operator for the $i$th
vector. This property allows for addition of angular momentum in the normal
way, through sums over magnetic quantum numbers and Clebsch-Gordan
coefficients. Now that the hyperspherical harmonics are defined, it is useful
to examine a simple example applying them.

\section{Jacobi Coordinate systems and Kinematic rotations}
\label{App:Jac_coords}
Here we detail the variety of coordinate systems that are used to describe the
four-fermion problem in the adiabatic hyperspherical framework and the
necessary transformations to describe one set in terms of another. The
coordinate systems used here are not only needed to describe correlations
between particles, they also allow the system to be reduced in dimensionality
by removing the center of mass motion and moving into a body fixed frame.
H-type Jacobi coordinates are constructed by considering the separation vector
for two two-body subsystems, and the separation vector between the centers of
mass of those two subsystems. K-type Jacobi coordinates are constructed in an
iterative way by first constructing a three body coordinate set, and then taking the separation vector between the fourth
particle and the center of mass of the three particle sub-system.

\subsection{Jacobi coordinates: H-trees vs K-trees: fragmentation and symmetry
considerations}
\label{HKtreeJacobi}

The first and most obvious symmetry in the four-body Hamiltonian 
is that of translational symmetry. By describing the system in
the center of mass frame, the dimensionality of the system can be reduced from
$d=12$ to $d=9$. This is done with the use of Jacobi coordinates. In the
interest of brevity, We consider here only those coordinates directly relevant
to the four-fermion problem. These coordinates may be broken into two sets,
H-type and K-type, shown schematically in Fig. \ref{Htype_Ktype}.

H-type Jacobi coordinates are constructed by considering the separation vector
for two two-body subsystems, and the separation vector between the centers of
mass of those two subsystems, i.e.%
\begin{align}
\bm{\rho}_{1}^{H\sigma}  &  =\sqrt{\dfrac{\mu_{ij}}{\mu}}\left(  \bm{r}%
_{i}-\bm{r}_{j}\right)  ,\nonumber\\
\bm{\rho}_{2}^{H\sigma}  &  =\sqrt{\dfrac{\mu_{kl}}{\mu}}\left(  \bm{r}%
_{k}-\bm{r}_{l}\right)  ,\label{Jac_coord_Htype}\\
\bm{\rho}_{3}^{H\sigma}  &  =\sqrt{\dfrac{\mu_{ij,kl}}{\mu}}\left(
\dfrac{m_{i}\bm{r}_{i}+m_{j}\bm{r}_{j}}{m_{i}+m_{j}}-\dfrac{m_{k}\bm{r}%
_{k}+m_{l}\bm{r}_{l}}{m_{k}+m_{l}}\right)  ,\nonumber\\
\bm{\rho}_{cm}  &  =\dfrac{\left(  m_{1}\bm{r}_{1}+m_{2}\bm{r}_{2}%
+m_{3}\bm{r}_{3}+m_{4}\bm{r}_{4}\right)  }{m_{1}+m_{2}+m_{3}+m_{4}%
},\nonumber\\
\mu_{ij}  &  =\dfrac{m_{i}m_{j}}{m_{i}+m_{j}},\mu_{ij,kl}=\dfrac{\left(
m_{i}+m_{j}\right)  \left(  m_{l}+m_{k}\right)  }{m_{1}+m_{2}+m_{3}+m_{4}%
}.\nonumber
\end{align}
Here the superscript $\sigma$ enumerates the 24 different H-type coordinates
that may be obtained through particle permutation, $\bm{\rho}_{cm}$ is the
position of center of mass of the four-body system, and $\mu$ is an arbitrary
reduced mass for the four-body system. The prefactors in each Jacobi vector,
which are given in terms of the various reduced masses in the problem, are
chosen to give the so-called mass scaled Jacobi vectors. The kinetic energy in
these coordinates can be written as%
\[
-\sum_{i=1}^{4}\dfrac{\hbar^{2}}{2m_{i}}\nabla_{r_{i}}^{2}=-\dfrac{\hbar^{2}%
}{2M}\nabla_{\rho_{cm}}^{2}-\dfrac{\hbar^{2}}{2\mu}\sum_{j=1}^{3}\nabla
_{\rho_{j}}^{2},
\]
where $M$ is the total mass of the four particles. The reduced mass, $\mu$,
can be chosen to preserve the differential volume element for the full 3D
problem, ensuring that $d^{3}\rho_{1}^{\sigma}d^{3}\rho_{2}^{\sigma}d^{3}%
\rho_{3}^{\sigma}d^{3}\rho_{cm}=d^{3}r_{1}d^{3}r_{2}d^{3}r_{3}d^{3}r_{4}$:%
\[
\mu=\left(  \dfrac{m_{1}m_{2}m_{3}m_{4}}{m_{1}+m_{2}+m_{3}+m_{4}}\right)
^{1/3}.
\]
Physically, the H-type coordinates are useful for describing correlations
between two particles, for example a two body bound state or a symmetry
between two particles, or two separate two-body correlations.

When two particles coalesce [e.g. when $\bm{r}_{i}=\bm{r}_{j}$ in Eq.
(\ref{Jac_coord_Htype})], the H-type coordinate system reduces to a three body
system with two of the four particles acting like a single particle with the
combined mass of its constituents:
\begin{align*}
\bm{\rho}_{1}^{H\sigma}  &  =0,\\
\bm{\rho}_{2}^{H\sigma}  &  =\sqrt{\dfrac{\mu_{kl}}{\mu}}\left(  \bm{r}%
_{k}-\bm{r}_{l}\right)  ,\\
\bm{\rho}_{3}^{H\sigma}  &  =\sqrt{\dfrac{\mu_{ij,kl}}{\mu}}\left(  \bm
{r}_{i}-\dfrac{m_{k}\bm{r}_{k}+m_{l}\bm{r}_{l}}{m_{k}+m_{l}}\right)  .
\end{align*}
Locating these ``coalescence points'' on the surface of the hypersphere is
crucial for accurately describing the interactions between particles, and the
coordinate reduction described above will prove useful for the construction of
a variational basis set.


K-type Jacobi coordinates are constructed in an iterative way by first
constructing a three body coordinate set, and then
taking the separation vector between the fourth particle and the center of
mass of the three particle sub-system, yielding%
\begin{align}
\bm{\rho}_{1}^{K\sigma}  &  =\sqrt{\dfrac{\mu_{ij}}{\mu}}\left(  \bm{r}%
_{i}-\bm{r}_{j}\right)  ,\nonumber\\
\bm{\rho}_{2}^{K\sigma}  &  =\sqrt{\dfrac{\mu_{ij,k}}{\mu}}\left(
\dfrac{m_{i}\bm{r}_{i}+m_{j}\bm{r}_{j}}{m_{i}+m_{j}}-m_{k}\bm{r}%
_{k}\right)  ,\label{Jac_coord_Ktype}\\
\bm{\rho}_{3}^{K\sigma}  &  =\sqrt{\dfrac{\mu_{ijk,l}}{\mu}}\left(
\dfrac{m_{i}\bm{r}_{i}+m_{j}\bm{r}_{j}+m_{k}\bm{r}_{k}}{m_{i}+m_{j}+m_{k}%
}-m_{l}\bm{r}_{l}\right)  ,\nonumber\\
\bm{\rho}_{cm}  &  =\dfrac{\left(  m_{1}\bm{r}_{1}+m_{2}\bm{r}_{2}%
+m_{3}\bm{r}_{3}+m_{4}\bm{r}_{4}\right)  }{m_{1}+m_{2}+m_{3}+m_{4}%
},\nonumber\\
\mu_{ij}  &  =\dfrac{m_{i}m_{j}}{m_{i}+m_{j}},\mu_{ij,k}=\dfrac{\left(
m_{i}+m_{j}\right)  m_{k}}{m_{i}+m_{j}+m_{k}},\nonumber\\
\mu_{ijk,l}  &  =\dfrac{\left(  m_{i}+m_{j}+m_{k}\right)  m_{l}}{m_{1}%
+m_{2}+m_{3}+m_{4}}.\nonumber
\end{align}
Again $\sigma$ enumerates the 24 different K-type coordinates that result from
particle permutations. Examining Fig. \ref{Htype_Ktype} shows that K-type
Jacobi coordinate systems are useful for describing correlations between three
particles within the four particle system. In the four fermion system, there
are no weakly bound trimer states meaning that K-type Jacobi coordinates will
not be used here, but the methods described in this report can be easily
generalized to include these type of states. Unless explicitly stated all
Jacobi coordinates from here on will be of the H-type, and for notational
simplicity, we will drop the $H$ superscripts.

\subsubsection{Coalescence points and permutation symmetry}

The proper description of coalescence points is crucial for describing
two-body interactions, but they are also important for describing points of
symmetry. For instance if two identical fermions are on top of one another it
is known that the wave function must vanish at this point owing to the
anti-symmetry of fermionic wave functions. Here, we are concerned with four
fermions in two different \textquotedblleft spin" states. Away from a p-wave
resonance, the interactions between identical fermions can be neglected for
low energy collisions. This means that there are two types of coalescence
points that must be described; two \textquotedblleft symmetry" points, when
two fermions of the same type are on top of each other, and four
\textquotedblleft interaction" points, places where two distinguishable
fermions interact via an s-wave potential.

It might be tempting at this point to choose a single Jacobi coordinate system
and then try to describe the interactions and symmetries in the same
coordinates, but this leads to problems. For instance if it is assumed that
particles 1 and 3 are spin up and particles 2 and 4 are spin down one might
start with coordinates that are simple to anti-symmetrize the system in:%
\begin{align}
\bm{\rho}_{1}^{s}  &  =\sqrt{\dfrac{4^{1/3}}{2}}\left(  \bm{r}_{1}-\bm
{r}_{3}\right)  ,\nonumber\\
\bm{\rho}_{2}^{s}  &  =\sqrt{\dfrac{4^{1/3}}{2}}\left(  \bm{r}_{2}-\bm
{r}_{4}\right)  ,\label{symm_coords}\\
\bm{\rho}_{3}^{s}  &  =\sqrt{4^{1/3}}\left(  \dfrac{\bm{r}_{1}+\bm{r}_{3}%
}{2}-\dfrac{\bm{r}_{2}+\bm{r}_{4}}{2}\right)  ,\nonumber
\end{align}
where it has been assumed that all of the particle masses are equal,
$m_{1}=m_{2}=m_{3}=m_{4}=m$ leaving $\mu=m/4^{1/3}$. The generalization to
distinguishable fermions of different masses is clear. We refer to this Jacobi
coordinate system as the symmetry coordinates for fairly obvious reasons. If a
permutation of two identical fermions is considered, for instance 1 and 3, the
transformation is simple:%
\begin{align}
P_{13}\bm{\rho}_{1}^{s}  &  =-\bm{\rho}_{1}^{s},\label{Eq:perm13symm1}\\
P_{13}\bm{\rho}_{2}^{s}  &  =\bm{\rho}_{2}^{s},\label{Eq:perm13symm2}\\
P_{13}\bm{\rho}_{3}^{s}  &  =\bm{\rho}_{3}^{s}. \label{Eq:perm13symm3}%
\end{align}
Similarly for the exchange of particle 2 and 4,%
\begin{align}
P_{24}\bm{\rho}_{1}^{s}  &  =\bm{\rho}_{1}^{s},\label{Eq:per24symm1}\\
P_{24}\bm{\rho}_{2}^{s}  &  =-\bm{\rho}_{2}^{s},\label{Eq:per24symm2}\\
P_{24}\bm{\rho}_{3}^{s}  &  =\bm{\rho}_{3}^{s}. \label{Eq:per24symm3}%
\end{align}
\qquad The points where two identical fermions coalesce are also simply
described by taking either $\bm{\rho}_{1}^{s}\rightarrow0$ or $\bm{\rho}%
_{2}^{s}\rightarrow0$.

The symmetry coordinates are not suited for describing an interaction between
two distinguishable fermions, for instance 1 and 2. This interaction occurs
around the point $\bm{r}_{1}=\bm{r}_{2}$. In the symmetry coordinates this
means that%
\[
\bm{\rho}_{1}^{s}=\bm{\rho}_{2}^{s}-\sqrt{2}\bm{\rho}_{3}^{s}.
\]
This equation describes a 6 dimensional sheet in the 9 dimensional space,
something that is not easy to describe directly in any basis set. To get
around this problem we introduce two more Jacobi coordinate systems that are
useful for describing interactions,%
\begin{align}
\bm{\rho}_{1}^{i1}  &  =\sqrt{\dfrac{4^{1/3}}{2}}\left(  \bm{r}_{1}-\bm
{r}_{2}\right)  ,\nonumber\\
\bm{\rho}_{2}^{i1}  &  =\sqrt{\dfrac{4^{1/3}}{2}}\left(  \bm{r}_{3}-\bm
{r}_{4}\right)  ,\label{interact_coords1}\\
\bm{\rho}_{3}^{i1}  &  =\sqrt{4^{1/3}}\left(  \dfrac{\bm{r}_{1}+\bm{r}_{2}%
}{2}-\dfrac{\bm{r}_{3}+\bm{r}_{4}}{2}\right)  ,\nonumber
\end{align}
and%
\begin{align}
\bm{\rho}_{1}^{i2}  &  =\sqrt{\dfrac{4^{1/3}}{2}}\left(  \bm{r}_{1}-\bm
{r}_{4}\right)  ,\nonumber\\
\bm{\rho}_{2}^{i2}  &  =\sqrt{\dfrac{4^{1/3}}{2}}\left(  \bm{r}_{3}-\bm
{r}_{2}\right)  ,\label{interact_coords2}\\
\bm{\rho}_{3}^{i2}  &  =\sqrt{4^{1/3}}\left(  \dfrac{\bm{r}_{1}+\bm{r}_{4}%
}{2}-\dfrac{\bm{r}_{2}+\bm{r}_{3}}{2}\right)  .\nonumber
\end{align}
The superscript $i1$ and $i2$ in Eqs. (\ref{interact_coords1}) and
(\ref{interact_coords2}) indicate that these Jacobi coordinates are appropriate
for interactions between distinguishable fermions. For instance, a coalescence
point between particles 1 and 2 is described by $\bm{\rho}_{1}^{i1}%
\rightarrow0$. Another benefit of these coordinates is that they are well
suited to describing a dimer wavefunction. If particles 2 and 3 are in a
weakly bound molecule then the wavefunction for that molecule is only a
function of $\bm{\rho}_{2}^{i2}$.

Using combinations of these three coordinate systems, $\bm{\rho}_{j}^{s}$,
$\bm{\rho}_{j}^{i1}$ and $\bm{\rho}_{j}^{i2}$, can describe all of the
possible two-body correlations of the fermionic system. This assumes that the
system in question is that of four equal mass fermions in two internal states
with s-wave interactions only. However, the method used is quite general. For
instance, K-type coordinates can be chosen to describe the possible three-body
correlations that can arise due to Efimov physics in bosonic systems
\cite{efimov1970,efimov1973,Braaten2006physrep}.

\subsection{Kinematic rotations}

Since we use different Jacobi systems to describe different types of
correlations, a method of transforming between different sets of coordinates
is needed. In the above section, equal mass particles are considered;
extension to arbitrary masses is fairly straightforward. To describe the
kinematic rotations we keep the masses arbitrary and specify for equal masses
later. It is convenient here to deal with transforming all of the Jacobi
coordinates at once. Thus the matrices whose columns are made of the Jacobi
vectors are used:%
\begin{align}
\bm{\varrho}^{s}  &  =\left\{  \bm{\rho}_{1}^{s},\bm{\rho}_{2}^{s}%
,\bm{\rho}_{3}^{s}\right\} \nonumber\\
\bm{\varrho}^{i1}  &  =\left\{  \bm{\rho}_{1}^{i1},\bm{\rho}_{2}^{i1}%
,\bm{\rho}_{3}^{i1}\right\} \label{coord_mats}\\
\bm{\varrho}^{i2}  &  =\left\{  \bm{\rho}_{1}^{i2},\bm{\rho}_{2}^{i2}%
,\bm{\rho}_{3}^{i2}\right\}  .\nonumber
\end{align}
The transformation that takes one coordinate system to another cannot stretch
or shrink the differential volume element, and thus it must be a unitary
transformation. Further, the transformation cannot mix the Cartesian
components of the Jacobi vector, i.e. $\rho_{x}^{i1}$ has no part of $\rho
_{y}^{s}$ in it. This means that the transformation will be a unitary matrix
that acts from the right, e.g.%
\begin{equation}
\bm{\varrho}^{i1}=\bm{\varrho}^{s}\bm{U}_{s\rightarrow i1}.
\label{tranform_Eq}%
\end{equation}

The matrices that perform these operations are called kinematic rotations
\cite{aquilanti1997qmh,kuppermann1997rsr,littlejohn1999qdk}, and they will be
put to extensive use in the calculations that follow. In truth,
transformations between coordinates systems that do not require an inversion
should be considered, but the general principle still holds if improper
rotations are included. Note that all of the matrix elements must be real, so
that the inverse transformation is given merely by the transpose.

We employ a direct \textquotedblleft brute force" method of finding these
matrices where the system of equations given in Eq. (\ref{Jac_coord_Htype}) are
solved for $\bm{r}_{1}$, $\bm{r}_{2},$ $\bm{r}_{3}$ and $\bm{r}_{4}$ in a
given Jacobi system. These normal lab-fixed coordinates can then be inserted
into the definition of the Jacobi coordinates that we wish to describe. The
kinematic rotation can then be extracted from the resulting relations.
Following this procedure gives

\begin{align}
\bm{U}_{s\rightarrow i1}  &  =\left[
\begin{array}
[c]{ccc}%
\tfrac{m_{3}}{m_{1}+m_{3}}\sqrt{\tfrac{\mu_{12}}{\mu_{13}}} & -\tfrac{m_{1}%
}{m_{1}+m_{3}}\sqrt{\tfrac{\mu_{34}}{\mu_{13}}} & \sqrt{\tfrac{\mu_{13}}%
{\mu_{12,34}}}\\
-\tfrac{m_{4}}{m_{2}+m_{4}}\sqrt{\tfrac{\mu_{12}}{\mu_{24}}} & \tfrac{m_{2}%
}{m_{2}+m_{4}}\sqrt{\tfrac{\mu_{34}}{\mu_{24}}} & \sqrt{\tfrac{\mu_{24}}%
{\mu_{12,34}}}\\
\sqrt{\tfrac{\mu_{12}}{\mu_{13,24}}} & \sqrt{\tfrac{\mu_{34}}{\mu_{13,24}}} &
\tfrac{m_{1}m_{4}-m_{2}m_{3}}{\left(  m_{1}+m_{2}\right)  \left(  m_{3}%
+m_{4}\right)  }\sqrt{\tfrac{\mu_{12,34}}{\mu_{13,24}}}%
\end{array}
\right]  ,\label{Kin_rot_s_i1}\\
\bm{U}_{s\rightarrow i2}  &  =\left[
\begin{array}
[c]{ccc}%
\tfrac{m_{3}}{m_{1}+m_{3}}\sqrt{\tfrac{\mu_{14}}{\mu_{13}}} & -\tfrac{m_{1}%
}{m_{1}+m_{3}}\sqrt{\tfrac{\mu_{23}}{\mu_{13}}} & \sqrt{\tfrac{\mu_{13}}%
{\mu_{12,34}}}\\
\tfrac{m_{4}}{m_{2}+m_{4}}\sqrt{\tfrac{\mu_{14}}{\mu_{24}}} & -\tfrac{m_{2}%
}{m_{2}+m_{4}}\sqrt{\tfrac{\mu_{23}}{\mu_{24}}} & -\sqrt{\tfrac{\mu_{24}}%
{\mu_{14,23}}}\\
\sqrt{\tfrac{\mu_{14}}{\mu_{13,24}}} & \sqrt{\tfrac{\mu_{23}}{\mu_{13,24}}} &
\tfrac{m_{1}m_{2}-m_{3}m_{4}}{\left(  m_{2}+m_{3}\right)  \left(  m_{1}%
+m_{4}\right)  }\sqrt{\tfrac{\mu_{14,23}}{\mu_{13,24}}}%
\end{array}
\right]  ,\label{Kin_rot_s_i2}\\
\bm{U}_{i1\rightarrow s}  &  =\left[  \bm{U}_{s\rightarrow i1}\right]
^{T};\text{ }\bm{U}_{i2\rightarrow s}=\left[  \bm{U}_{s\rightarrow
i2}\right]  ^{T},\label{Kin_rot_inv}\\
\bm{U}_{i1\rightarrow i2}  &  =\bm{U}_{i1\rightarrow s}\bm{U}_{s\rightarrow
i2}=\left[  \bm{U}_{s\rightarrow i1}\right]  ^{T}\bm{U}_{s\rightarrow
i2};\text{ }\bm{U}_{i2\rightarrow i1}=\left[  \bm{U}_{i1\rightarrow
i2}\right]  ^{T}. \label{Kin_rot_i1_i2}%
\end{align}
The same method can be used to find the kinematic rotations to other Jacobi
systems, for instance to K-type coordinates.

\section{Implementation of Correlated Gaussian basis set expansion}
\label{Sec:Impl}

\subsection{Symmetrization of the basis functions and evaluation of the
matrix elements} \label{ASymm}

The CG basis functions take the form
\begin{equation}
\label{Basis1I2}
\Phi_A(\bm{x}_{1},\bm{x}_{2},...,\bm{x}_{N})=
\mathcal{S}\left\{\exp(-\frac{1}{2}
\bm{x}^T.\bm{A}.\bm{x})\right\}.
\end{equation}
The symmetrization operator $\mathcal{S}$ can be expanded in a set
of simple particle permutations,
\begin{equation}
\ket{\mathcal{S}(A)}=\sum^{N_p}_{i=1}
\mbox{sgn}(P_i) \ket{P_i(A)}.
\end{equation}
Here, $N_{p}$ is the number of permutations that characterize the
symmetry $\mathcal{S}$. Each of these permutations, $P_i$, has a
sign associated, $\mbox{sgn}(P_i)$, and is a given rearrangement of
the spatial coordinates
\begin{equation}
P_i(\Phi_A(\bm{x}_1,...,\bm{x}_N)=\Phi_A(\bm{x}_{P_i(1)},...,\bm{x}_{P_i(N)})
\end{equation}
The label $i$ characterizes the rearrangement. This rearrangement of
the spatial coordinates is equivalent to a rearrangement of the
interparticle widths $\{d_{ij}\}$ (or the $\{\alpha_{ij}\}$),
\begin{equation}
P_k(\{d_{ij}\})=\{d_{P_k(ij)}\}.
\end{equation}
Therefore, permutation operations can be easily applied and become
transformations of the matrix $A$.

In general the evaluation of the symmetrized matrix elements of an
operator $O$ is,
\begin{equation}
\braket{\mathcal{S}(A)|O|\mathcal{S}(B)}=\sum^{N_p}_{i=1}\sum^{N_p}_{i'=1}
\mbox{sgn}(P_i) \mbox{sgn}(P_{i'})\braket{P_{i'}(A)|O|P_i(A)},
\end{equation}
which implies an $N_p^2$ evaluation of unsymmetrized matrix
elements. Fortunately, if $\mathcal{S}(O)=O$ then,
\begin{equation}
\label{permut} \braket{\mathcal{S}(A)|O|\mathcal{S}(B)}=N_p
\braket{A|O|\mathcal{S}(B)}=N_p \braket{\mathcal{S}(A)|O|B}.
\end{equation}
This property significantly reduces the numerical demands since the
left hand side of Eq.(\ref{permut}) implies $N_p^2$ permutations,
while the right-hand side only implies $N_p$ permutations. All
operators of the Hamiltonian are invariant under the $\mathcal{S}$,
operator and their matrix elements obey Eq.(\ref{permut}).

To obtain density profiles and pair-correlation functions, we use
the delta function operator. A single delta function operator in a
given coordinate is not invariant under this transformation; for
this reason, the computational evaluation is more expensive.
Alternatively, we can create a similar operator as a sum of delta
functions. If the sum of delta functions reflects the symmetry of
the problem, the new operator would be invariant under
$\mathcal{S}$.

The permutation operator clearly depends on the symmetry properties of the constituent particles. For identical bosons and fermions,
\begin{equation}
\mathcal{S}=\sum^{N_p}_{i=1}\alpha_i P_i,
\end{equation}
where $N_p=N!$ and $\alpha_i=1$ for bosons and $\alpha_i=(-1)^p$;
$p=0,1$ is the parity of the operator $P_i$. For two-component
systems (boson-boson, fermion-fermion, or a Bose-Fermi mixture),
\begin{equation}
\mathcal{S}=\sum^{N_{p_1}}_{i_1=1}\sum^{N_{p_2}}_{i_2=1}\alpha_{i_1}
\alpha_{i_2} P_{i_1}P_{i_2},
\end{equation}
where $N_{p_1}=N_1!$, $N_{p_2}=N_2!$, and $N_1$ and $N_2$ are the
number of particles in component 1 and 2, respectively.

The symmetrization operation, if it involves a permutation with a
negative sign, can significantly reduce the accuracy of matrix
elements.  In certain cases, the unsymmetrized matrix elements can
be almost identical. Because of the negative sign of the
permutation, the symmetrized matrix elements can become a
subtraction of very similar numbers. Therefore, accuracy is reduced.
These basis functions are usually unphysical, so it is convenient to
eliminate them. To do this, we evaluate
$|\braket{\mathcal{S}(A)|\mathcal{S}(A)}|/\max(|\braket{P_i(A)|P_i(A)}|)$.
If this is a small number, then the accuracy of the matrix elements
is reduced.  So, in general, we introduce a tolerance of the order
of $10^{-3}$ to determine whether to keep or discard the basis
functions.

\subsection{Evaluation of unsymmetrized basis functions} \label{AEvBasisFunc}

For convenience, we introduce the following simplify notation,
\begin{equation}
\label{BasisDef}
\braket{\bm{x}_1,\cdots,\bm{x}_N|A}=\exp(-\frac{1}{2}\bm{x}^T.\bm{A}.\bm{x}).
\end{equation}

As a simple example, consider the overlap matrix element
\begin{equation}
\label{Over} \braket{A|B}=\int
d\bm{x}_1..d\bm{x}_N\exp(-\frac{1}{2}
\bm{x}^T.(A+B).\bm{x}).
\end{equation}
Since the matrix $A+B$ is real and symmetric, there exists a set of
eigenvectors $\bm{y}=\{\bm{y}_1,...,\bm{y}_N\}$ with
eigenvalues $\{\beta_1,...,\beta_N\}$ that diagonalize the matrix.
In this set of coordinates, Eq.~(\ref{Over}) takes the simple form,
\begin{equation}
\braket{A|B}=(4\pi)^N\int_0^\infty dy_1 y_1^2
e^{-\beta_1 y_1^2/2}...\int_0^\infty dy_N y_N^2 e^{-\beta_N y_N^2/2}
=\left(\frac{(2\pi)^N}{\det(A+B)}\right)^{3/2}.
\end{equation}
Here, we used the product $\beta_1.\beta_2...\beta_N=\det(A+B)$.
These basics steps can be followed to evaluate the remaining matrix
elements.

To evaluate the kinetic energy, we use the following property,
\begin{equation}
\label{BasisKinProp} \braket{A|-\frac{\hbar^2}{2 m}
\nabla^2_{\bm{x}_i} |B}=\frac{\hbar^2}{2
m}\braket{\nabla_{\bm{x}_i} A |\nabla_{\bm{x}_i}B}.
\end{equation}
This property can be simply proven by applying an integration by
parts. Also, it simplifies the matrix element evaluation and
provides an expression which is symmetric in $A$ and $B$. Then, the
matrix element takes the form,
\begin{equation}
\label{BasisKin} \braket{A|-\frac{\hbar^2}{2 m}\sum_i^N
\nabla^2_{\bm{x}_i} |B}=\frac{\hbar2}{2 m}
3\,\mbox{Tr}((A+B)^{-1} A.B) \braket{A|B}
\end{equation}
Another important matrix element, which is similar to
Eq.(~\ref{BasisKin}), is
\begin{equation}
\label{BasisTrap} \braket{A|\bm{x}^T
C\bm{x}|B}=\frac{\hbar^2}{2 m} 3\,\mbox{Tr}((A+B)^{-1} C)
\braket{A|B}.
\end{equation}
Here, $C$ is an arbitrary matrix. This matrix element is used to
calculate the trapping potential energy. In such case, $C=m\omega^2
I/2$, where $I$ is the identity matrix.

Finally, we calculate the matrix element for a two-body central
force:
\begin{equation}
\label{BasisPot} \braket{A|V(\bm{r}_i-\bm{r}_j)|B}=\int d^3r
V(\bm{r}) \braket{A|\delta(b_{ij}^T\bm{x}-r)|B} =
G_{c_{ij}}[V] \braket{A|B},
\end{equation}
where $\bm{r}_i-\bm{r}_j=b_{ij}^T \bm{x}$,
$c^{_1}_{ij}=b_{ij}^T (A+B)^{-1}b_{ij}$, and $G_c[V]$ is the
Gaussian transform of the potential
\begin{equation}
\label{gausstransf} G_c[V]=\left(\frac{c}{2\pi}\right)^{3/2}\int
d^3r V(\bm{r}) e^{-c r^2/2}.
\end{equation}
These matrix elements are enough to describe few-body systems.

\subsection{Jacobi vectors and CG matrices} \label{AJac}

Here, we present the construction of the matrices that
characterize the basis functions in terms of the widths $d_{ij}$. In
the following $\bm{r}=\{\bm{r}_1,...,\bm{r}_N\}$
correspond to Cartesian coordinates, while
$\boldsymbol{\rho}=\{\boldsymbol{\rho}_1,...,\boldsymbol{\rho}_{N-1}\}$
correspond to mass-scaled Jacobi coordinates. First, consider the
basis function with the center of mass included
\begin{equation}
\ket{A}=\Psi_0(R_{CM})\exp\left(-{\sum_{j\ge i}
\frac{(\bm{r}_1-\bm{r}_2)^2}{2d_{ij}^2}}\right)=\exp(-\frac{1}{2}
\bm{r}^T.\bm{A}.\bm{r}).
\end{equation}
In the equal-mass case for $N$ particles, it is more convenient to
simply use Cartesian coordinates. The ground-state--center-of-mass
wave function of particles in a harmonic trap takes, conveniently, a
Gaussian form $\Psi_0(R_{CM})=e^{-N R_{CM}^2/2a_{ho}^2}$. Thus,
$\Psi_0(R_{CM})$ can be written as
$\Psi_0(R_{CM})=e^{-\bm{r}^T.M^{CM}.\bm{r}/2}$, where
$M^{CM}$ is the center-of-mass matrix whose matrix elements are
$M^{CM}_{kl}=1/(N a_{ho}^2)$ for all $k$ and $l$. Then, for each
interparticle distance $r_{ij}$, there exists a matrix $M^{(ij)}$ so
that $r_{ij}^2=\bm{r}^T.M^{(ij)}.\bm{r}$. The matrix
elements of the $M^{(ij)}$ matrices are
$M^{(ij)}_{ii}=M^{(ij)}_{jj}=1$, $M^{(ij)}_{ij}=M^{(ij)}_{ji}=-1$;
the rest are zero, yielding
\begin{equation}
\label{MatConsWCM} A=M^{CM}+\sum_{j\ge i} \frac{1}{d_{ij}^2}
M^{(ij)}.
\end{equation}

In some cases it is important to include the center-of-mass motion.
For example, this allows one to extract single-particle observables
such as density profiles.

If the center of mass is not included, then Eq.~(\ref{MatConsWCM})
can be written as,
\begin{equation}
A=\sum_{j> i} \frac{1}{d_{ij}^2} M^{(ij)}.
\end{equation}

Next, we present the mass-scaled Jacobi vectors
(see Fig.~\ref{Htype_Ktype}) and the corresponding form of the matrices
$M^{(ij)}$.





Using the H-tree Jacobi coordinates defined in~\ref{HKtreeJacobi}, Eq.~(\ref{Jac_coord_Htype}), the interparticle separation distances can be written:
\begin{gather}
\bm{r}_1-\bm{r}_2=\boldsymbol{\rho}_1/d_1,\\
\bm{r}_1-\bm{r}_3=\sqrt{\frac{\mu_3}{\mu}}\left(\boldsymbol{\rho}_3
+\frac{\mu_1 d_3}{m_1 d_1}\boldsymbol{\rho}_1 -\frac{\mu_2 d_3}{m_3 d_2}\boldsymbol{\rho}_2 \right),\\
\bm{r}_1-\bm{r}_4=\sqrt{\frac{\mu_3}{\mu}}\left(\boldsymbol{\rho}_3
+\frac{\mu_1 d_3}{m_1 d_1}\boldsymbol{\rho}_1 +\frac{\mu_2 d_3}{m_4 d_2}\boldsymbol{\rho}_2 \right),\\
\bm{r}_2-\bm{r}_3=\sqrt{\frac{\mu_3}{\mu}}\left(\boldsymbol{\rho}_3
-\frac{\mu_1 d_3}{m_2 d_1}\boldsymbol{\rho}_1 +\frac{\mu_2 d_3}{m_3 d_2}\boldsymbol{\rho}_2 \right),\\
\bm{r}_2-\bm{r}_4=\sqrt{\frac{\mu_3}{\mu}}\left(\boldsymbol{\rho}_3
-\frac{\mu_1 d_3}{m_2 d_1}\boldsymbol{\rho}_1 +\frac{\mu_2 d_3}{m_4 d_2}\boldsymbol{\rho}_2 \right),\\
\bm{r}_3-\bm{r}_4=\boldsymbol{\rho}_2/d_2.
\end{gather}

For both the $N=3$ and $N=4$ systems, the interparticle distances
can be written in terms of the Jacobi vectors
\begin{equation}
\bm{r}_i-\bm{r}_j=\sum_k c_k^{(ij)}\boldsymbol{\rho}_k.
\end{equation}
Now we can write the matrices $M^{(ij)}$ in these Jacobi vectors
that describe an interparticle distance. The matrix elements of
these matrices are simply  $M^{(ij)}_{kl}= c_k^{(ij)}c_l^{(ij)}$.


\subsection{Selection of the basis set} \label{SelectBasisSet}

There are different strategies for selecting a basis set. If the
numbers of dimensions of the system we are studying is not that
large, then we can try to generate a large basis set that is
complete enough to describe several eigenstates at different
interaction strengths.

The Gaussian widths $d_{ij}$ are selected randomly and cover a range
of values from $d_0$ to the trap length $a_{ho}$. Specifically, the
$d_{ij}$ are selected randomly using a Gaussian distribution of
range 1 and then scaled to three different distances: $d_0$, an
intermediate distance $\sqrt{d_0 a_{ho}}$, and $a_{ho}$. These three
distances are fixed once the interparticle potential range $d_0$ is
fixed.

The basis set selection depends on the correlation we want to
describe. So, the selection process changes depending whether the
particles are bosons or fermions. For fermions, when there is no
trimer formation, basis functions with more than two particles close
together are not important.

For example, the algorithm for the selection of the basis functions
for a two-component four-fermion system divides the basis into three
parts: the first subbasis generates $d_{ij}$, which are all of the
order of $a_{ho}$; they are useful for describing weakly interacting
states. The second subbasis generates two $d_{ij}$ of the order of
$d_0$ or $\sqrt{d_0 a_{ho}}$ and the rest of the order of $a_{ho}$;
they are useful to describe dimer-dimer states. The third subbasis
has one $d_{ij}$ of the order of $d_0$ or $\sqrt{d_0 a_{ho}}$ and
the rest of the order of $a_{ho}$. They are useful to describe
dimer--two-free-atom states.

%

\subsection{Controlling Linear dependence} \label{ALinDep}

A large basis set is usually needed to describe several eigenstates in a wide range of interactions.
Since the  basis set is over complete and the basis
functions are chosen semi-randomly, the resulting basis can have linear dependence
problems.  In our implementation we eliminate the linear dependence by reducing the size of basis set.



To do this, we first diagonalize the overlap matrix and then
eliminate the eigenstates with negative or low eigenvalues. The
remaining eigenstates form an orthonormal basis set. Finally, we
transform the Hamiltonian to the new orthonormal basis set.

The threshold for the elimination can be selected automatically
taking into account the lowest eigenvalue. If the lowest eigenvalue
$O_1$ is small and positive, the tolerance can be selected as, for
example, $10^3 O_1$. If $O_1$ is negative and the magnitude is
large, then the basis set has a lot of linear dependence, and it is
more convenient to change the initial basis set.

\subsection{Stochastical variational method} \label{SVM}

The SVM has been developed in the context of nuclear physics to
solve few-body
problems~\cite{varga1995psf,varga1996svm,varga1997sfb}. It allows a
systematically improvement of the basis set. A detailed discussion
of the implementation of the SVM will not be presented here but can
be found in Refs.~\cite{sorensen2005cmb,suzuki1998sva}. In the
following, we present the main concepts of the SVM.

The SVM is based on the variational nature of the spectrum obtained
by a basis set expansion. Consider a basis set of size $D$ with
eigenvalues $\{\epsilon_1,...,\epsilon_D\}$, if we add a new basis
function then the new eigenvalues $\{\lambda_1,...,\lambda_{D+1}\}$
obey $\lambda_1\le \epsilon_1\le\lambda_2
...\epsilon_D\le\lambda_{D+1}$. Here, we assume that both set of
eigenvalues are arranged in increasing order. Thus, by adding a new
basis, all the $D$ eigenvalues should decrease or remain the same.
Therefore, the lower the new eigenvalues are the better the
improvement of the basis set. Thus we can test the utility of the
added basis function by considering the improvement in the
eigenvalues.

In most cases, we are not interested in improving the complete
spectrum. To select which states or energies we want to improve, we
can construct an appropriate minimization function. This function
would depend only on the energies we want to improve and is
minimized by the SVM.

In order to optimize the basis set, the SVM utilizes a trial an
error procedure. Starting from an initial basis set of size $D$,
several basis functions are selected stochastically and added, one
at a time, to the basis set. For each $D+1$ basis set, the new
eigenvalues are evaluated. The basis function that produces the best
improvement of the selected energies is kept while the remaining
basis functions are discarded. The initial basis function is then
increased by one and the trial an error procedure is repeated.

If this procedure is continued indefinitely
the size of the basis set has become large and the calculations
become forbiddingly slow. Therefore, it is convenient to increase
the basis up to a reasonable size and then continue the optimization
process without increasing the basis size. This optimization can be
carried out by a refinement process. Instead of adding a new basis
function, we test the importance of the basis functions of the basis
set. The trial and error procedure is then applied to each of the
functions of the basis set.

For the SVM procedure to be efficient, the evaluation of both the
matrix elements and the eigenvalues need to be fast and accurate. It
is particularly important to obtain very accurate matrix elements
because the improvement due to a single basis function is usually
very small and can only be evaluated reliably if the matrix element
are very accurate. The matrix element evaluation in the CG and CGHS
is both fast and accurate making these methods particularly suitable
for SVM optimization.

Also, the evaluation of the eigenvalues can be significantly speeded
up in the trial and error procedure. The basis functions are added
or replaced one by one which allowing us to reduce the evaluation of
the eigenvalues to a root finding procedure. This root finding
procedure is much faster than any diagonalization procedure.

The SVM automatically takes care of the selection of the basis
function. Also, it tries to avoid linear dependence in the basis set
by constraining the normalized overlap between any two basis
function, i.e., $O_{12}/\sqrt{O_{11}O_{22}}$, to be below some
tolerance $O_{max}$. The tolerance $O_{max}$ is usually selected
between 0.95 and 0.99. For example, the size of the basis set of
$N=3$ and $4$ can be increased up to 700 and 8000 respectively
without introducing significant linear dependence.

\section{Dimer-Dimer Relaxation Rates}
\label{App:DD_relax}

\begin{figure}[htbp]
\begin{center}
\includegraphics[width=2.5in]{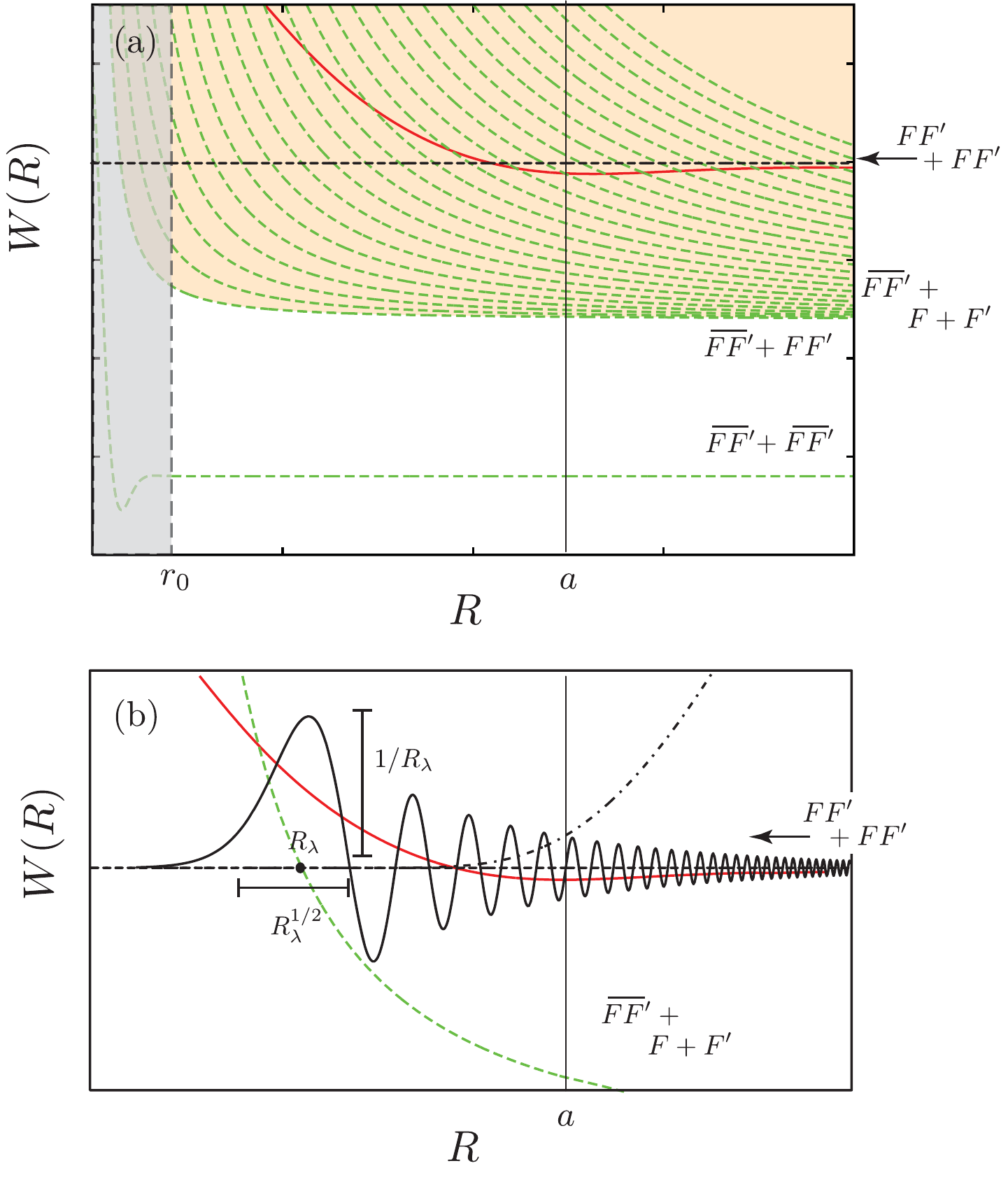}
\end{center}
\caption{(a){} A schematic of the channels involved involved inthe dimer-dimer
relazaxtion process is shown. The dimer-dimer potential (red curve) goes
through an inifinite number of crossing with deeply bound states (green dashed
curves). (b) The hyperradial behavior of the outgoing wavefunction is shown.}%
\label{relax_chann}%
\end{figure}

In this appendix I present the derivation of the dimer-dimer relaxation rate
used in Section 6.5. This process occurs when the two dimers collide causing
at least one of the dimers to relax to a deeply bound state. The difference of
the binding energies is then releases as kinetic energy. This process can be
pictured in the hyperspherical picture as an infinite series of very closely
spaced crossings between the dimer-dimer channel and channels consisting of a
deeply bound dimer and two free particles. This near continuum of crossings is
shown schematically in Fig. \ref{relax_chann}(a).

Using Fermi's golden rule between the initial dimer-dimer state and the final
states gives%
\begin{equation}
V_{rel}^{dd}\propto\sum_{\lambda}\left\vert \left\langle \Psi_{dd}\left(
R;\Omega\right)  \left\vert V\left(  R,\Omega\right)  \right\vert
\Psi_{\lambda}\left(  R,\Omega\right)  \right\rangle \right\vert
^{2},\label{Eq:Fermigrule}%
\end{equation}
where $\Psi_{dd}\left(  R;\Omega\right)  $ is the dimer-dimer wavefunction,
$V\left(  R,\Omega\right)  $ is the interaction potential and $\Psi_{k}$ is
the $\lambda$th deeply bound dimer state. I now will assume that the
dimer-dimer wavefunction is approximated by
\begin{equation}
\Psi_{dd}\left(  R;\Omega\right)  \approx F_{dd}\left(  R\right)  \Phi
_{dd}\left(  R;\Omega\right)  ,\label{Eq:singlechannass}%
\end{equation}
where $\Phi_{dd}\left(  R;\Omega\right)  $ is the dimer-dimer hyperangular
channel function and $F_{dd}\left(  R\right)  $ is the hyperradial
wavefunction resulting from the single channel approximation. I further assume
that the outgoing deeply bound dimer wavefunction can be written as%
\begin{equation}
\Psi_{k}\left(  R,\Omega\right)  \approx\psi\left(  r_{12}\right)
\theta_{\lambda}\left(  \vec{r}_{34},\vec{r}_{12,34}\right)
\label{Eq:deep_d_ass}%
\end{equation}
where $\psi\left(  r_{12}\right)  $ is the wavefunction for an s-wave deeply
bound dimer and $\theta_{\lambda}\left(  \vec{r}_{34},\vec{r}_{12,34}\right)
$ is the free space behavior of the resulting three particle system.

Examining one of the terms from the sum in Eq. (\ref{Eq:Fermigrule}) with a
single two-body interaction gives%
\begin{equation}
V_{rel}^{dd\left(  \lambda\right)  }\propto\left\vert \int F_{dd}\left(
R\right)  \Phi_{dd}\left(  R;\Omega\right)  V_{23}\left(  r_{23}\right)
\psi\left(  r_{12}\right)  \theta_{\lambda}\left(  \vec{r}_{34},\vec
{r}_{12,34}\right)  dRd\Omega\right\vert ^{2},\label{Eq:Fermigrule1term}%
\end{equation}
where $V_{rel}^{dd\left(  k\right)  }$ is the contribution to the relaxation
rate by the $\lambda$th term in Eq. (\ref{Eq:Fermigrule})$.$ The first thing to
notice is in this is that the factor $V_{23}\left(  r_{23}\right)  \psi\left(
r_{12}\right)  $ is non-zero only when particles $1,2,$ and $3$ are in close
proximity, and when particles $1,2,$ and $3$ are in close proximity the
remaining degrees of freedom are simplified as well,
\begin{equation}
\vec{r}_{34}\approx C\vec{r}_{12,34}.
\end{equation}
This means that the wavefunction $\theta\left(  \vec{r}_{34},\vec{r}%
_{12,34}\right)  $ can be rewritten as
\begin{equation}
\theta_{\lambda}\left(  \vec{r}_{34},\vec{r}_{12,34}\right)  \approx
G_{\lambda}\left(  R\right)  f_{\lambda}\left(  \Omega\right)  ,
\end{equation}
where $G_{\lambda}$ and $f_{\lambda}$ are the hyperradial and hyperangular
behavior associated with the $\lambda$th outgoing channel. A further
simplification can be made by realizing that $f_{\lambda}\left(
\Omega\right)  $ must be independent of $\Omega$ when particles $1,2,$ and $3$
are in close proximity because the total wavefunction must have zero spatial
angular momentum. Rewriting (\ref{Eq:Fermigrule1term}) with these
simplifications yields%
\begin{equation}
V_{rel}^{dd\left(  \lambda\right)  }\propto\left\vert \int F_{dd}\left(
R\right)  G_{k}\left(  R\right)  \int\Phi_{dd}\left(  R;\Omega\right)
V_{23}\left(  r_{23}\right)  \psi\left(  r_{12}\right)  d\Omega dR\right\vert
^{2}.
\end{equation}
The hyperangular integral is approximated by the probability that
three-particles are close to each other in the dimer-dimer channel function.

And example of $G_{\lambda}$ is shown in Fig. \ref{relax_chann}(b). Away from
the classical turning point $G_{\lambda}$ oscillates very rapidly. This fast
oscillation will generally cancel out meaning the main contribution to the
hyperradial integral is from the region near the classical turning point
$R_{\lambda}.$ Putting this all together yields%
\begin{equation}
V_{rel}^{dd}\propto\sum_{\lambda}\dfrac{\left\vert F_{dd}\left(  R_{\lambda
}\right)  \right\vert ^{2}}{R_{\lambda}}\mathcal{F}\left(  R_{\lambda}\right)
,\label{Eq:Vrelcontib}%
\end{equation}
where $\mathcal{F}\left(  R_{\lambda}\right)  $ is the probability that three
particles are in close proximity in the dimer-dimer channel function. The
final step in this derivation is to turn the sum over $\lambda$ into an
integral over $R_{\lambda}$,%
\begin{equation}
V_{rel}^{dd}\propto\int\rho\left(  R_{\lambda}\right)  \dfrac{\left\vert
F_{dd}\left(  R_{\lambda}\right)  \right\vert ^{2}}{R_{\lambda}}%
\mathcal{F}\left(  R_{\lambda}\right)  dR_{\lambda}%
\end{equation}
where $\rho\left(  R_{\lambda}\right)  $ is the, nearly constant, density of
states. This is possible due to the near-continuum nature of the outgoing
states. By inserting the WKB approximation wavefunction for $F_{dd}\left(
R\right)  $, the result of Eq. (\ref{Eq:Relaxation_rate}) is obtained, i.e.%
\begin{equation}
V_{rel}^{dd}\propto\int\rho\left(  R\right)  \dfrac{P_{WKB}\left(  R_{\lambda
}\right)  }{\kappa\left(  R\right)  R}\mathcal{F}\left(  R\right)
dR.\label{Eq:Vrelresult}%
\end{equation}


\begin{thebibliography}{161}
\bibitem[{Jav()}]{Javfootnote}
\bibinfo{note}{The $(3N-7)$ numerical integration results from the following
  reasoning: initially we have $3N$ numerical integration but 3 dimensions are
  removed by decoupling the center of mass motion, 3 dimension are removed
  fixing the Euler angles and 1 dimension is removed fixing $R$.}
\end{thebibliography}

\end{document}